\documentclass[letterpaper,twocolumn,10pt]{article}
\usepackage{usenix-2020-09}
\usepackage{epsfig,endnotes}
\usepackage[english]{babel}
\usepackage{blindtext}
\usepackage{adjustbox}
\usepackage{array}

\usepackage[ruled]{algorithm2e}
\usepackage{tikz}
\usepackage{enumitem,kantlipsum}
\usepackage{mathrsfs,amsmath}
\usepackage{subfigure}
\usepackage{mathtools}
\usepackage{multirow}
\usepackage{diagbox}
\usepackage{flushend} 
\usepackage{url}
\usepackage{pifont}
\usepackage{color}
\usepackage{xcolor}

\newcommand{\ours}{\textsf{CurvingLoRa}\xspace}
\newcommand{\parabreak}{\vspace*{1.00ex minus 0.25ex}\noindent}

\definecolor{mypink}{cmyk}{0, 0.7808, 0.4429, 0.1412}

\begin{document}

\title{CurvingLoRa to Boost LoRa Network Capacity via Concurrent Transmission}
\author{
{\rm Chenning Li}\\
Michigan State University
\and
{\rm Xiuzhen Guo}\\
Tsinghua University
\and
{\rm Longfei Shangguan}\\
Microsoft and University of Pittsburgh
\and
{\rm Zhichao Cao}\\
Michigan State University
\and
{\rm Kyle Jamieson}\\
Princeton University

} 

\maketitle

\begin{abstract}
\noindent{}LoRaWAN has emerged as an appealing technology to connect IoT devices but it functions without explicit coordination among transmitters, which can lead to many packet collisions as the network scales. State-of-the-art work proposes various approaches to deal with these collisions, but most functions only in high signal-to-interference ratio (SIR) conditions and thus does not scale to many scenarios where weak receptions are easily buried by stronger receptions from nearby transmitters.  In this paper, we take a fresh look at LoRa's physical layer, revealing that its underlying linear chirp modulation fundamentally limits the capacity and scalability of concurrent LoRa transmissions. We show that by replacing linear chirps with their non-linear counterparts, we can boost the capacity of concurrent LoRa transmissions and empower the LoRa receiver to successfully receive weak transmissions in the presence of strong colliding signals. Such a non-linear chirp design further enables the receiver to demodulate fully aligned collision symbols --- a case where none of the existing approaches can deal with.
We implement these ideas in a holistic LoRaWAN stack based on the USRP N210 software-defined radio platform.
Our head-to-head comparison with two state-of-the-art research systems and a standard LoRaWAN baseline demonstrates that \ours improves the network throughput by 1.6--7.6$\times$ while simultaneously sacrificing neither power efficiency nor noise resilience.
An open-source dataset and code will be made available before publication.

\end{abstract}

\section{Introduction}
\label{sec-intro}

As we gradually reach a cyber-physical world where everything near and far is  connected wirelessly, a fundamental question worth discussing is which wireless technologies are best suited for achieving this goal. While Wi-Fi and cellular networks have proved their success in provisioning high-throughput wireless connectivity in small geographic areas, a remaining challenge is connecting those low-power IoT devices deployed in wide areas. Most of these devices are powered by batteries and thus require minimal communication overhead.

LoRaWAN~\cite{alliance2020lorawan} has emerged as a competitive solution because of its ability to communicate over a large distance. Central to LoRaWAN is a dedicated PHY-layer design that leverages Chirp Spread Spectrum (CSS) modulation to facilitate packet decoding in extremely harsh SNR conditions (which can be as low as $-20$~dB \cite{LoSee}). 
Coupling with a long-term duty cycling mechanism, a deployed LoRa node can last for a few years with a single dry-cell battery.
These dual merits of low-power and long-range make LoRaWAN an attractive solution for IoT connectivity outdoors.

Unlike Wi-Fi, which uses {\it carrier sensing}~\cite{cali2000dynamic} to avoid packet collisions, LoRaWAN functions without explicit coordination due to its stringent power budget. It instead adopts the least restrictive MAC protocol---ALOHA~\cite{abramson1970aloha}---that allows participating nodes to transmit immediately once they wake up.\footnote{LoRaWAN recently released a new feature called Channel Activity Detection
(CAD) that allows the receiver to scan the channel before transmitting. CAD however incurs extra power consumption and thus may not be applicable to rural deployments where battery replacement is usually infeasible.} Such {\it laissez-faire} transmission inevitably causes packet collision when multiple LoRa nodes transmit simultaneously, resulting in retransmissions that can drain the battery of collision nodes and crowd the precious wireless spectrum on the unlicensed band. 
Packet collisions are exacerbated with increased network size, thus reducing throughput and hence fundamentally challenging the scalability of LoRa networks in real deployment.

\begin{figure}[t]
    \centering
    \includegraphics[width=3.2in]{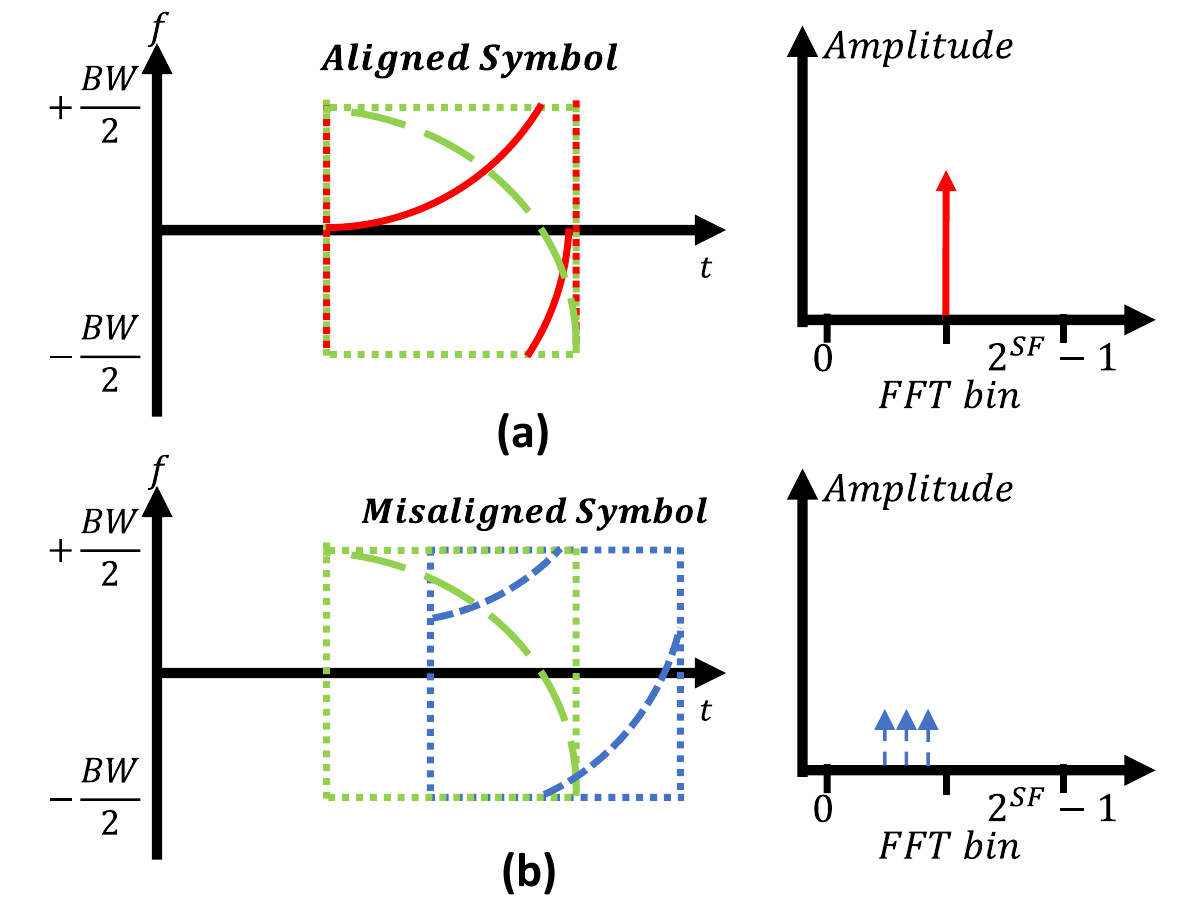}
    \vspace{-4mm}
    \caption{An illustration of \ours's energy converging and scattering effect. (a): The energy of a non-linear chirp symbol will converge to a specific frequency point when it aligns with the down-chirp. (b): The energy will spread into multiple FFT bins when this non-linear chirp mismatches with the down-chirp.}
    \label{fig:intro}
    \vspace{-4mm}
\end{figure}


\parabreak{}In this paper, we take a fresh look at the physical layer design of LoRaWAN and reveal that the underlying {\it linear chirp based modulation} fundamentally limits the capacity and scalability of concurrent LoRa transmissions.
We present \ours, a simple but effective PHY-layer design to boost the LoRa network capacity by simply replacing the standard linear chirp with its nonlinear chirp counterpart.

\ours is based on a unique property of non-linear chirps, which we term the {\it energy scattering and converging effect}. When a non-linear up-chirp symbol misaligns with the non-linear down-chirp during demodulation, their multiplication will {\it spread} the power of the non-linear up-chirp symbol into multiple FFT bins where the associated energy peaks are inherently weak. 
Such energy scattering effect will show up as long as the non-linear up-chirp is not well aligned with the down-chirp (The theoretical analysis is in Appendix \ref{appendix-2}). 
In contrast, when this non-linear up-chirp is well aligned with the down-chirp, its signal power will {\it converge} to a specific frequency point, leading to a strong energy peak after FFT (shown in Figure~\ref{fig:intro}).

This allows the receiver to manipulate the signal-to-interference ratio (SIR\footnote{Defined as the ratio between the power of the targeted LoRa chirp and the power of remaining concurrent LoRa chirps.}) of each collision symbol for reliable demodulation. 
In contrast, the power of linear chirps will be always {\it converging} to a single frequency point regardless of its alignment with the down-chirp in demodulation process. This energy converging effect fundamentally limits the decodability of linear chirps in the presence of collisions.

We analyze the performance of non-linear chirp and compare it with its linear chirp counterpart in various SNRs, SIRs, and symbol overlapping ratio conditions.
We show that such a non-linear chirp remarkably improve the transmission concurrency while retaining both the high power efficiency and strong noise resilience as linear chirp does (\S\ref{sec-design}).
We then design a holistic PHY-layer to realize non-linear chirp modulation and demodulation (\S\ref{sec-system}) and implement it on software-defined radios for evaluation.
The experimental results show that compared to two state-of-the-art systems~\cite{mLoRa,tong_combating_2020}, \ours can effectively improve the network throughput by $1.6-6.6\times$ and $2.8-7.6\times$ in an indoor and outdoor deployment, respectively. We make following contributions:

\begin{itemize}[leftmargin=*]
    \item We reveal that the PHY-layer design of LoRaWAN fundamentally limits the transmission concurrency and propose a simple but effective solution. Our approach takes the advantage of power scattering effect of non-linear chirps to enable LoRa concurrent transmissions in extreme SNR, SIR, and symbol overlapping ratio conditions.
    \item Through both theoretical analysis and experimental validation, we demonstrate that \ours outperforms both the current practice and the standard LoRaWAN while sacrificing neither the power efficiency nor noise resilience. These desired properties make non-linear chirp a potential complementary solution to its linear chirp counterpart. 
    \item We design a holistic PHY-layer and implement it on software defined radios platform to evaluate \ours in various real-world scenarios. The results confirm  that the \ours can greatly improve the network throughput.
\end{itemize}

\section{Related Work}
\label{sec-related-work}


Prior work on resolving LoRa collisions has followed a common theme: exploring the unique features of collided LoRa symbols in the time domain~\cite{mLoRa,oct,xia2019ftrack}, frequency domain~\cite{colora,tong_combating_2020,xu_fliplora_2020,eletreby2017empowering}, or both~\cite{hu2020sclora,shahid2021concurrent}.
For instance, mLoRa~\cite{mLoRa} observes that collisions usually start with a stretch of interference free bits on the packet header. The receiver can thus decode these uncontaminated bits first and then leverage successive interference cancellation~\cite{gollakota2009interference,mollanoori2013uplink} to decode the collided bits iteratively.
Choir~\cite{eletreby2017empowering} uses the frequency variation caused by oscillator imperfection to map bits to each individual LoRa transmitter.
FTrack~\cite{xia2019ftrack} jointly exploits the distinct tracks on the frequency domain and misaligned symbol edges in the time domain to separate collisions.
By combining spectra obtained from different parts within each symbol, CIC~\cite{shahid2021concurrent} exploits the sub-symbols that provide both time and frequency resolution to cancel out the interference under low SNR conditions (\emph{e.g.}, above $-10$~dB).

While above ideas have demonstrated their efficacy, they still face two scalability issues that fundamentally challenge their applicability in practice:
First, the vast majority of these approaches do not scale to a large number of concurrent transmissions. For instance, mLoRa~\cite{mLoRa} and FTrack~\cite{xia2019ftrack} barely support up to three concurrent transmissions to maintain a symbol error rate less than 0.1. While Choir~\cite{eletreby2017empowering} improved over the above methods, it does not scale to more than ten devices. Although NScale~\cite{tong_combating_2020} can support tens of concurrent transmissions, it requires the overlap ratio between different symbols to be lower than a rigid threshold, which is unlikely to be held in practice given laissez-faire LoRa transmissions.
Second, none of the foregoing approaches scale well to near-far deployment scenarios. This is due to the fact that after dechirping, the weak reception from a remote transmitter produces a tiny FFT peak that is likely to be buried by strong FFT peaks from LoRa nodes that are closer to the receiver. 

Although successive interference cancellation (SIC) can be leveraged to deal with this near-far issue, it functions only in high SNR conditions due to the following reasons.
First, due to Carrier Frequency Offset (CFO) and Sampling Frequency Offset (SFO), the phase of received chirp symbols is likely to distort by a certain degree.
This phase distortion is critical to the success of signal recovery in SIC but is difficult to estimate in low SNR conditions~\cite{colora}.
Second, SIC suffers from hardware imperfections~\cite{tong_combating_2020}, which is common on low-end IoT devices. The symbol recovering error accumulates gradually and is likely to fail the SIC in the end. In addition, the impact of ambient RF noise on SIC, particularly the parameter estimation for signal reconstruction, gets exacerbated at low SNR conditions.
Instead of leveraging new features on time or frequency domain to combat LoRa collisions, \ours addresses this issue from a fresh new angle and designs LoRa signals to facilitate concurrent transmissions. 


Nonlinear frequency modulation has been widely used in radar systems. Lesnik \emph{et al.} \cite{lesnik2008modification} demonstrate that using nonlinear frequency modulation can enhance signal sensitivity.
Doerry \emph{et al.} \cite{doerry2006generating} and Benson \emph{et al.} \cite{benson2015modern} details the way to build non-linear chirp receivers.
Kahn \emph{et al.} \cite{khan2013performance} and 
Hosseini \emph{et al.} \cite{hosseini2019nonlinear, hosseini2021nonlinear} use nonlinear chirps in Multi user orthogonal chirp spread spectrum (MU-OCSS) communication system to mitigate the multiple access interference problem.
Wang \emph{et al.} \cite{wang2015non} propose to use non-linear chirps for communication systems of binary orthogonal keying mode.
In contrast to these efforts, \ours explores a new possibility of using non-linear chirps to improve reception of concurrent LoRa-like signals.

\section{Motivation}
\label{sec-background}

In this section, we briefly introduce the LoRa physical layer and then analyze the pros and cons of linear chirps (\S\ref{subsec-background-2}). A discussion on the limitations of resolving linear-chirp LoRa collisions follows (\S\ref{subsec-background-3}).

\label{subsec-background-1}
\begin{figure}[t]
\centering
\subfigure[Baseline up-chirp]{ 
	\includegraphics[width=0.33\columnwidth]{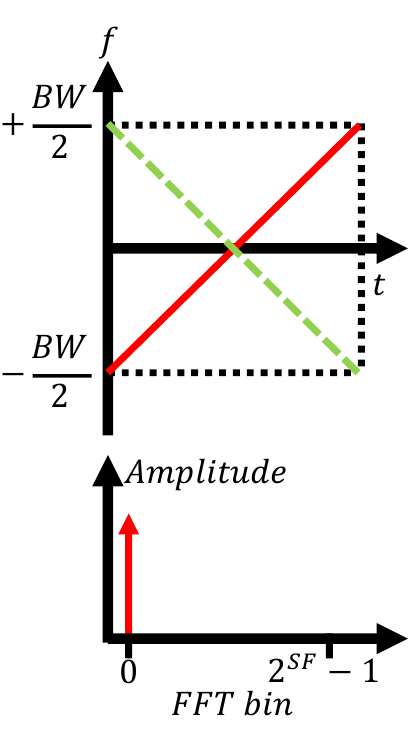}
        \label{subfig-baseline-chirp}
}
\hspace{-0.06\columnwidth}
\subfigure[Shifted up-chirp]{
	\includegraphics[width=0.33\columnwidth]{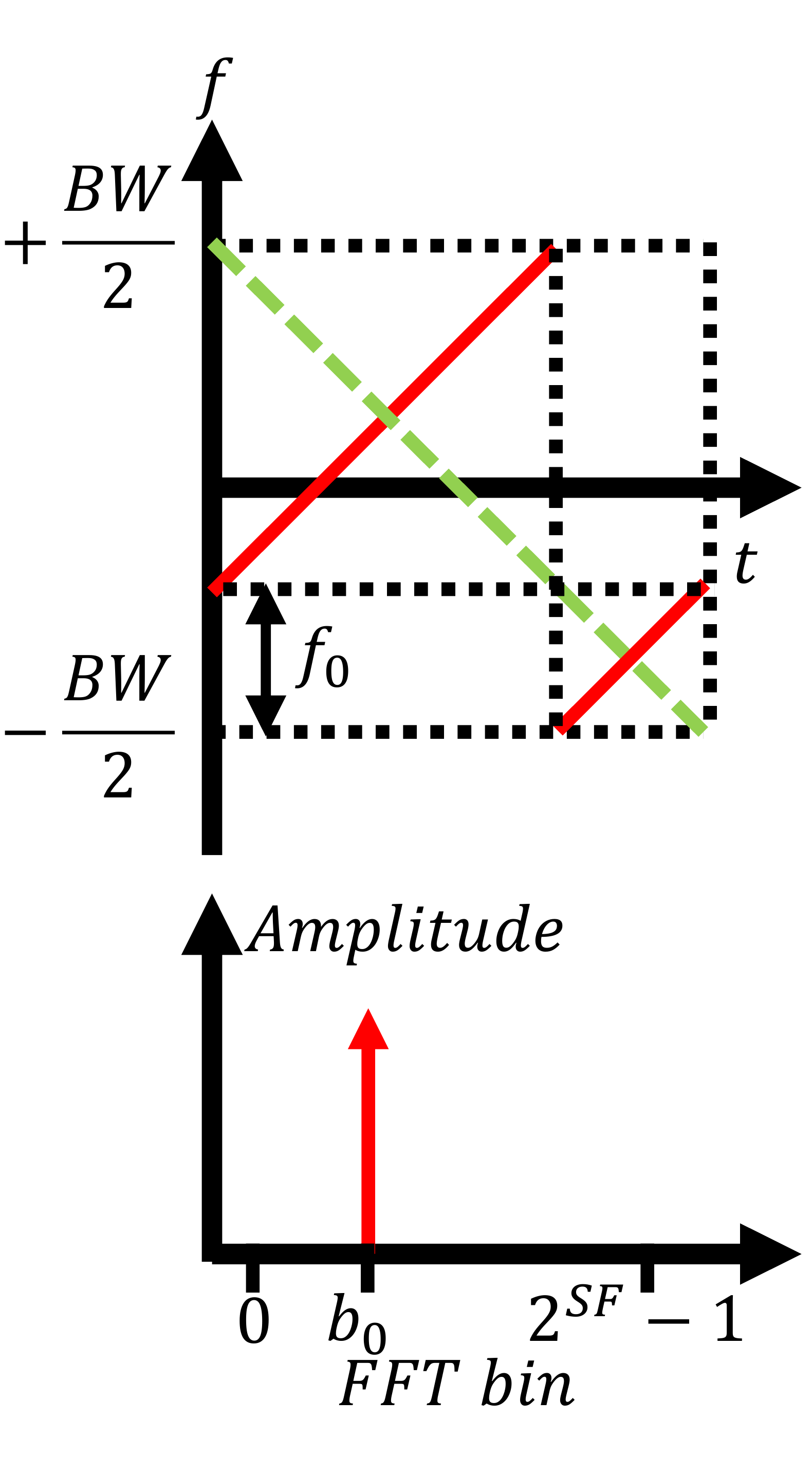}
        \label{subfig-shifted-chirp}
}
\hspace{-0.06\columnwidth}
\subfigure[Symbol collision]{
	\includegraphics[width=0.33\columnwidth]{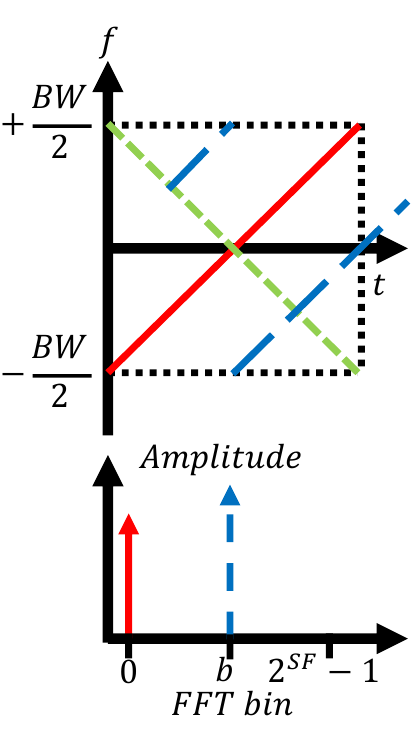}
        \label{subfig-symbol-collision}       
}
\vspace{-3mm}
\caption{LoRa PHY-layer design. 
{\normalfont 
(a): the multiplication of an up-chirp and a down-chirp leads to an energy peak on a specific FFT. (b): The position of this energy peak varies with the initial frequency offset of the incoming up-chirp. (c): Two collided symbols have separate energy peaks on FFT bins.}}
\label{fig-lora-preliminary}
\vspace{-3mm}
\end{figure}

\begin{figure*}[t] \label{fig:sic}
    \centering
    \subfigure[SER fluctuates with SNR in the absence of collisions.]{
    \includegraphics[width=0.495\columnwidth]{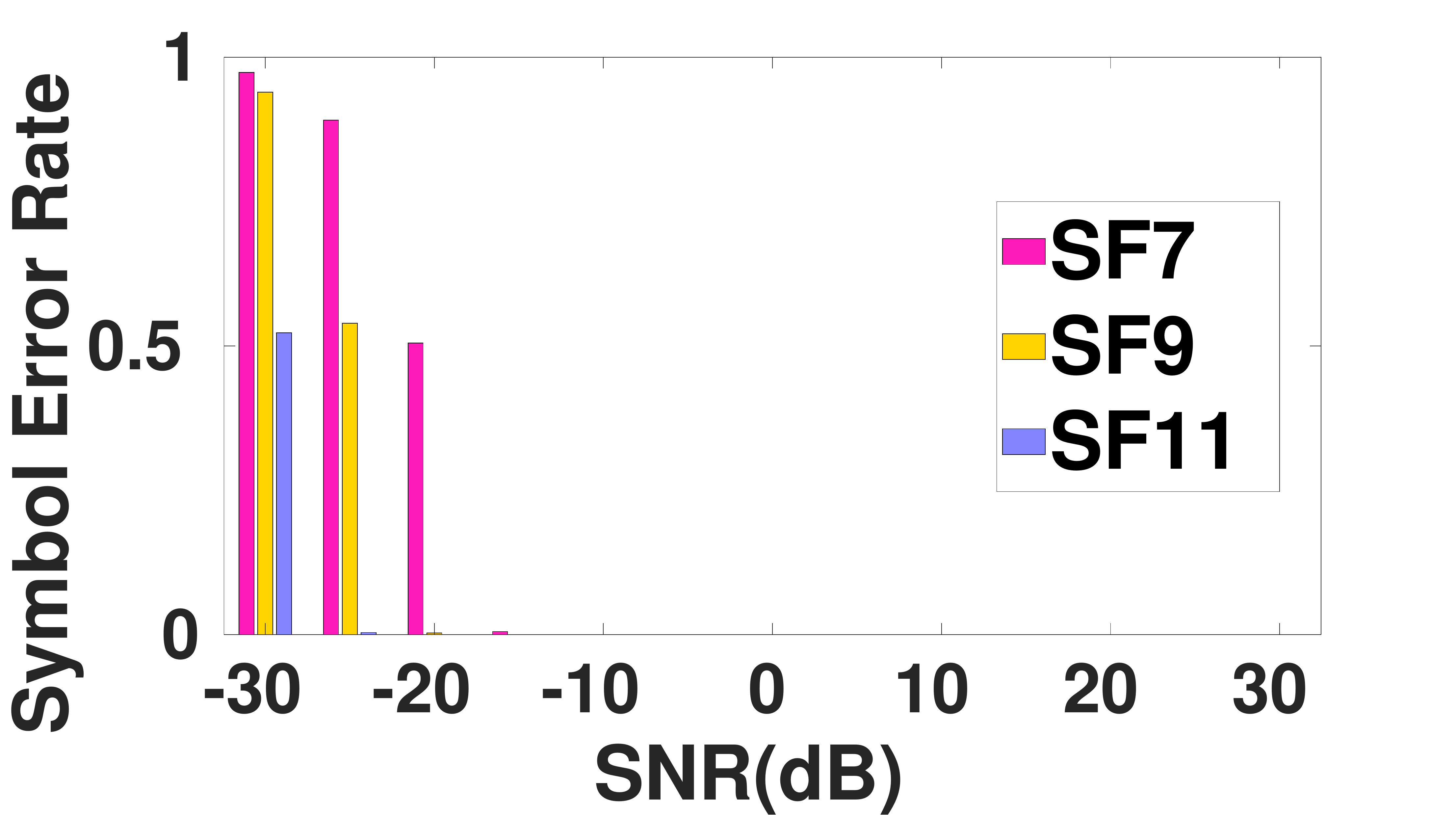}
    \label{subfig-sic-1}
    }
    \hspace{-0.01\columnwidth}
     \subfigure[SER fluctuates with SNR in the presence of collisions.]{
    \includegraphics[width=0.495\columnwidth]{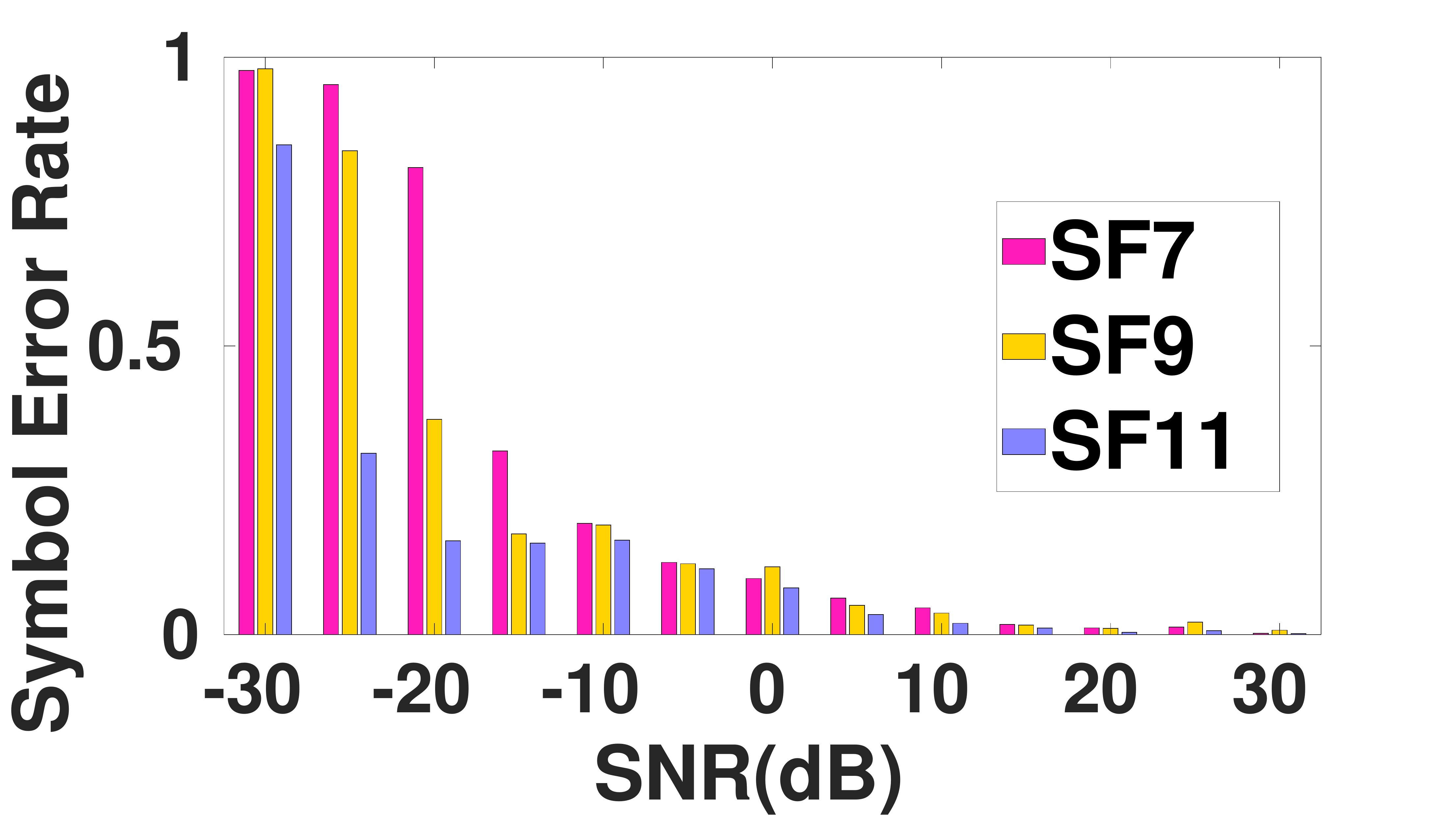}
    \label{subfig-sic-2}
    }
   \hspace{-0.01\columnwidth}
     \subfigure[SER fluctuates with SIR in the presence of collisions.]{
    \includegraphics[width=0.495\columnwidth]{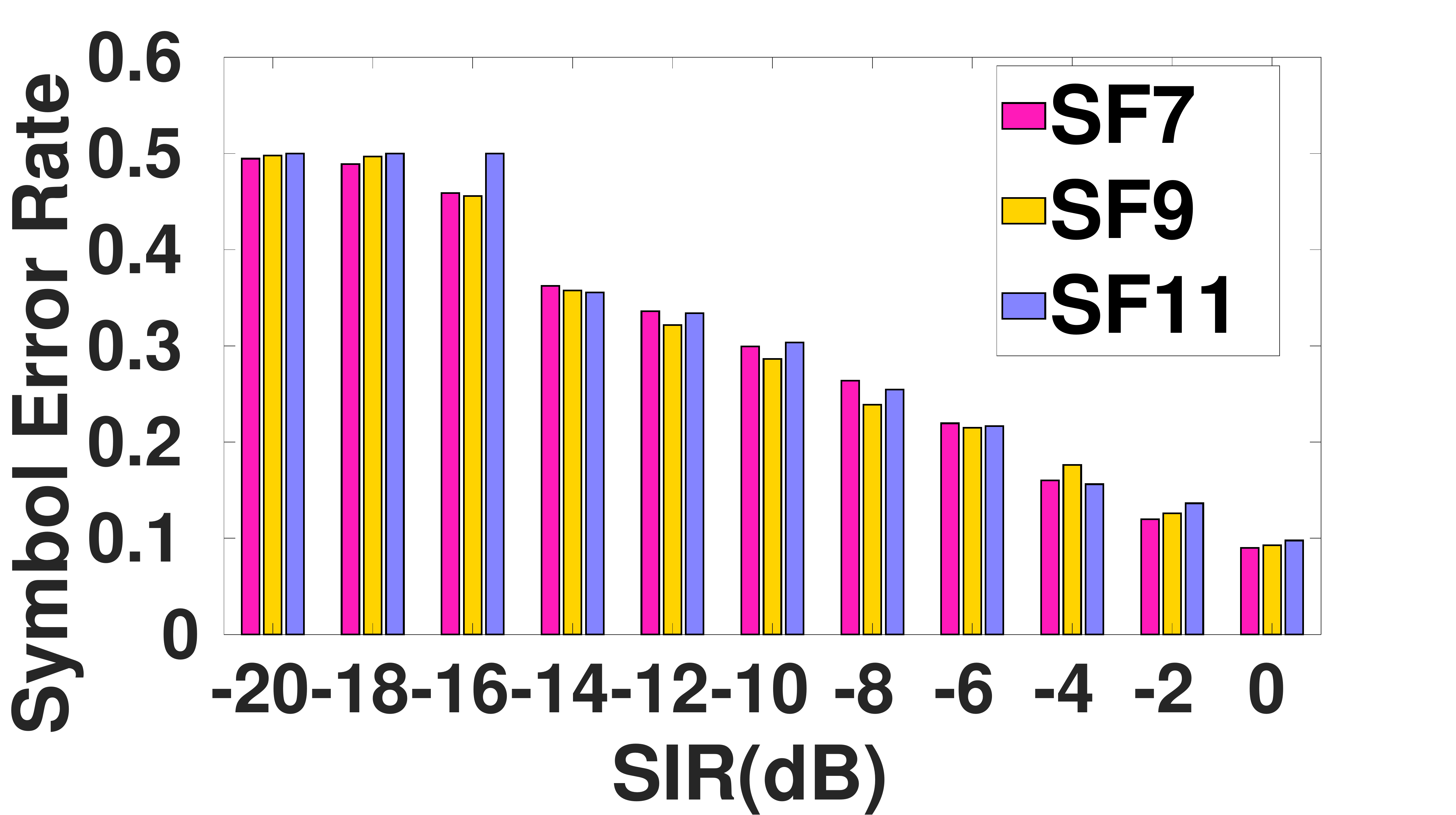}
    \label{subfig-sic-3}
    }
    \hspace{-0.01\columnwidth}
    \subfigure[SER fluctuates with symbol offset in the presence of collisions.]{
    \includegraphics[width=0.495\columnwidth]{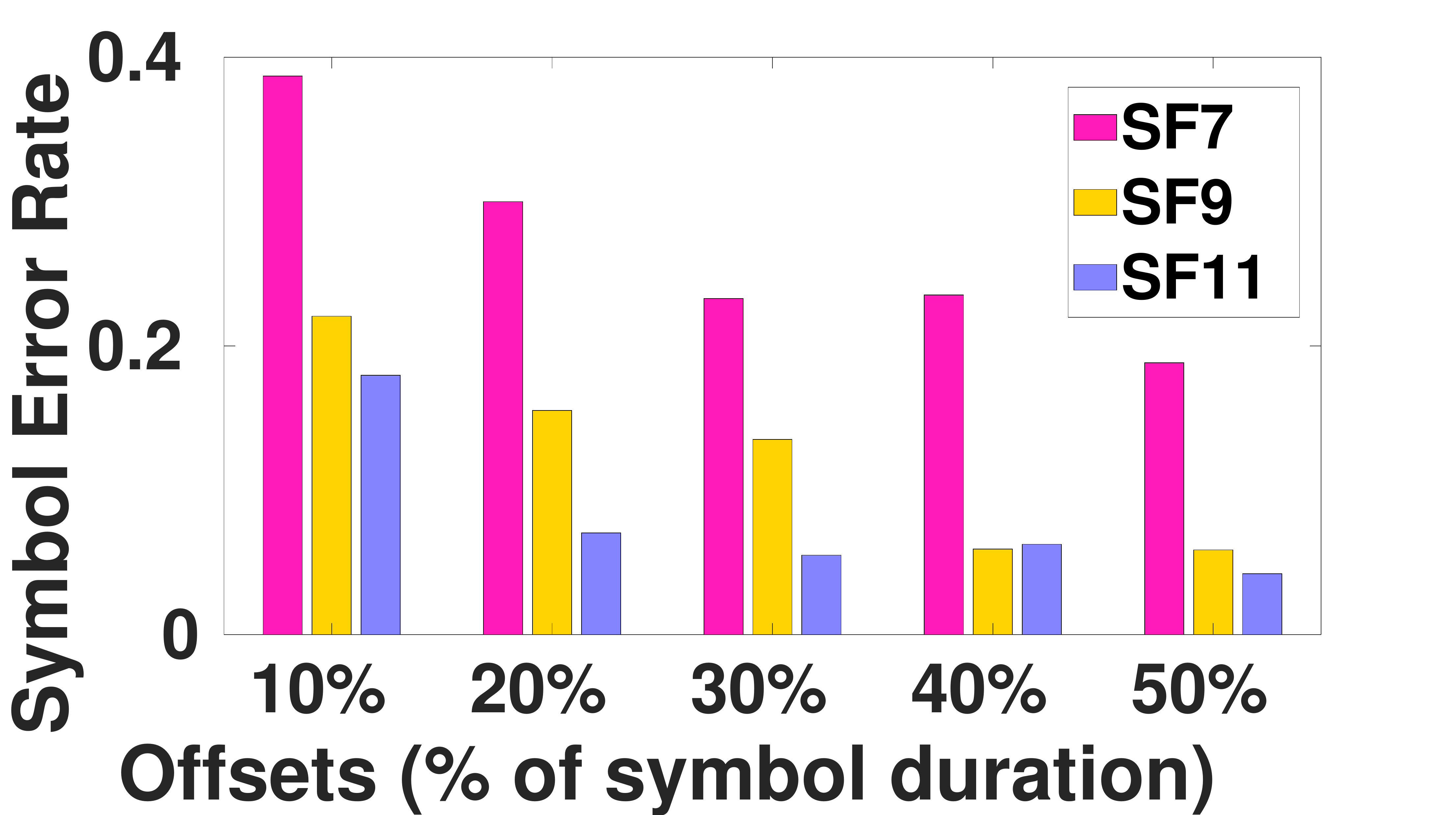}
    \label{subfig-sic-4}	
    }
    \vspace{-4mm}
    \caption{Evaluation of the successive interference cancellation-based collision resolving method~\cite{mLoRa,eletreby2017empowering} in various settings.}
    \vspace{-4mm}
\end{figure*}

\paragraph{LoRa Physical Layer.}
LoRa modulates data with chirp spread spectrum (CSS)~\cite{berni1973utility, eletreby2017empowering}.
The transmitter encodes bits by varying the initial frequency offset of a standard up-chirp.\footnote{A chirp signal whose frequency grows linearly from $-{BW}/{2}$ to ${BW}/{2}$.}
For instance, bits `00' are encoded by an up-chirp with zero initial frequency offset, while bits `01' are encoded by shifting the initial frequency by $f_0$. The frequency component beyond ${BW}/{2}$ will be wrapped to $-{BW}/{2}$, ensuring full bandwidth occupancy.
Upon the reception of a LoRa packet, the receiver multiplies each chirp symbol with a standard down-chirp (i.e., the conjugate of a standard up-chirp). The multiplication leads to an FFT peak in the frequency domain, which allows the receiver to demodulate LoRa symbols based on the position of FFT peaks.
Figure~\ref{fig-lora-preliminary}(a)-(b) shows an example.

\subsection{The Pros and Cons of Linear Chirp}
\label{subsec-background-2}

In essence, the aforementioned \emph{dechirp} converges the power of each LoRa symbol to a specific frequency point (\emph{i.e.}, an energy peak on an FFT bin), which allows the LoRa chirp to be decodable in extremely low SNR conditions (\emph{i.e.}, $-20$~dB).
As more LoRa nodes get involved, we are expected to see packet collisions at the receiver since LoRa nodes abide by the least-restrictive MAC protocol ALOHA.
To solve packet collisions, LoRaWAN~\cite{alliance2020lorawan} stipulates a set of spreading factors (SF) from seven to 12 and different bandwidths (\emph{i.e.}, 125/250/500~KHz). Therefore, LoRa packets with different SFs or bandwidths can transmit concurrently on the same frequency band.
The receiver uses down-chirps with different SFs to disambiguate these concurrent transmissions.
However, the capacity of this regulation is limited: it supports only 18 pairs of SF\&BW combinations~\cite{hessar2018netscatter}.

Collision happens when two concurrent transmissions use the same SF and bandwidth.
In this case, we are expected to see two energy peaks in two separate FFT bins, as shown in Figure~\ref{subfig-symbol-collision}.
In practice, due to the near-far issue, one transmission (e.g., packet $A$ in red) may experience a stronger attenuation than the other (e.g., packet $B$ in blue).
Hence the energy peak of $A$ tends to be weaker than that of $B$ in FFT bins.
Accordingly, the receiver will take $A$ as noise and demodulate $B$ only.
When $A$ and $B$ experience similar attenuations, the receiver can reliably demodulate neither of them because their individual energy peak may bury each other across different symbols.
In a nutshell, when two LoRa packets collide, only the strongest transmission can be correctly demodulated by LoRaWAN. 
Otherwise, no transmissions can be reliably demodulated.


\subsection{Resolving Linear-Chirp LoRa Collisions}
\label{subsec-background-3}

Section \ref{sec-related-work} overviews the current practice on resolving LoRa collisions and explains their pros and cons. In this section, we implement a state-of-the-art SIC-based system, mLoRa~\cite{mLoRa}, and examine its performance in various SNR and SIR conditions. Noting that we also compare with other SOTA systems in the evaluation part.
Figure~\ref{subfig-sic-2} shows the symbol error rate (SER) in different SNR conditions.
We observe that to maintain a low symbol error rate (e.g., <0.1), the SNR of the received signal should be higher than $5dB$ -- $25dB$ higher than the minimum SNR requirement for demodulating a collision-free LoRa transmission.
Such a high SNR requirement sets a strong barrier to mLoRa's adoption since LoRa transmissions tend to be very weak after attenuation over a long distance.
We then plot the symbol error rate measured in different SIR settings in Figure~\ref{subfig-sic-3}.
The result shows that the symbol error rate grows dramatically with the decreasing SIR, indicating that mLoRa~\cite{mLoRa} cannot demodulate the targeting LoRa transmission in the presence of strong concurrent LoRa transmission.
We further evaluate the impact of the symbol offset on SER.
The result (Figure~\ref{subfig-sic-4}) shows that the SER grows with decreasing symbol offset, confirming our analysis. 

\vspace{1mm}
\noindent
\textbf{Remarks.} The above analysis reveals that the linear chirp in LoRaWAN does not scale to concurrent LoRa transmissions.
Although the state-of-the-arts have proposed various approaches to resolve LoRa collisions, most of them function only in good SNR or SIR conditions and thus sacrifice the precious processing gain brought by the chirp modulation.


\section{Analysis: Non-linear vs. Linear Chirps}
\label{sec-design}

We now show that by replacing the linear chirp with its non-linear counterpart, we can boost the capacity of concurrent transmissions (\S\ref{subsec-design-1}) while allowing the receiver to demodulate collision signals in severe SIR conditions (\S\ref{subsec-design-2}).
In addition, we show by both theoretical analysis and empirical validation that such a non-linear chirp design sacrifices neither noise resilience (\S\ref{subsec-design-3}) nor power efficiency (\S\ref{subsec-design-4}).



\begin{figure}[t]
    \centering
    \subfigure[]{ 
	    \includegraphics[width=3.2in]{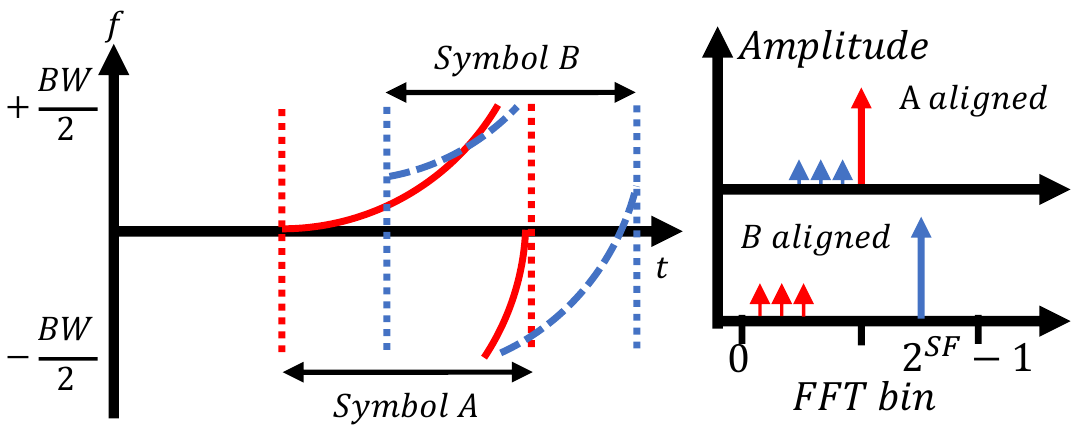}
        \label{subfig-nonlinear-vs-linear-sir-1}
    }\\\vspace{-4.2mm}
    \subfigure[]{
    	\includegraphics[width=3.2in]{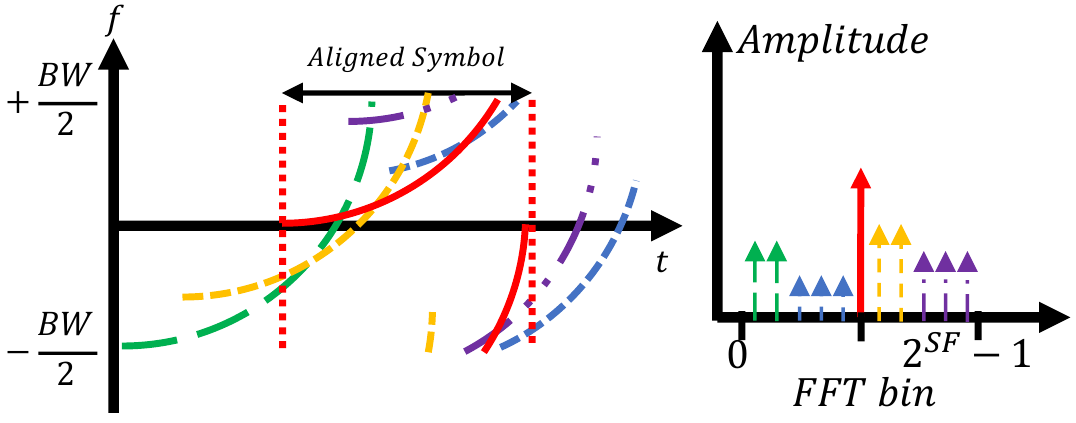}
    	\label{subfig-nonlinear-vs-linear-sir-2}
    }
    \vspace{-4mm}
    \caption{(a): The receiver takes advantage of the power scattering effect to demodulate two collision symbols. (b): an illustration of demodulating five collision symbols.}
    \label{fig:demodulation}
    \vspace{-4mm}
\end{figure}

\begin{figure*}[t]
    \centering
    \subfigure[L. chirp collisions in time domain]{
    \includegraphics[width=0.515\columnwidth]{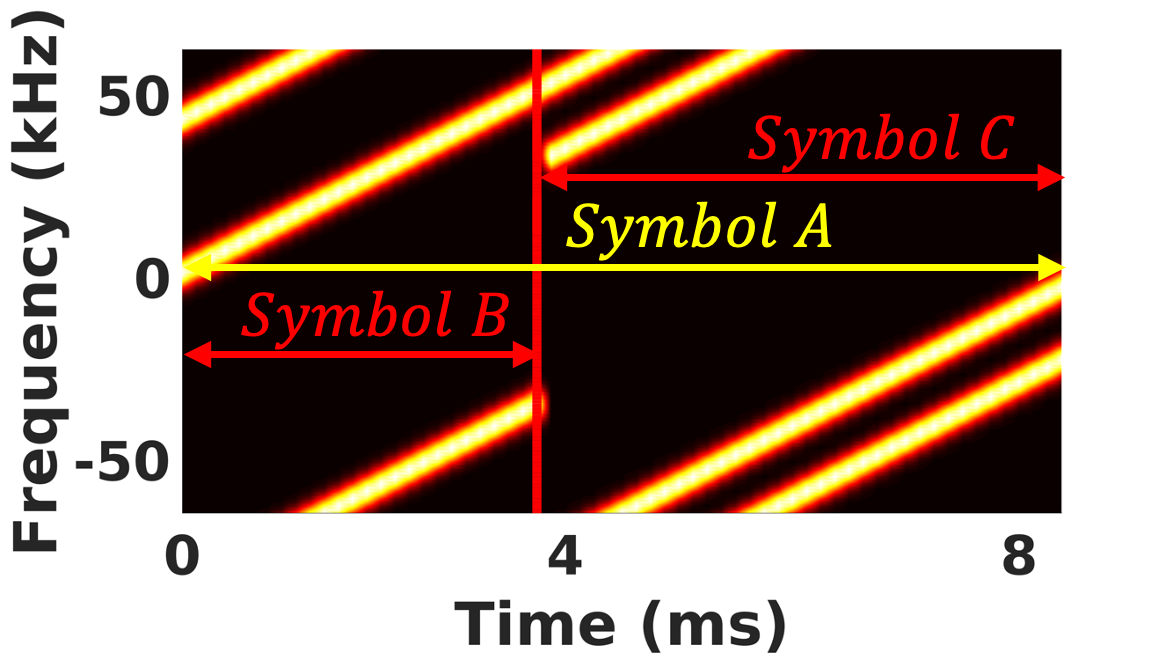}
    \label{subfig-nonlinear-vs-linear-pre-1}
    }
      \subfigure[L. chirp collisions in freq.]{
    \includegraphics[width=0.475\columnwidth]{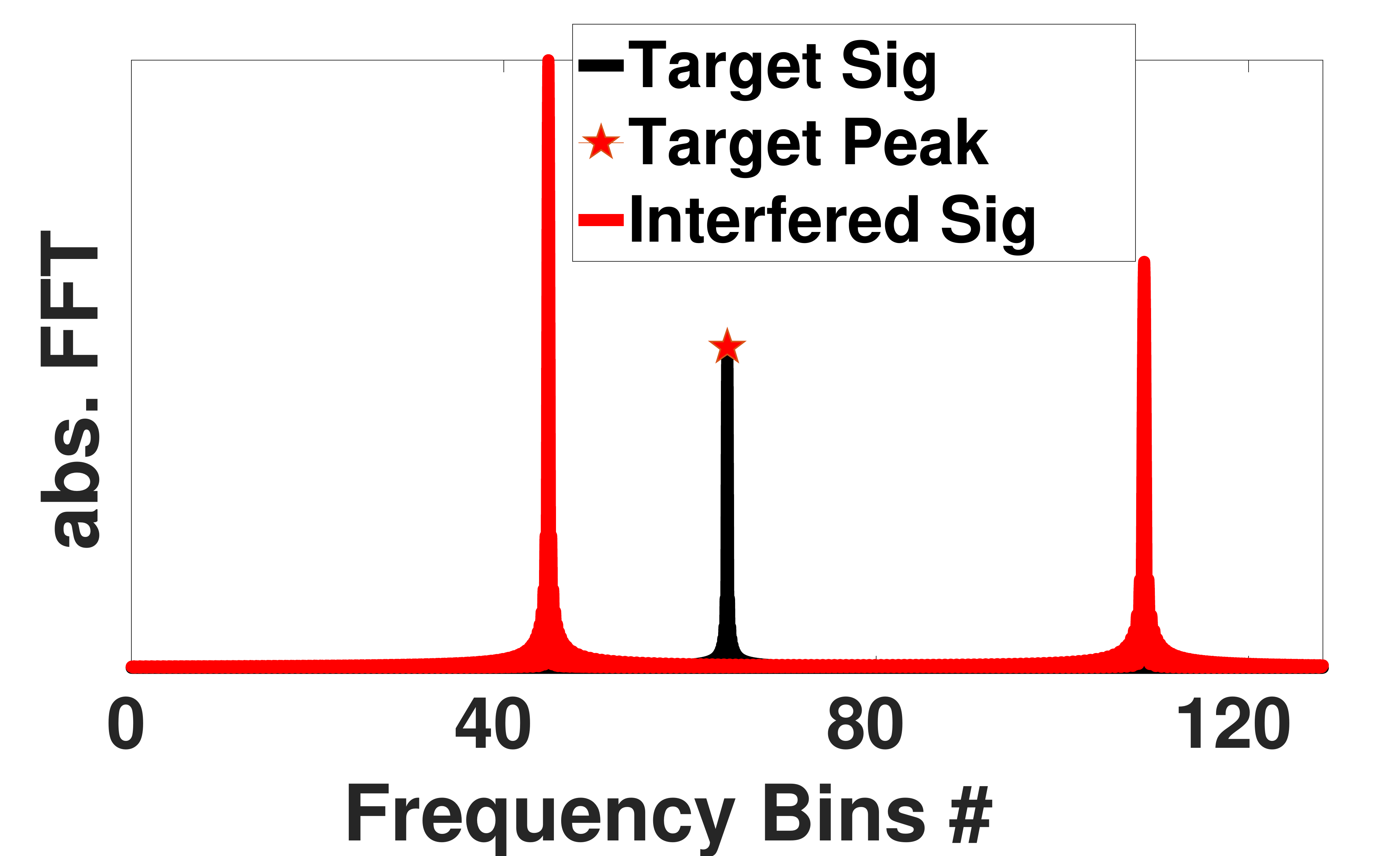}
    \label{subfig-nonlinear-vs-linear-pre-2}
    }
      \subfigure[NL. chirp collisions in time domain]{
    \includegraphics[width=0.515\columnwidth]{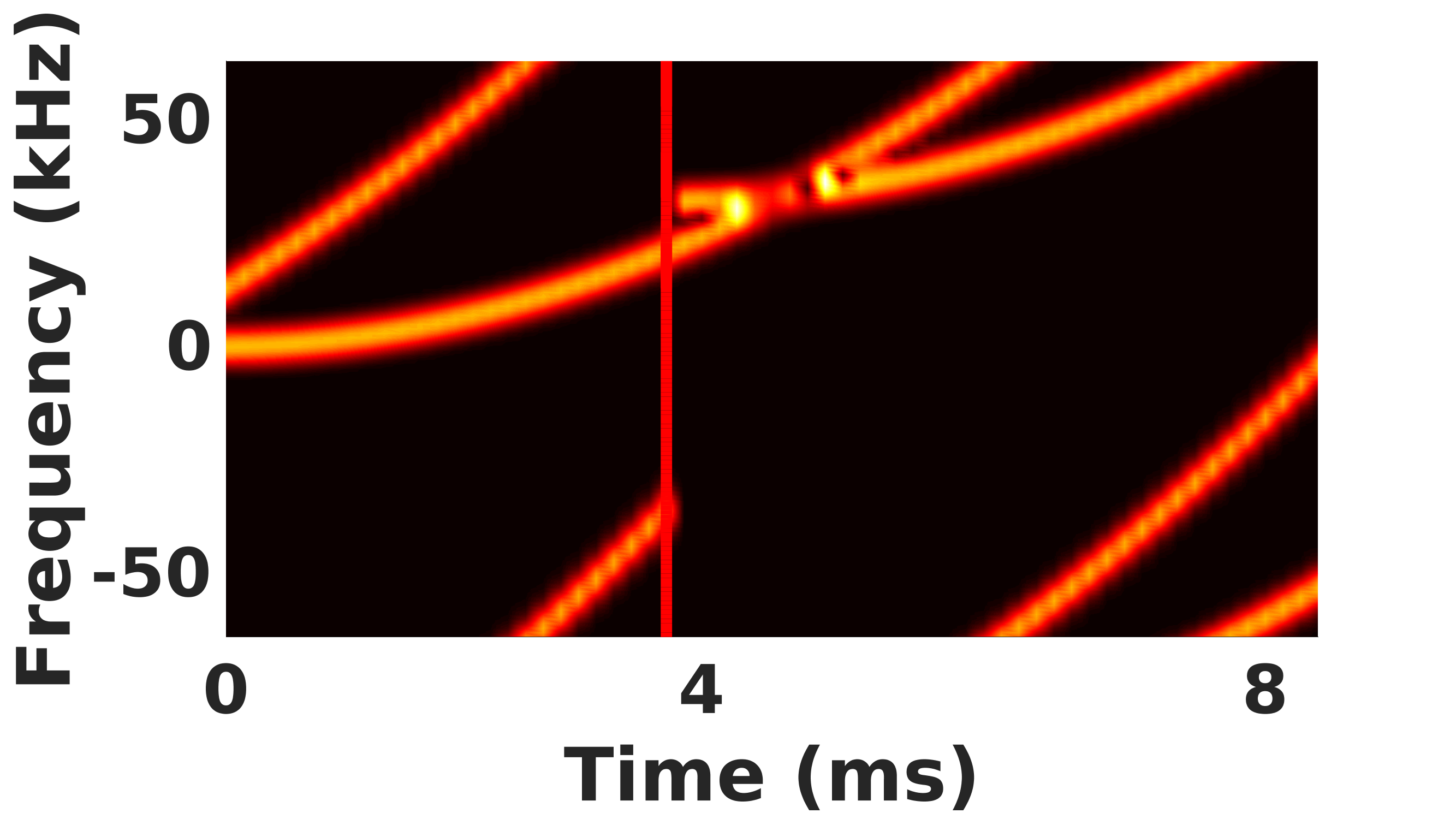}
    \label{subfig-nonlinear-vs-linear-pre-3}
    }
    \subfigure[NL. chirp collisions in freq.]{
    \includegraphics[width=0.475\columnwidth]{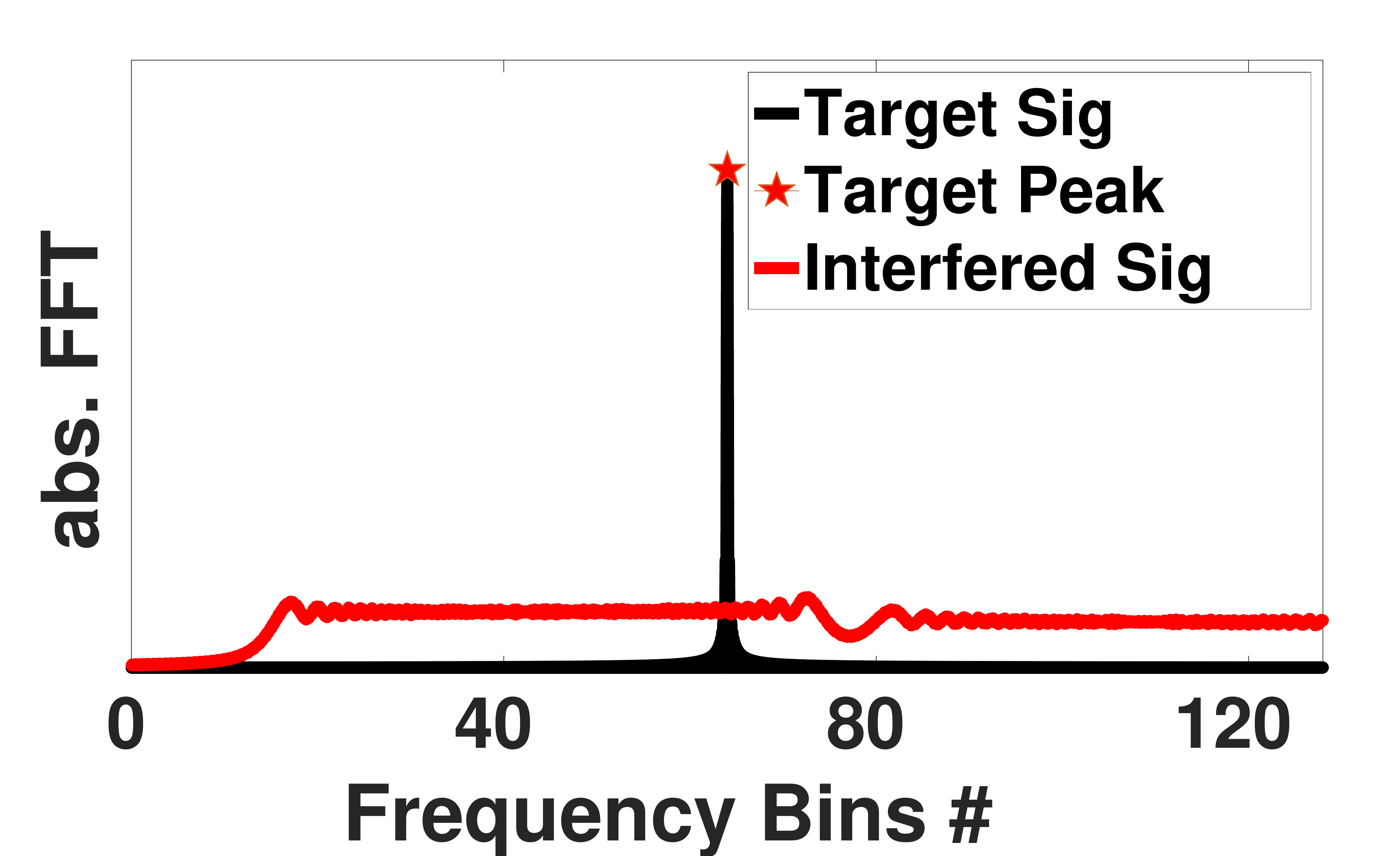}
    \label{subfig-nonlinear-vs-linear-pre-4}	
    }
    \caption{Comparison of linear and non-linear chirps on resolving the near-far issue. 
    {\normalfont
    (a): Three linear chirps with different SNR collide at the receiver. (b): Due to the near-far issue, the energy peak (in black) of the weak reception is overwhelmed by the energy peaks (in red) of strong collision symbols. (c): Three non-linear chirps in the same collision situations. (d): The spectral power of strong receptions is spread over multiple frequency points, making the corresponding FFT peaks (in red) significantly lower than the converged power (in black) of the weak reception.}}
    \label{fig-nonlinear-vs-linear-pre}
    \vspace{-4mm}
\end{figure*}

\subsection{Non-linear Chirps Meet Collisions}
\label{subsec-design-1}

We define a non-linear up-chirp as a signal whose frequency grows non-linearly from $-BW/2$ to $BW/2$. The non-linear function can be either polynomial, logarithmic, exponential, or trigonometric.
The receiver operates dechirp to demodulate non-linear chirp symbols.



\vspace{1mm}
\noindent
Considering two collision symbols $A$ and $B$, as shown in Figure~\ref{fig:demodulation}(a).
The receiver takes a sliding window approach to demodulate incoming signals.
As aforementioned, when symbol $A$ aligns with the down-chirp in the current observing window, we are expected to see a strong energy peak (termed as peak $A$) on the associated FFT bin.
At the same time, the energy of symbol $B$ will be spread over multiple, clustered FFT bins due to its misalignment with the down-chirp.
Compared to peak $A$, the amplitude of these clustered energy peaks is significantly weaker.
The receiver inherently takes these clustered FFT peaks as noise and demodulates symbol $A$.
As the observing window slides, symbol $B$ will align with the down-chirp at a time.
Consequently, the dechirp converges the power of symbol $B$ to a specific frequency point while spreading the power of symbol $A$ into multiple frequency bins instead.
The receiver then takes symbol $A$ as noise and demodulates $B$.
When it comes with two collision packets with each consisting of tens of symbols, following the same principle, the receiver slides the observing window step by step and demodulates their individual symbols alternatively.

\vspace{1mm}
\noindent
\textbf{Remarks.} In essence, when the collision symbols are not strictly aligned, there will be only one symbol that aligns with the down-chirp in each observing window.
This indicates that each time only one symbol gets its energy accumulated while the energy of all the other colliding signals is being scattered over multiple FFT bins, as shown in Figure~\ref{fig:demodulation}(b).
Hence the receiver can easily pick up each individual symbol on separated observing windows and decode them chronologically.

\subsection{Accounting for the Near-far Effect}
\label{subsec-design-2}

The above section explains the basic idea of non-linear chirp and its unique energy scattering effect, we next demonstrate that this energy scattering effect can be leveraged to address the {\it near-far issue} where weak receptions are buried by strong receptions from nearby transmitters.

Consider a large number of colliding transmitters where some are physically closer to the receiver compared to the others.
When linear chirps are adopted, the power of strong receptions {\it converges} to a specific frequency point where the associated energy peak is easily distinguishable after dechirp.
The weak receptions from remote transmitters, however, have significantly weaker energy peaks.
The receiver thus takes those weak receptions as noise.
Figure~\ref{subfig-nonlinear-vs-linear-pre-1} shows a snapshot where two linear chirp packets with a distinguishable SIR (-10dB) collide at the receiver.
Suppose the current observing window aligns with the symbol $A$ in yellow of the weak packet.
Due to the near-far issue, symbol $B$ and $C$ in red produce stronger energy peaks on associated FFT bins even they both are not aligned with the current observing window (shown in Figure~\ref{subfig-nonlinear-vs-linear-pre-2}).
Hence the receiver cannot demodulate symbol $A$ successfully.
In contrast, when non-linear chirps are adopted (shown in Figure~\ref{subfig-nonlinear-vs-linear-pre-3}), the power of strong reception symbols $B$ and $C$ are both {\it scattered} into multiple FFT bins after dechirp.
Due to such an energy scattering effect, the energy peaks induced by these strong symbols become lower than the accumulated energy peak induced by the weak symbol $A$. This allows the receiver to demodulate symbol $A$ in the presence of strong collision symbols $B$ and $C$ (shown in Figure~\ref{subfig-nonlinear-vs-linear-pre-4}).

\begin{figure}[t]
\centering
\subfigure[SER of linear and non-linear chirps in different SIR conditions.]{ 
	    \includegraphics[width=0.22\textwidth]{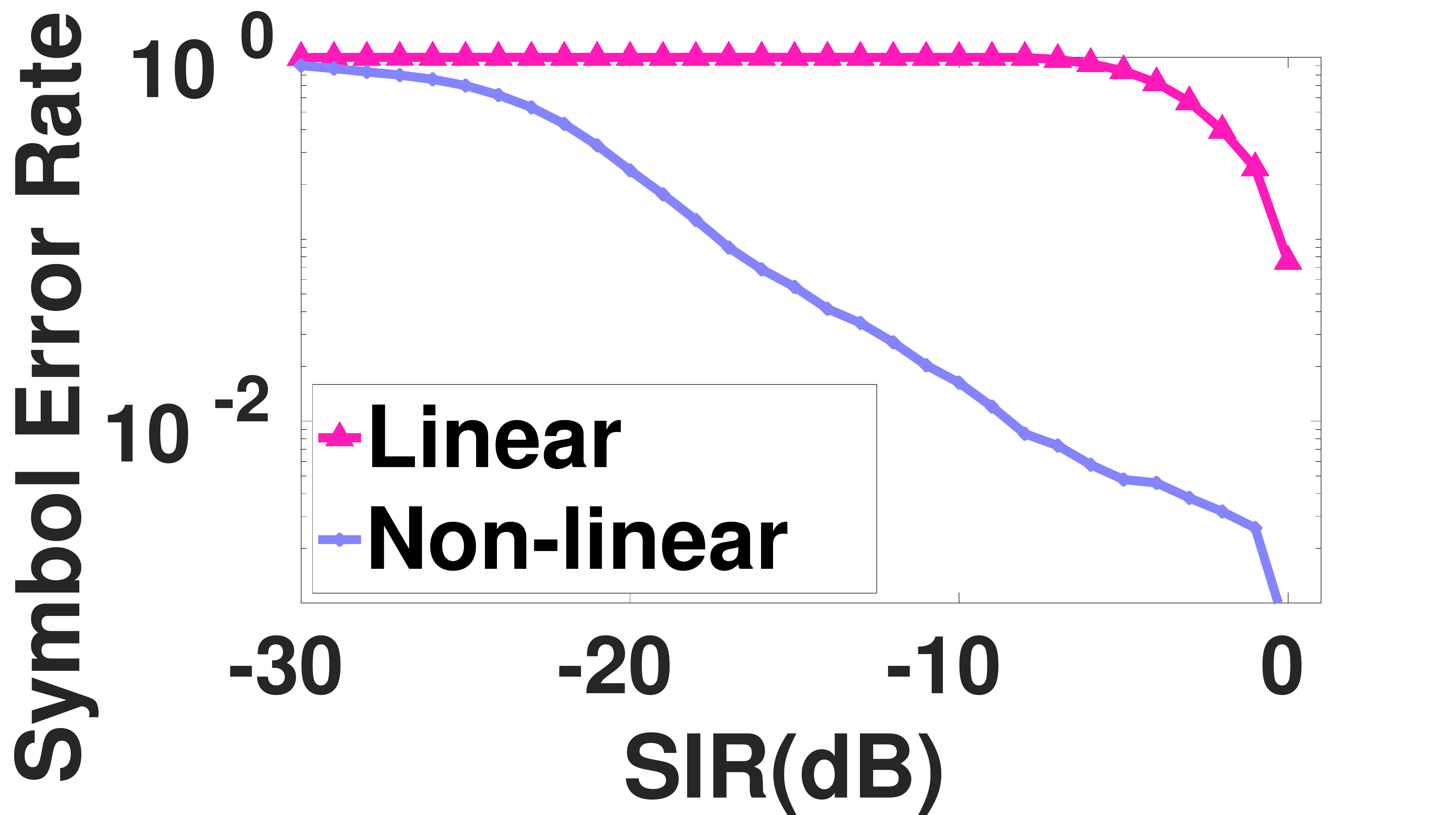}
        \label{subfig-nonlinear-vs-linear-sir-1}
}
\hspace{-0.02\columnwidth}
\subfigure[The SIR threshold for linear and non-linear chirps to achieve 1\% SER.]{
	\includegraphics[width=0.22\textwidth]{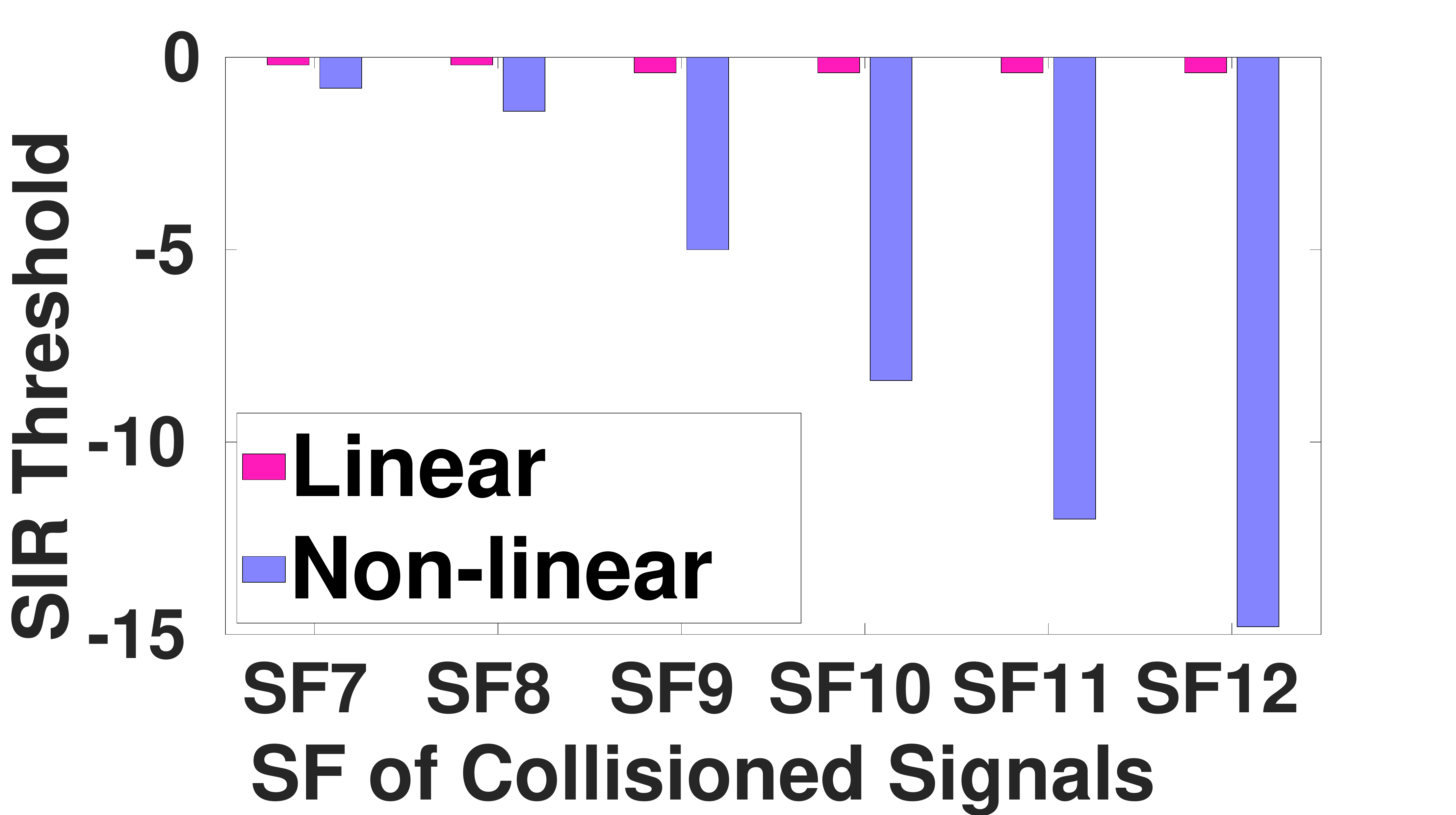}
	\label{subfig-nonlinear-vs-linear-sir-2}
}
\vspace{-3mm}
\caption{Comparing the SER of linear and non-linear chirps in various near-far conditions.}
\label{fig-nonlinear-vs-linear-sir}
\vspace{-4mm}
\end{figure}

\begin{figure*}[t]
    \centering
    \subfigure[Two collided linear chirps]{
    \includegraphics[width=0.53\columnwidth]{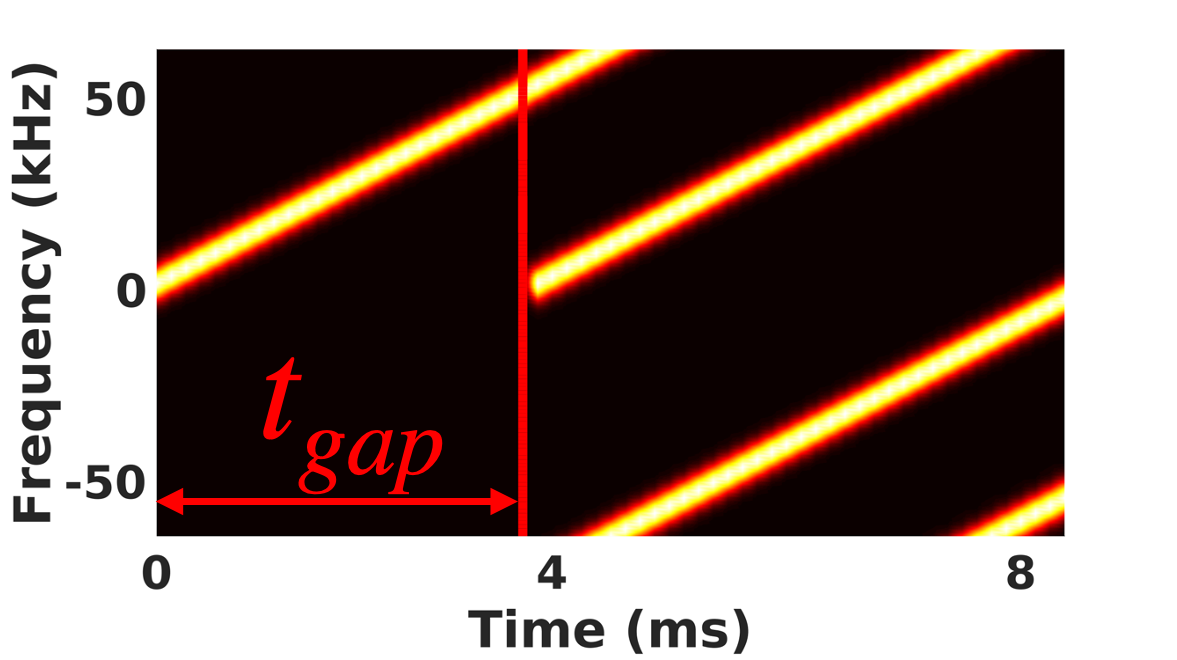}
    \label{subfig-nonlinear-vs-linear-tgap-1}
    }
    \hspace{-0.06\columnwidth}
      \subfigure[SIR=-1dB]{
    \includegraphics[width=0.49\columnwidth]{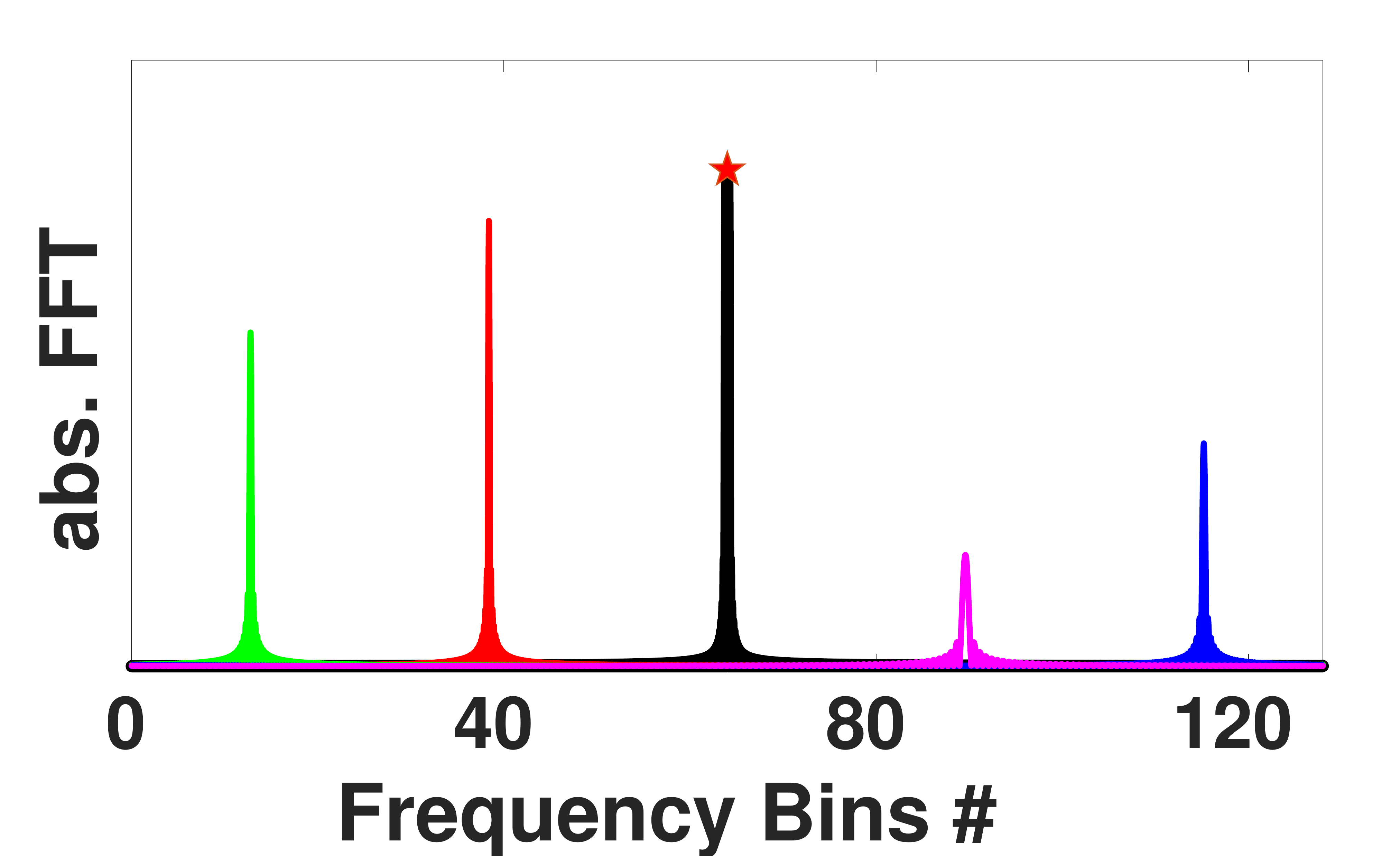}
    \label{subfig-nonlinear-vs-linear-tgap-2}
    }
    \hspace{-0.06\columnwidth}
      \subfigure[SIR=-5dB]{
    \includegraphics[width=0.49\columnwidth]{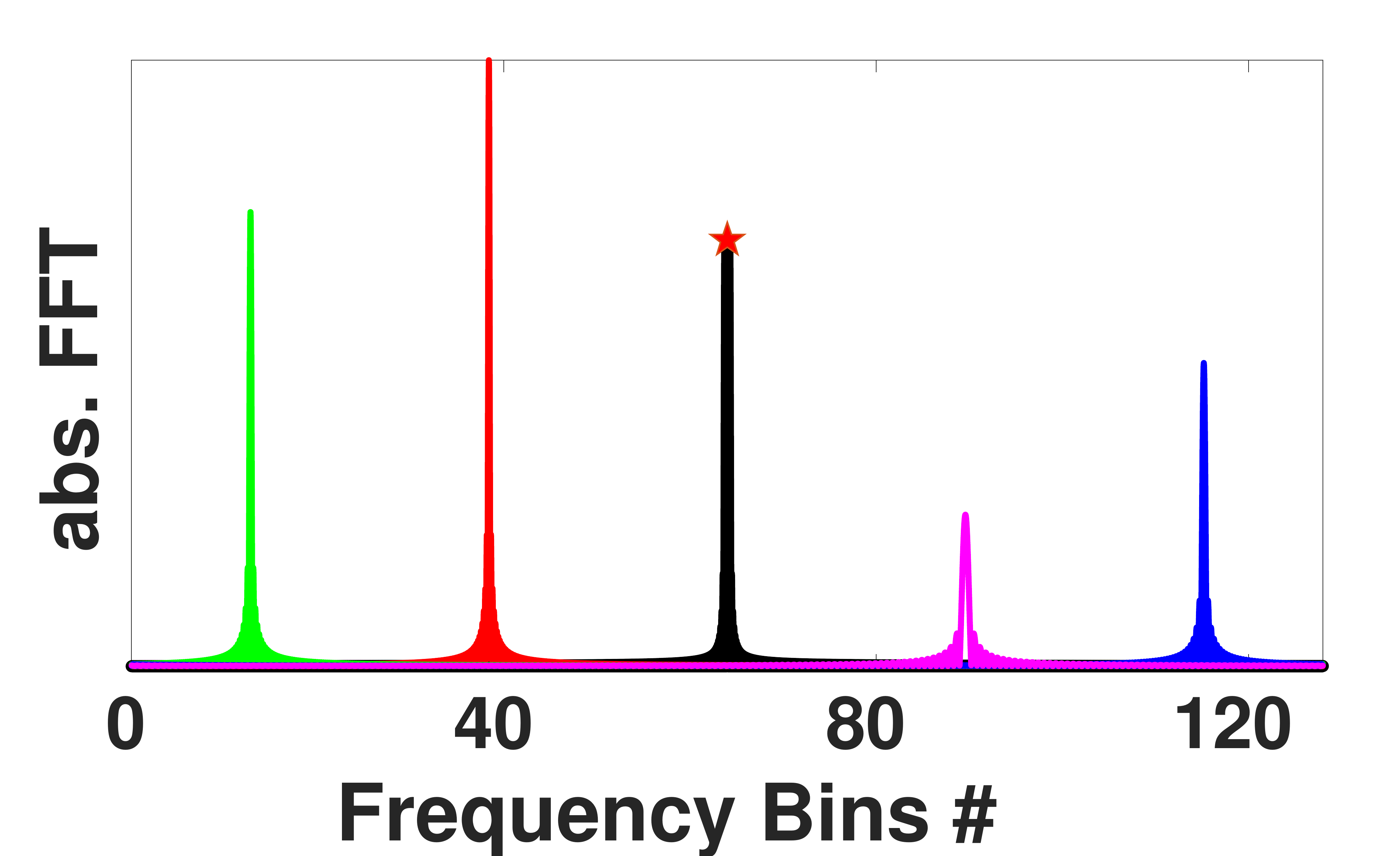}
    \label{subfig-nonlinear-vs-linear-tgap-3}
    }
    \hspace{-0.06\columnwidth}
    \subfigure[SIR=-10dB]{
    \includegraphics[width=0.49\columnwidth]{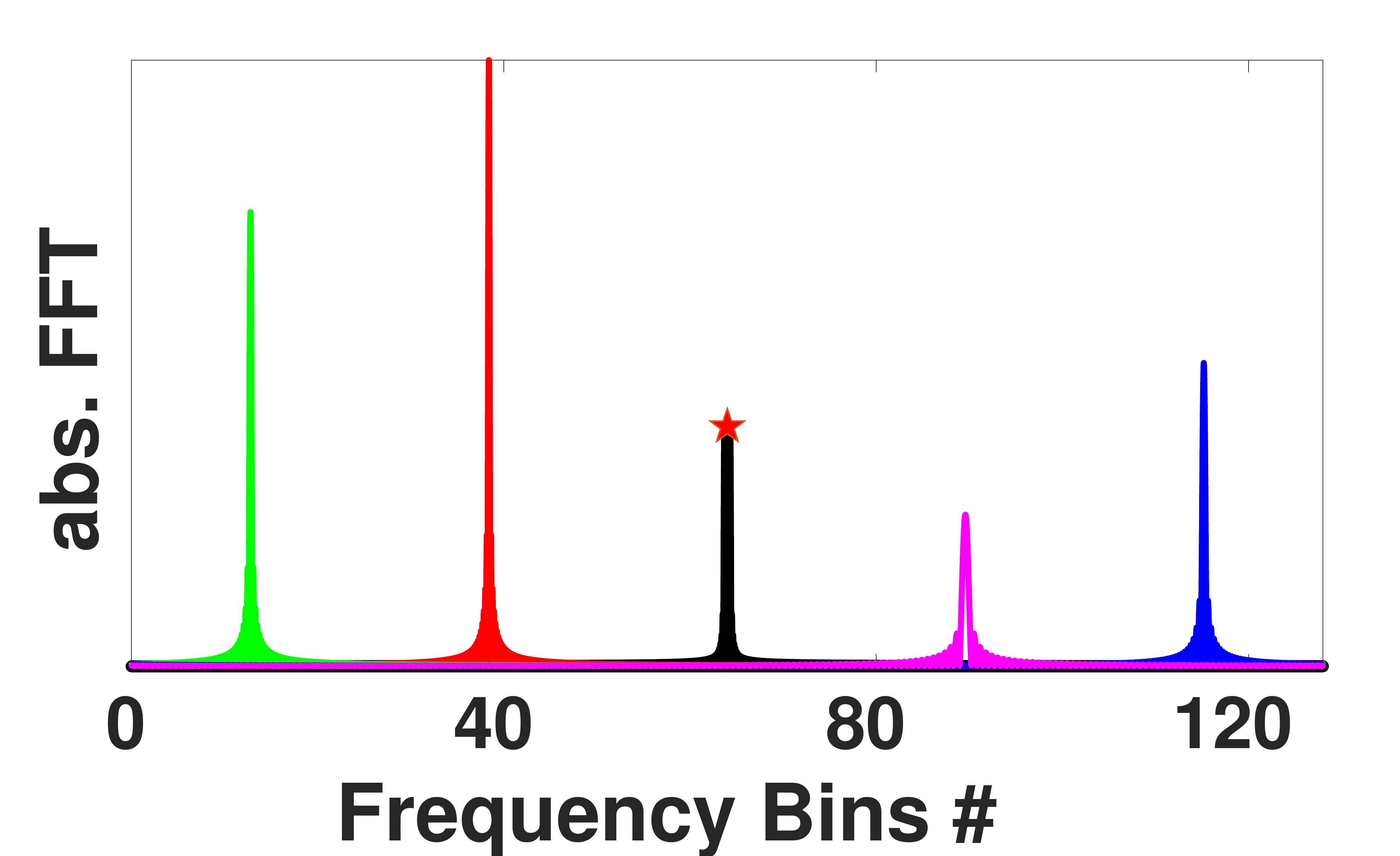}
    \label{subfig-nonlinear-vs-linear-tgap-4}	
    }\vspace{-3mm}
    \subfigure[Two collided non-linear chirps]{
    \includegraphics[width=0.53\columnwidth]{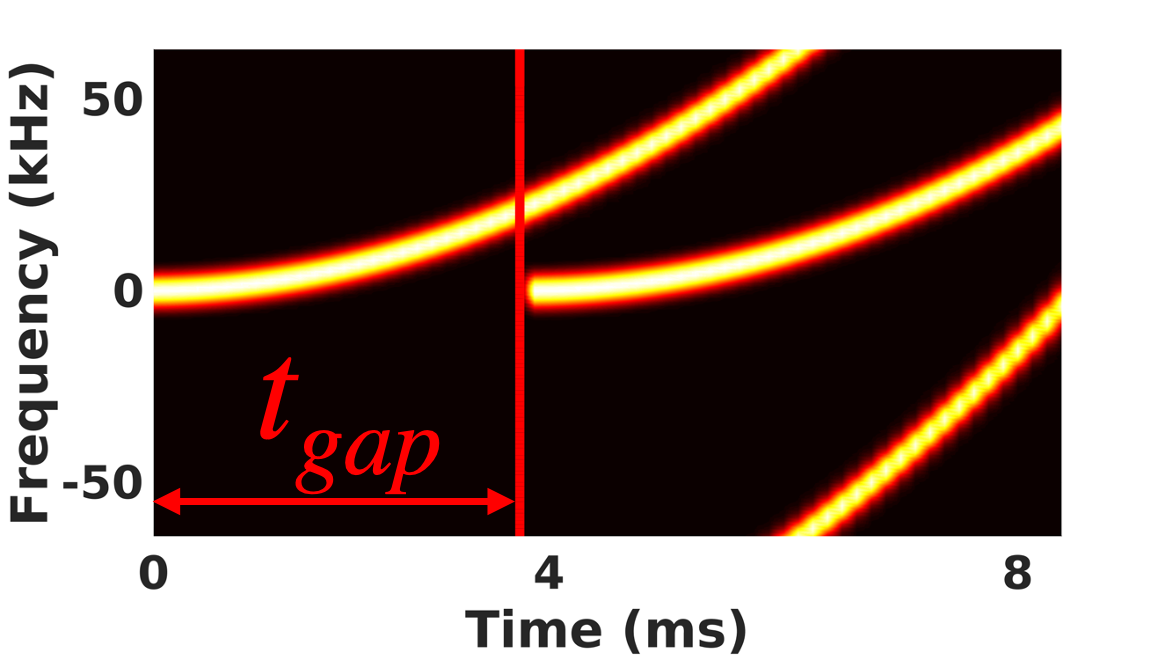}
    \label{subfig-nonlinear-vs-linear-tgap-5}
    }
    \hspace{-0.06\columnwidth}
      \subfigure[SIR=-1dB]{
    \includegraphics[width=0.49\columnwidth]{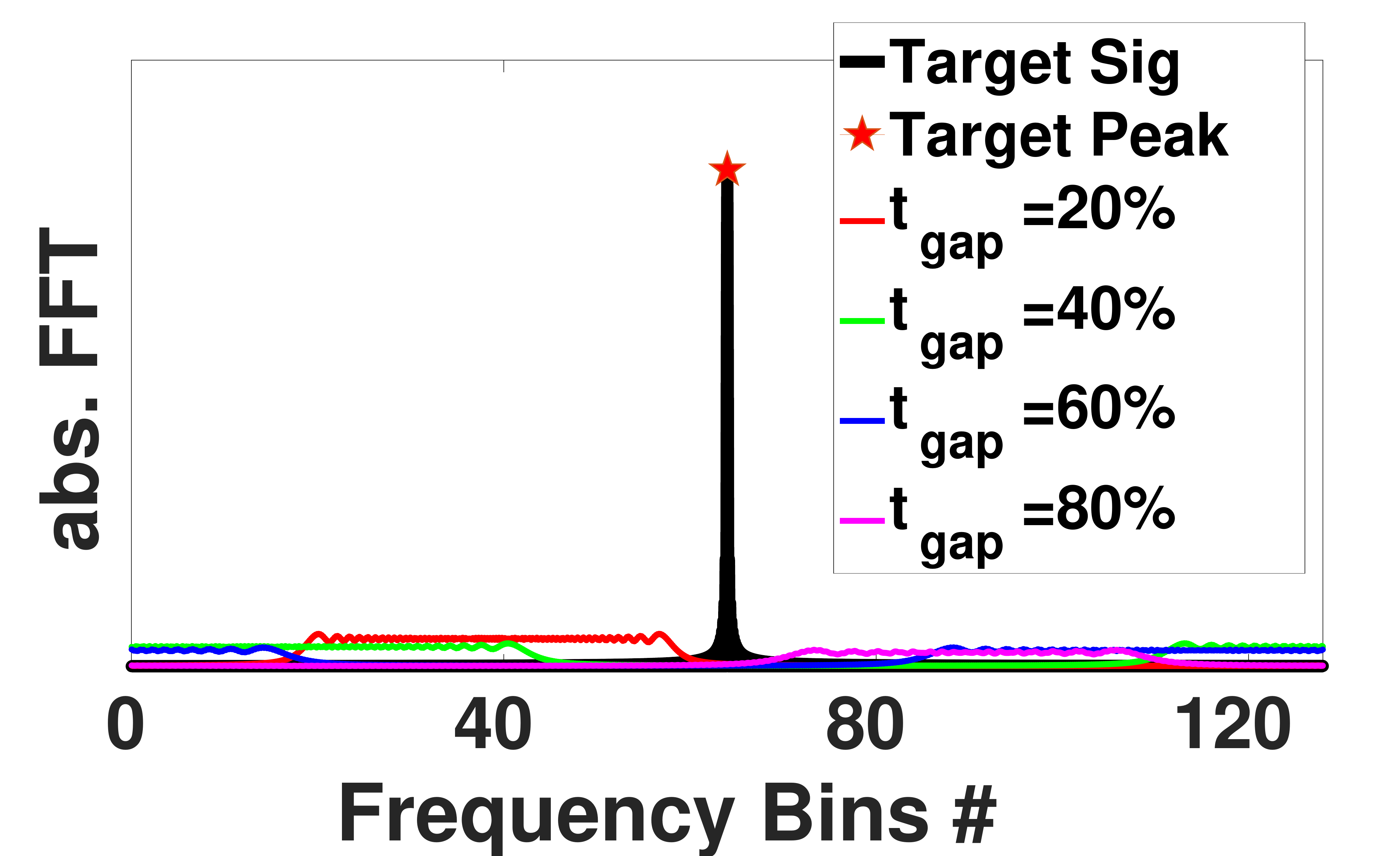}
    \label{subfig-nonlinear-vs-linear-tgap-6}
    }
    \hspace{-0.06\columnwidth}
      \subfigure[SIR=-5dB]{
    \includegraphics[width=0.49\columnwidth]{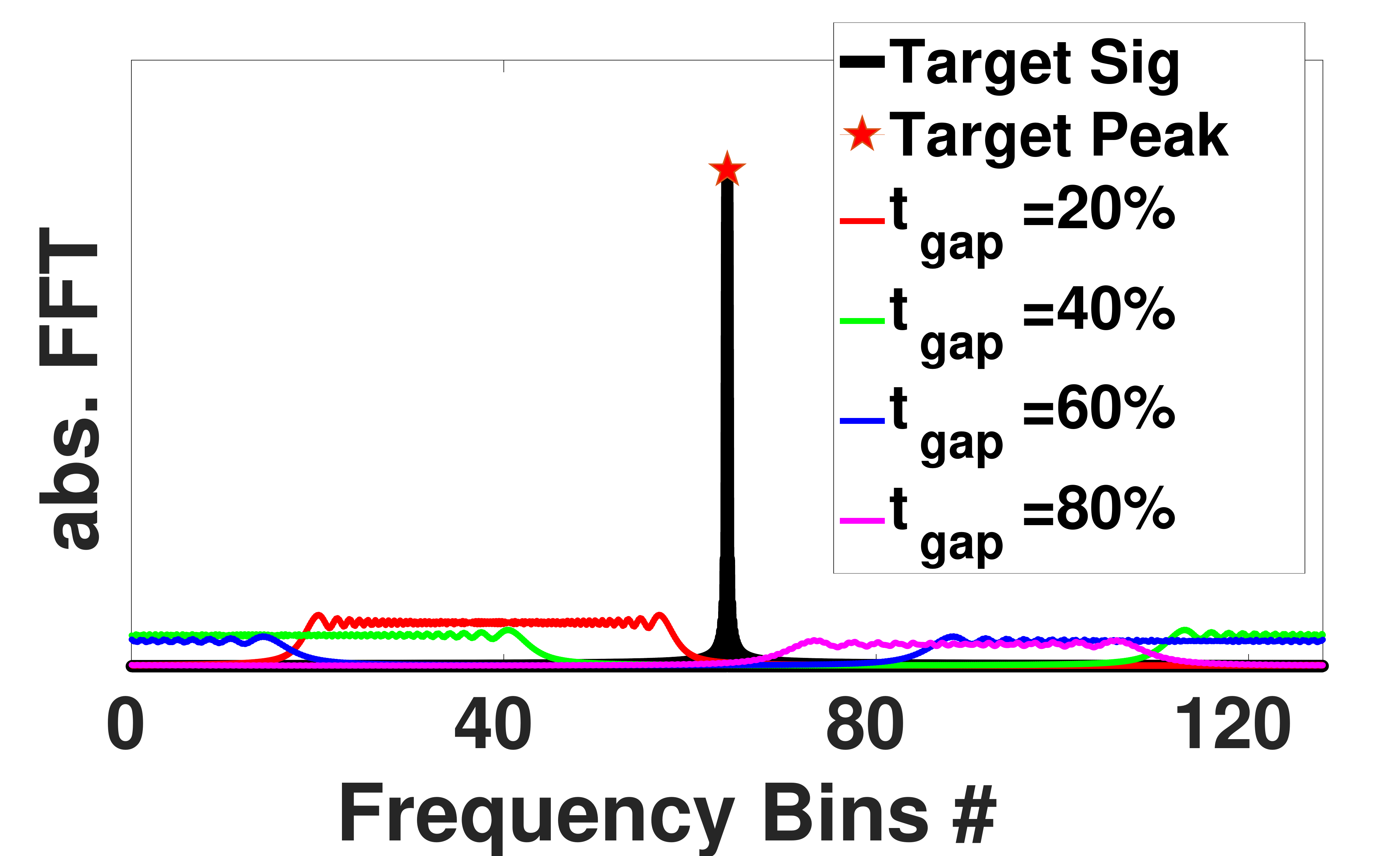}
    \label{subfig-nonlinear-vs-linear-tgap-7}
    }
    \hspace{-0.06\columnwidth}
    \subfigure[SIR=-10dB]{
    \includegraphics[width=0.49\columnwidth]{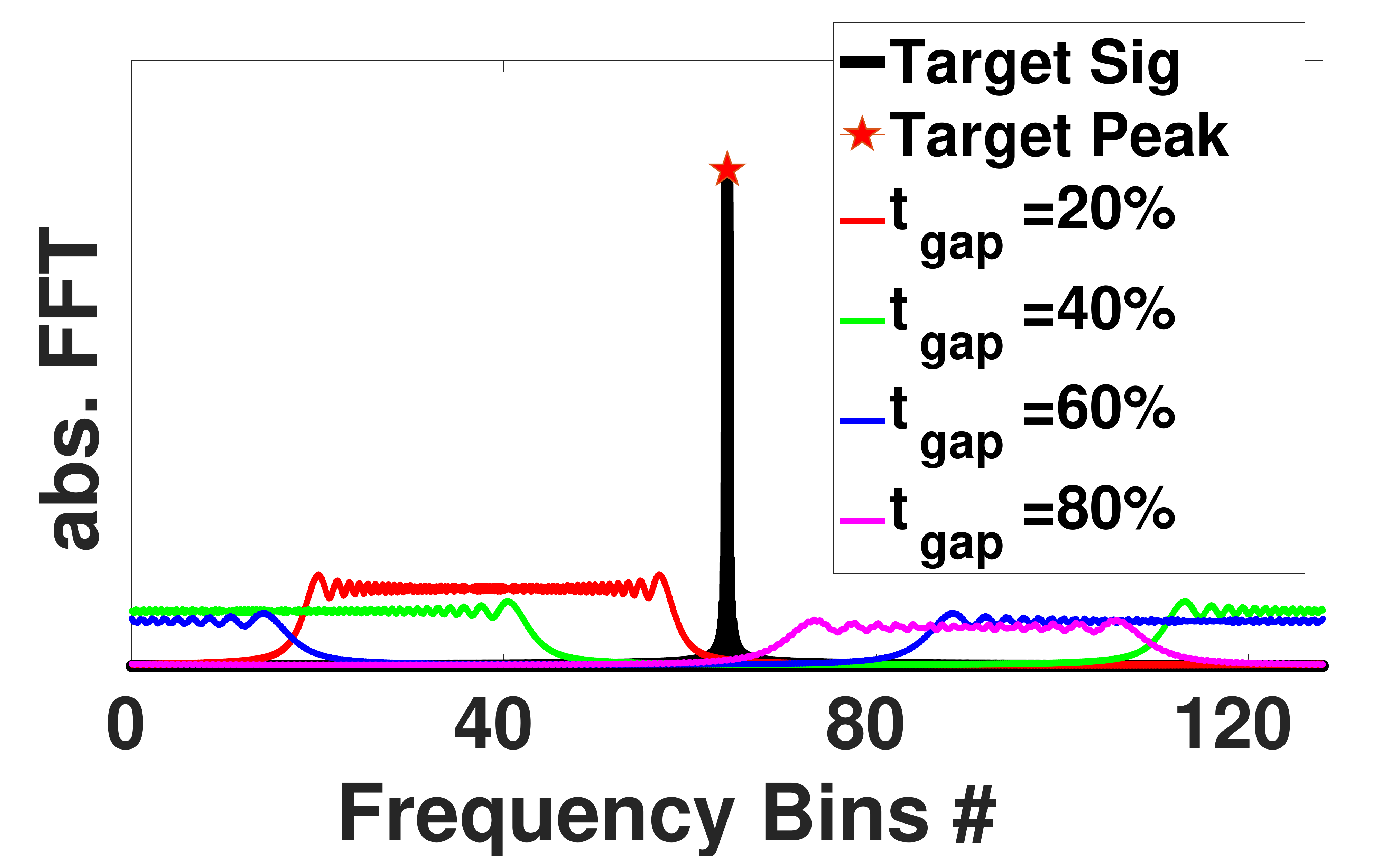}
    \label{subfig-nonlinear-vs-linear-tgap-8}
    }
    \caption{Examining the resilience to symbol time offset in different symbol-to-interference ratio settings.}
    \label{fig-nonlinear-vs-linear-tgap}
    \vspace{-4mm}
\end{figure*}

\vspace{1mm}
\noindent
\textbf{Validation.} To demonstrate the effectiveness of non-linear chirps on resolving the near-far issues, we compare the symbol error rate of non-linear and linear chirps in different signal-to-interference ratio (SIR) settings.
The spreading factor, bandwidth, and sampling rate are set to 10, 125KHz, and 1MHz, respectively.

Figure~\ref{subfig-nonlinear-vs-linear-sir-1} shows the results. 
In accordance with our analysis, we observe that the linear chirp fails to demodulate the targeting symbol in the presence of strong concurrent transmissions (i.e., SER=100\% when SIR<$-5dB$). 
The SER then drops to around 10\% when the power of the targeting symbol is comparable to that of colliding symbols (i.e., SIR=$0dB$).
In contrast, the receiver can successfully demodulate the weak non-linear symbols as long as the SIR is higher than $-25dB$.
We also found that the non-linear chirp can still achieve $10\%$+ SER in an extreme case where the colliding signal is $20dB$ stronger than the targeting signal (i.e., SIR=$-20dB$).
The SER drops dramatically to 1\% as the SIR decreases from $-20dB$ to $-10dB$.
We further evaluate the SER in different spreading factor settings.
Figure~\ref{subfig-nonlinear-vs-linear-sir-2} shows the minimum SIR required by each type of chirps in order to achieve less than 1\% symbol error rate.
We observe that the linear chirp requires a minimal SIR of around $0dB$ in all six spreading factor settings.
In contrast, the non-linear chirps require a minimal SIR less than $0dB$, and the SIR requirement drops dramatically with increasing spreading factor. 
These results clearly demonstrate that the non-linear chirp by its own design is more scalable to near-far issues than its linear chirp counterpart.

\vspace{1mm}
\noindent
\textbf{The impact of symbol time offset.} We define $t_{gap}$ as the symbol time offset between two colliding symbols $A$ and $B$ (shown in Figure~\ref{subfig-nonlinear-vs-linear-tgap-1}).
Suppose the current observing window aligns with symbol $A$, then after dechirping, the power of the interfering symbol $B$ will be scattered into multiple FFT bins.
The amplitude of these scattered FFT peaks  is proportional to 1-$t_{gap}$ because only those signal samples that are within the overlapping window will contribute to the energy peaks.
Hence a smaller $t_{gap}$ will lead to stronger interfering peaks.
We vary $t_{gap}$ and plot the energy peaks in Figure~\ref{fig-nonlinear-vs-linear-tgap}.

Figure~\ref{subfig-nonlinear-vs-linear-tgap-2} shows the energy peaks of linear chirps. When SIR=$-1dB$, the targeting peak is still distinguishable from the interfering  peak across all four $t_{gap}$ settings.
When SIR drops to $-5dB$ (shown in Figure~\ref{subfig-nonlinear-vs-linear-tgap-3}), the interfering peak grows dramatically with decreasing $t_{gap}$. It finally surpasses the targeting peak when $t_{gap}$ drops to 20\%.
When SIR grows to -10$dB$ (shown in Figure~\ref{subfig-nonlinear-vs-linear-tgap-4}), the interfering peak easily exceeds the targeting peak in 3/4 $t_{gap}$ settings.
In contrast, when a non-linear chirp is adopted, we merely observe tiny energy peaks induced by the interfering symbol $B$. The targeting peak is easily distinguishable even in the case that the colliding symbol B is almost aligned with the targeting symbol A (i.e., $t_{gap}$=20\%) across all three SIR settings, as shown in  Figure~\ref{subfig-nonlinear-vs-linear-tgap-6}-(h).
These results manifest that the non-linear chirp is robust to symbol time offset. In \S\ref{subsec-system-1} we further demonstrate that by adopting different forms of non-linear chirps, the receiver can even demodulate two well-aligned collision symbols (i.e., $t_{gap}$=0) --- a case that none of existing LoRa collision demodulation approaches can deal with.

\subsection{Noise Tolerance}
\label{subsec-design-3}

\begin{figure}[t]
\centering
\subfigure[SER fluctuates with SNR in various spreading factor settings.]{ 
	    \includegraphics[width=0.22\textwidth]{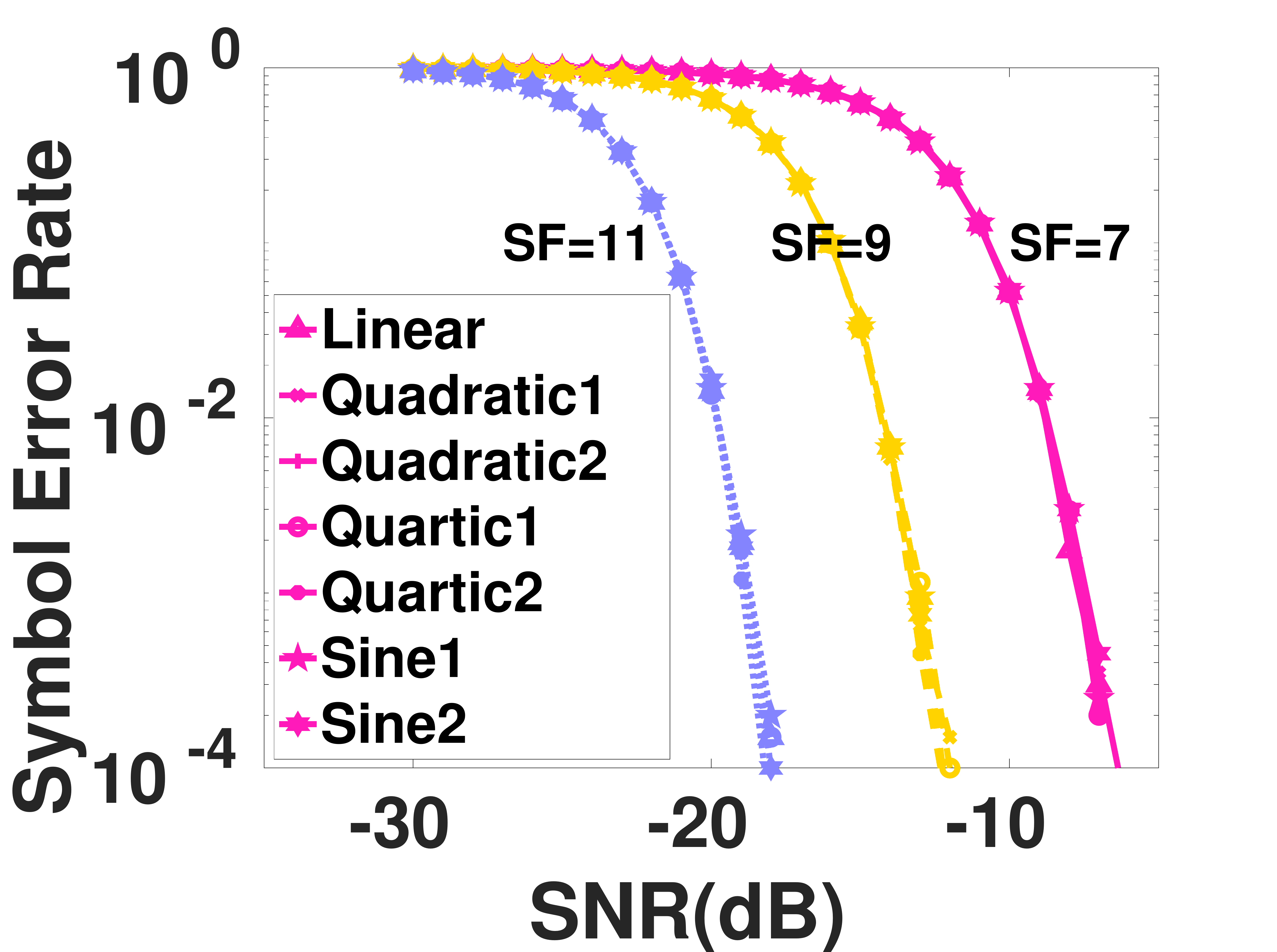}
        \label{subfig-nonlinear-vs-linear-noise-1}
}
\subfigure[The CDF of SER in various spreading factor settings.]{
	\includegraphics[width=0.22\textwidth]{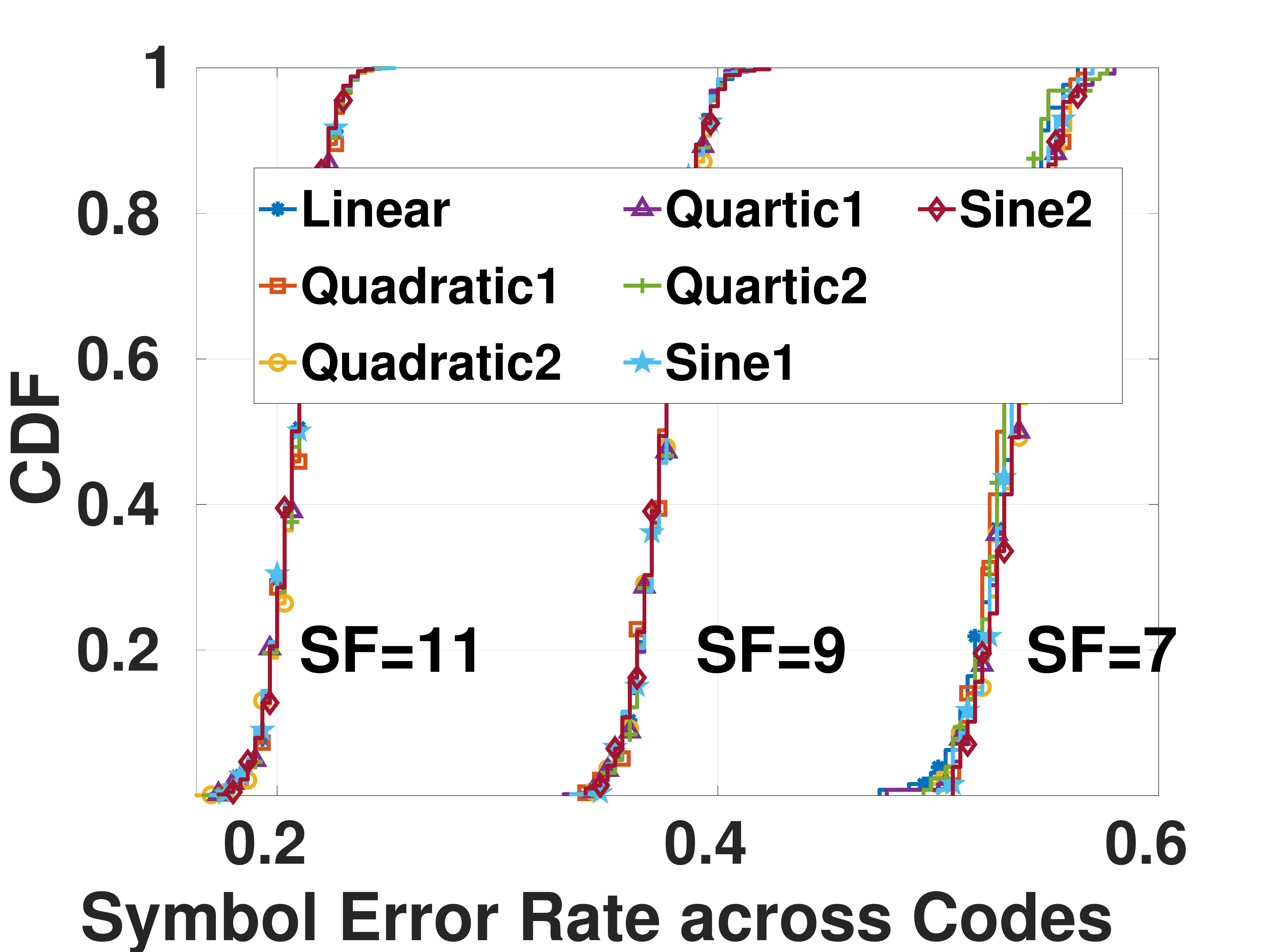}
        \label{subfig-nonlinear-vs-linear-noise-2}
}
\vspace{-3mm}
\caption{Comparing the SER of various types of non-linear chirps in different SNR conditions.}
\label{fig-nonlinear-vs-linear-noise}
\vspace{-4mm}
\end{figure}


We next demonstrate that the non-linear chirp achieves the same noise tolerance as the linear chirp counterpart does.



\vspace{1mm}
\noindent
\textbf{Validation.} We evaluate the noise tolerance of six non-linear chirps that cover a range of shapes and convexity (\S\ref{subsec-system-1}): 
\vspace{-0.15in}
\begin{table}[h]
\small
\centering
\begin{adjustbox}{width=1\columnwidth}
\begin{tabular}{ll}
\hline 
 (1): $quadratic1$---$f(t) = t^2$ & (2): $quadratic2$---$f(t) = -t^2+2t$ \\
 (3): $quartic1$---$f(t) = t^4$ & (4): $quartic2$---$f(t) = -t^4+4t^3-6t^2+4t$ \\
 (5): $Sine1$---$f(t) = sin(t),t\in[-\pi/2,\pi/2)$ & (6): $Sine2$---$f(t) = sin(t),t\in[-3\pi/8,3\pi/8)$ \\
\hline 
\end{tabular}
\end{adjustbox}
\vspace{-0.15in}
\end{table}

Figure~\ref{subfig-nonlinear-vs-linear-noise-1} shows the SER achieved by these chirps in three spreading factor settings.
We observe that all these six types of non-linear chirps achieve consistent symbol error rates with their linear chirp counterpart across all three different spreading factor settings.
In particular, when $SF=11$, the receiver achieves 1\% SER on both non-linear and linear chirps in an extremely low SNR condition (i.e., $-20dB$).
The minimal SNR (for achieving the same SER) then grows to $-14dB$ and further to $-9dB$ as the spreading factor drops to 9 and 7, respectively.
The result demonstrates that the non-linear chirp achieves superior noise tolerance as the linear chirp does.

Since the LoRa symbol varies with the chirp's initial frequency offset, given a certain type of non-linear chirp, one may worry that the SER of this chirp may not be consistent across different LoRa symbols.
To validate this concern, we generate different chirp symbols by varying the initial frequency offset of a standard up-chirp. We then compare its symbol error rate with linear-chirps in the same SF settings.
As shown in Figure~\ref{subfig-nonlinear-vs-linear-noise-2}, the linear and non-linear chirps achieve very similar SER in all three SF settings.



\subsection{Power Consumption}
\label{subsec-design-4}

Next, we show that the non-linear chirp generation consumes the same order of power as the linear chirp generation does.
We leverage Direct Digital Synthesis (DDS)~\cite{dds_tutorials}, a digital signal processing method to generate chirp signals. 
Compared to other analog frequency synthesis~\cite{analog_frequency} or voltage-controlled oscillator (VCO)~\cite{talla2017lora} based approaches, DDS is immune to both frequency and amplitude drifts and thus has been widely adopted for chirp signal generation in a radar system, e.g., frequency modulated continuous wave (FMCW) radars~\cite{dds_fmcw1, dds_fmcw2} and synthetic aperture radars (SAR)~\cite{dds_SRA1, dds_SRA2}. 



DDS works as follows. It first generates a reference signal at a constant frequency $f_{clk}$, and stores the signal samples in a local buffer, called a phase-amplitude mapping-table. Let $L$ be the length of this mapping-table.
To generate a desired chirp signal, DDS then accesses the mapping-table following the equation defined below: 
\begin{equation}
\begin{split}
\label{eq:dds_frequency}
\phi_i=\sum\limits_{m=1}^{i}f_i=\sum\limits_{m=1}^{i}(f_{i-1}+K_i \times \frac{f_{clk}}{2^L})=\sum\limits_{m=1}^{i}\sum\limits_{j=1}^{m}K_i \times \frac{f_{clk}}{2^L}
\end{split}
\end{equation}
where $\phi_i$ and $f_i$ represent the phase and frequency of the $i^{th}$ sampling point of the chirp signal to be generated, respectively. $K_i$ is the slope of this chirp signal, describing how its frequency changes over time.
$K_i$ is a constant value for linear chirps. It varies over time for non-linear chirps.
The transmitter then retrieves these signal samples from the mapping-table and generates the chirp signal accordingly.

\begin{figure}[t]
    \centering
    \includegraphics[width=3.5in]{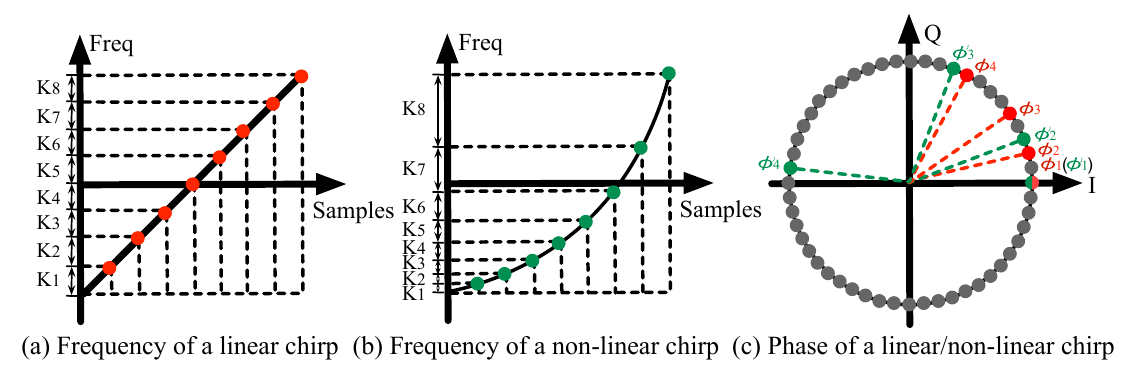}
    \vspace{-4mm}
    \caption{An illustration of DDS operation on generating the linear and non-linear chirps, respectively.}
    \label{fig:dds}
    \vspace{-4mm}
\end{figure}

Figure~\ref{fig:dds} describes DDS's high-level operations.
To generate a linear chirp (Figure~\ref{fig:dds}(a)), the transmitter sets $K_i$ to a constant value (i.e., 1) and accesses the frequency samples at index (1, 2, 3, 4, ...). The phase samples are retrieved at index (1, 3, 6, 10, ...).
In contrast, to produce a non-linear chirp (Figure~\ref{fig:dds}(b)), $K_i$ varies over time, e.g., ($K_1$=1, $K_2$=2, $K_3$=4, $K_4$=8, ...). The frequency and phase index then changes to (1, 3, 7, 15, ...) and (1, 4, 11, 26, ...), respectively. 

\vspace{1mm}
\noindent
\textbf{Validation.}  We prototype DDS on a Zynq-7000 FPGA~\cite{FPGA} and measure the power consumption of linear and non-linear chirp generation, respectively.
The FPGA board is equipped with an ultra low-power 12-bit ADC and a 256KB RAM.
The phase-amplitude mapping-table is generated by a 1~MHz clock signal. It stores $2^{12}$ sample points.
We then retrieve signal samples from this mapping table to generate chirps (bandwidth=125KHz, SF=7).
The sampling points of each chirp in total are 8192.
Our measurement study shows that the transmitter consumes the same order of power on generating the baseband of these two types of chirp signals: 315.6~$\mu W$ for non-linear chirps and 306.2~$\mu W$ for linear-chirps, respectively.
The up-conversion of baseband to RF band (900~MHz) consumes around 40~mW~\cite{upconverter} for both chirps.
Hence the total power consumption (baseband+RF) of the DDS-based approach is similar to commercial LoRa nodes~\cite{LoRa_Transceiver}.

\section{CurvingLoRa PHY-Layer}
\label{sec-system}
The above section shows a set of desirable properties of non-linear chirps, in this section, we describe the PHY-layer design on non-linear chirp modulation (\S\ref{subsec-system-1}), demodulation (\S\ref{subsec-system-2}), and the frame format for packet detection (\S\ref{subsec-system-3}).

\subsection{Modulation}
\label{subsec-system-1}

Similar to the standard linear chirp modulation in LoRa, \ours defines a base non-linear chirp and modulates it by varying its initial frequency offset. 

\noindent \textbf{Base non-linear chirp generation}. We define a base non-linear chirp as a monotonic curve growing from $(0, -\frac{BW}{2})$ to $(\frac{2^{SF}}{BW}, \frac{BW}{2})$, where the coordinate $(x,y)$ represents the (time, frequency) boundary of this chirp.
$SF$ and $BW$ denote the spreading factor and bandwidth, respectively.
Since a monotone nonlinear function can be approximated by the sum of a set of polynomial functions in time-frequency domain, the base non-linear chirp thus can be represents as:
\begin{equation}
\label{equ-nonlinear-baseline-up-chirp} 
    \begin{split}
    f_c(t) &= \sum^{n}_{i=0}k_it^i, t \in [0,\frac{2^{SF}}{BW}], f_c(t) \in [-\frac{BW}{2},\frac{BW}{2}] \\
    \end{split}
\end{equation}
where $k_i, i\in[0,n]$ are a set of coefficients to fit the non-linear curve into the range of symbol time and bandwidth. 
Notice that for a linear chirp, all these coefficients are zero except for $k_0=-\frac{BW}{2}$ and $k_1=\frac{BW^2}{2^{SF}}$.

To facilitate the coefficient configuration in different $BW$ and $SF$ settings, we further design a polynomial chirp function in a unified space ($[0,1]_x\times[0,1]_{f(x)}$) as follow:
\begin{equation} 
\label{equ-unified-space-function}
    \begin{split}
    f(x) &= \sum^{n}_{i=0}a_ix^i, x \in [0,1], f(x) \in [0,1] \\
    \end{split}
\end{equation} 
where $a_i,i\in[0,n]$ is the $i^{th}$ coefficient. The relationship between the coefficient defined in the unified space and that defined in the time-frequency domain (i.e., $k_i$ in Equation~\ref{equ-nonlinear-baseline-up-chirp}) can be represented as follows:
\begin{equation}
	\label{equ-coefficient-mapping}
	\begin{split}
		k_0 = BW \times a_0-\frac{BW}{2},\;
		k_i = \frac{BW^{i+1}}{2^{SF\times i}}a_i, i\in[1,n]
	\end{split}
\end{equation} 

Given $BW$ and $SF$, we can compute the coefficient $k_i$ for each polynomial term defined in Equation~\ref{equ-nonlinear-baseline-up-chirp} and generate the base up-chirp accordingly.
The down-chirp can be generated by conjugating the base up-chirp.

\noindent \textbf{Base non-linear chirp modulation}. Once we have a base non-linear chirp, the transmitter then varies the initial frequency offset of this base chirp to modulate data:
\begin{equation} 
\label{equ-shifted-nonlinear-chirps}
    \begin{aligned}
        h(t) &=  
    e^{j 2 \pi (f_0+f_c(t)) t} 
    \end{aligned}
\end{equation}
where $f_0$ is the initial frequency offset of this chirp.
In essence, given the same $BW$ and $SF$ configurations, \ours achieves the same link throughput with the standard linear-chirp based LoRa.

\begin{figure}[t]
    \centering
    \subfigure[Linear up-chirp]{ 
        \includegraphics[width=0.33\columnwidth]{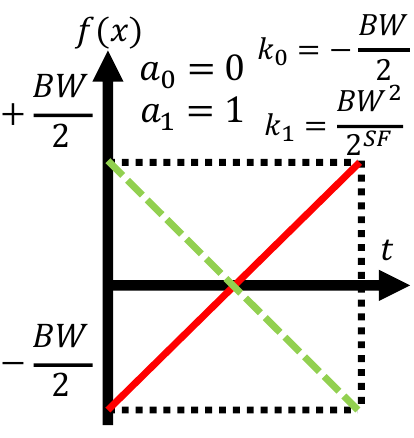}
        \label{subfig-linear-mapping}
        }
    \hspace{-0.06\columnwidth}
    \subfigure[Convex up-chirp]{
        \includegraphics[width=0.33\columnwidth]{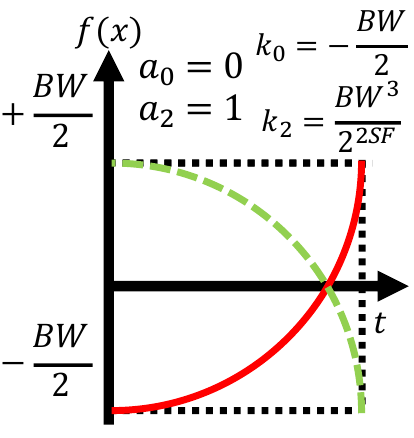}
        \label{subfig-convex-mapping}
        }
     \hspace{-0.06\columnwidth}
    \subfigure[Concave up-chirp]{
        \includegraphics[width=0.33\columnwidth]{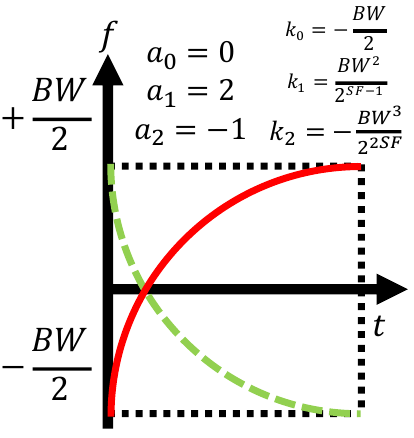}
        \label{subfig-concave-mapping}
        }
    \caption{An illustration of linear and non-linear chirps with the corresponding function parameters.
    }
    \label{fig-unified-symbol-modulation}
    \vspace{-4mm}
\end{figure}

\noindent \textbf{Modulation knobs}. By using different polynomial functions defined in the unified space (e.g., $f(x)=x$, $f(x)=x^2$ and $f(x)=2x-x^2$), the transmitter can easily build different base chirps.
Figure~\ref{fig-unified-symbol-modulation} shows a convex and a concave non-linear chirp produced by two different polynomial functions.
These different polynomial functions provide us another knob to boost the capacity of concurrent LoRa transmissions.
To understand the rationale behind this, let's consider a case where two LoRa transmissions (i.e., $S_A$ and $S_B$) are happenly well aligned at the receiver.
Let $S_R$ be their superposition.


\begin{figure}[t]
    \centering
    \subfigure[L. meets L.]{ 
        \includegraphics[width=0.30\columnwidth]{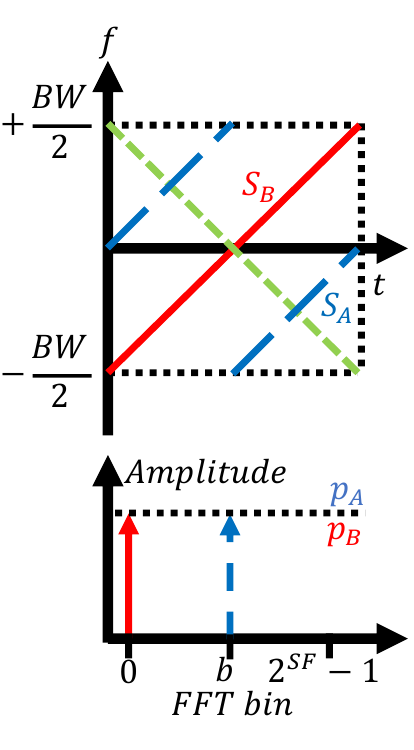}
        \label{subfig-knobs-1}
        }
    \subfigure[NL. meets NL.]{
        \includegraphics[width=0.30\columnwidth]{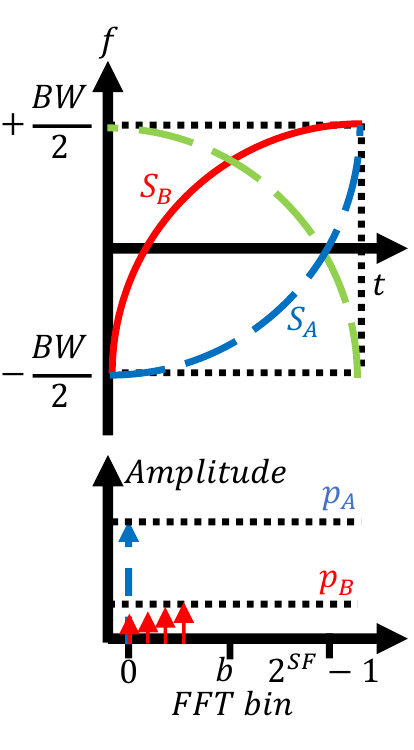}
        \label{subfig-knobs-2}
        }
    \subfigure[NL. meets L.]{
        \includegraphics[width=0.30\columnwidth]{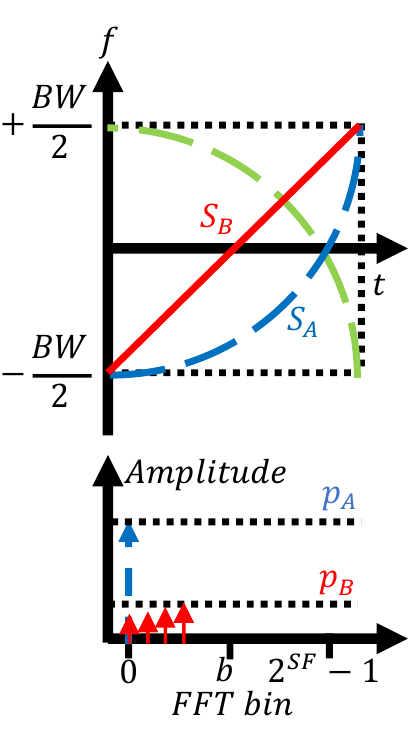}
        \label{subfig-knobs-3}       
        }
    \caption{An illustration of symbol collisions.
    {\normalfont
    (a): A linear chirp (L.) collides with another linear chirp. (b): A non-linear (NL.) chirp collides with another non-linear chirp. (c): A non-linear chirp collides with a linear chirp.}}
    \vspace{-4mm}
\label{fig-knobs}
\end{figure}

\noindent $	\bullet$ \textbf{Case one}: when both $S_A$ and $S_B$ are linear chirps, we are expected to see two separate energy peaks on FFT bins (shown in Figure~\ref{subfig-knobs-1}). In this case, all existing parallel decoding approaches~\cite{mLoRa,oct,xia2019ftrack,colora,tong_combating_2020,xu_fliplora_2020,eletreby2017empowering,hu2020sclora} fail to disambiguate the collision symbols as these two well-aligned symbols exhibit similar FFT peaks. 
    
\noindent $	\bullet$ \textbf{Case two}: when both $S_A$ and $S_B$ are non-linear chirps (i.e., generated by two different polynomial functions), the receiver can decode each symbol from their collision as follows.
The receiver first multiplies $S_R$ with the conjugate of $S_A$. As a result, the energy of $S_B$ will be spread over multiple FFT bins, whereas the energy of $S_A$ will concentrate on a single, isolated FFT bin, as shown in Figure~\ref{subfig-knobs-2}.
The receiver can easily pick up this energy peak and decode $S_A$. $S_B$ can be decoded similarly by replacing the down-chirp with the conjugate of $S_B$.

\noindent $	\bullet$ \textbf{Case three}: when one of the LoRa symbols is based on linear chirp and another is based on non-linear chirp, similar to case two, the receiver can alternate between different down-chirps to decode each of them accordingly, as shown in Figure~\ref{subfig-knobs-3}.

We have three takeaways from the above analysis: $i)$ the transmitters can use different types of non-linear chirps as an orthogonal approach to boost the concurrency of LoRa transmissions.
$ii)$ the non-linear chirp based LoRa nodes can  co-exist with those linear-chirp based legacy LoRa nodes. 
$iii)$ the adoption of different non-linear chirps also facilitates the demodulation of well-aligned collision symbols.

\subsection{Demodulation}
\label{subsec-system-2}

\begin{figure}[t]
    \centering
    \subfigure[with STO and CFO] {
    \includegraphics[width=0.29\columnwidth]{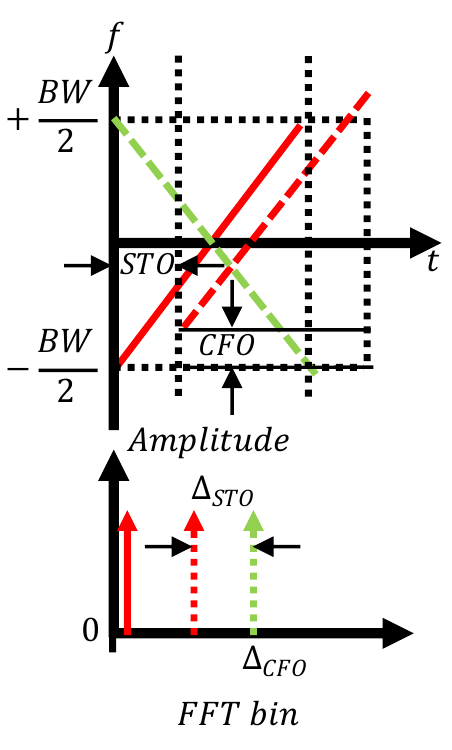}
    	\label{subfig-symbol-alignment}
    }
    \subfigure[with CFO]{ 
    \includegraphics[width=0.29\columnwidth]{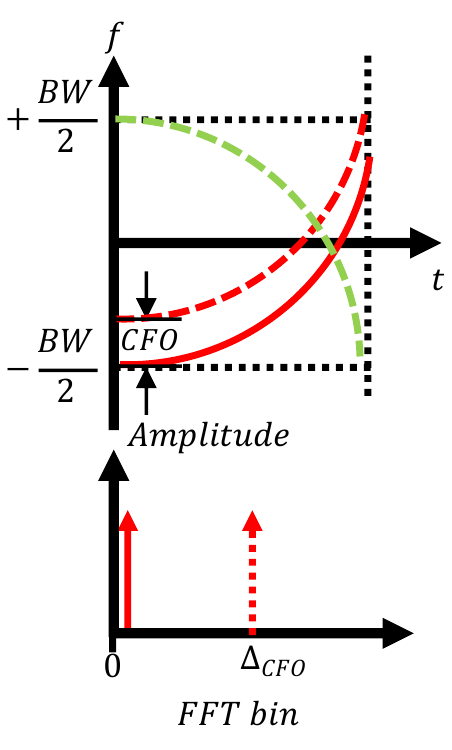}
        \label{subfig-fixCFO1}
    }
    \subfigure[with STO and CFO]{
        \includegraphics[width=0.29\columnwidth]{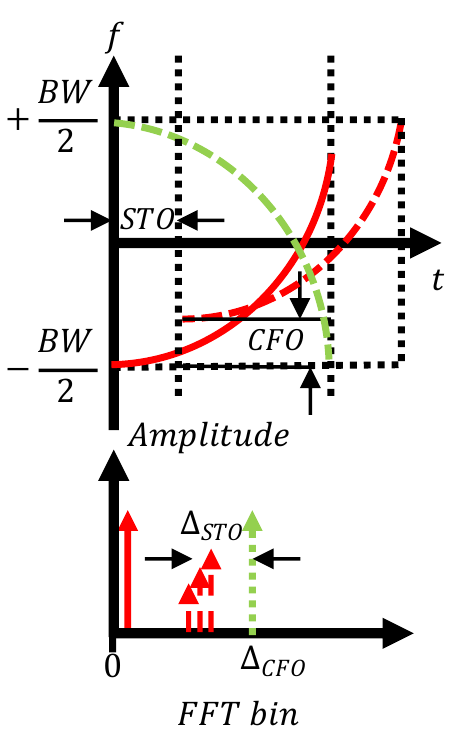}
        \label{subfig-fixCFO2}
    }
    \caption{The influence of symbol time offset (STO) and carrier frequency offset (CFO) on demodulation. 
    {\normalfont
    (a) both STO and CFO move the energy peak of a linear chirp from its desired FFT bin.
    (b) CFO moves the energy peak of a non-linear chirp from its desired FFT bin. 
    (c) STO spreads the power of a non-linear chirp into multiple FFT bins.}}
    \vspace{-4mm}
     \label{fig-fixCFO}
\end{figure}

Similar to linear chirp demodulation, the receiver operates dechirp to demodulate non-linear chirps.

\noindent \textbf{Accounting for the Misalignment.} 
Symbol alignment is critical to the demodulation performance, particularly for non-linear chirp demodulation, as the misalignment will spread the spectrum power of a chirp symbol into multiple frequency points, which fails the demodulation.
While this misalignment, in theory, is only caused by the symbol time offset (STO) between the incident chirp symbol and the down-chirp, in practice, it is also affected by the carrier frequency offset (CFO) caused by clock offset.

In linear chirp demodulation, the dechirp converges the spectrum power of each linear chirp symbol to a specific frequency point.
The existence of STO and CFO both renders the energy peak merely deviates from its desired position in FFT bins.
After the dechirp, the receiver can thus leverage the preamble to estimate such frequency shift and then correct the symbol by applying the estimated frequency shift to the energy peak.
However, such a post-processing approach cannot be directly applied to non-linear chirp, as the existence of STO will instead spread the spectrum energy into multiple FFT bins.
Hence the receiver has to align the chirp symbol with the down-chirp and compensate for the CFO before operating dechirp on each non-linear chirp symbol.


To better understand this issue, we take Figure~\ref{fig-fixCFO} as an example, where the receiver demodulates the linear chirp and non-linear chirp, respectively.
To align the incoming linear chirp shown in Figure~\ref{fig-fixCFO}(a), the LoRa receiver simply operates multiplication on these two chirps.
Due to the symbol time offset, the resulting FFT peak will be shifted from its desired bin by the amount of $\Delta_{STO}$.
CFO leads to an extra shift of the FFT peak $\Delta_{CFO}$.
By leveraging the preamble in the LoRa header, the receiver can easily estimate $\Delta_{CFO}$+$\Delta_{STO}$ and offset their impact on the energy peak.
In contrast, the multiplication of two misaligned non-linear chirps (shown in Figure~\ref{fig-fixCFO}(b)) spreads the spectrum energy into multiple FFT bins, as shown in Figure~\ref{fig-fixCFO}(b).
The existence of CFO further shift these FFT peaks and complicate the symbol alignment, as shown in Figure~\ref{fig-fixCFO}(c).


In \ours, we put a pair of conjugate chirps---a standard linear up-chirp followed by a standard linear down-chirp---as the pilot symbols of a LoRa packet to estimate the STO and CFO.
Suppose these two linear chirps
are well aligned with their conjugate counterpart in dechirping
process, respectively. 
The resulting two FFT peaks are supposed
to be superimposed at the same FFT bin in the absence of CFO.
On the contrary, the existence of $STO$ and $CFO$ will shift these two FFT peaks by the amount of $\Delta_{CFO}+\Delta_{STO}$ and $-(\Delta_{CFO}+\Delta_{STO})$ from their desired position.
The receiver then estimates the STO and CFO using the similar method as in NScale~\cite{tong_combating_2020} and offsets the symbol misalignment and carrier frequency offset accordingly.
It then operates dechirp on the corrected symbols to demodulate each symbol.

\subsection{Frame Format}
\label{subsec-system-3}

\begin{figure}
    \centering
    \includegraphics[width=0.45\textwidth]{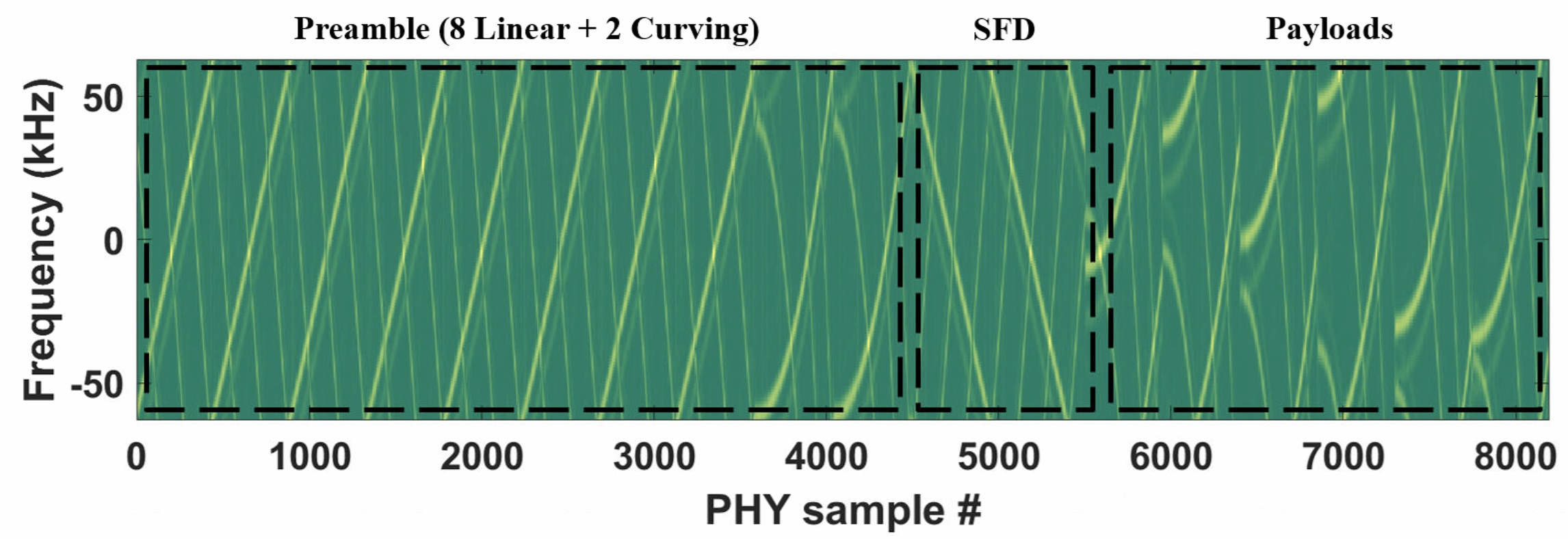}
    \vspace{-4mm}
    \caption{Packet format of \ours.}
    \label{fig-frame-format}
    \vspace{-4mm}
\end{figure}

A typical LoRa packet comprises multiple preamble symbols, 2 mandatory sync word symbols, 2.25 Start Frame Delimiter (SFD) symbols followed by a variable number of payload symbols~\cite{liando2019known,tong_combating_2020}.
Following the standard LoRa packet format, we encode the sync word symbols and payloads with non-linear chirps while retaining the linear chirps in preambles and SFDs, shown in Figure~\ref{fig-frame-format}.
The preamble contains 8 identical linear up-chirps for packet detection and alignment, followed by 2 non-linear chirps of sync word for configuration recognition of payloads.
The SFD consists of 2.25 standard down-chirps while the payload contains multiple chirp symbols with configurable length and chirp type. 
As mentioned in $\S$\ref{subsec-system-2}, a pair of up-chirp and down-chirp (i.e., pilot symbols) is needed to facilitate the symbol alignment.
Instead of putting an extra pair of such pilot symbols on the LoRa packet, we reuse the last linear up-chirp symbol in the LoRa preamble and the first linear down-chirp symbol of SFD as the pilots.
The use of linear chirp-based preamble may introduce the following two types of collisions:

\noindent \textbf{$\bullet$} Linear chirps collide with non-linear chirps when the preamble of one packet happenly aligns with the payload of another packet. In this case, the receiver can still leverage the energy scattering and converging effect to detect the preamble and further demodulate each individual signal (\S\ref{subsec-system-1}).

\noindent \textbf{$\bullet$} Linear chirps collide with linear chirps when the preamble of one packet happenly aligns with the preamble of another packet. In practice, however, this case rarely happens as the preamble contains only eight symbols, whereas the payload may last for hundreds of symbols~\cite{colora,tong_combating_2020}.

\section{Implementation}
\label{sec-implementation}

\vspace{1mm}
\noindent
\textbf{Hardware and software}. We implement \ours on software defined radios USRP N210 equipped with a UBX daughter board.
The modulation and demodulation are implemented based on UHD+GNURadio~\cite{url:gnu-radio}.
The transmitter and receiver work on the 904.0MHz ISM band, equipped with a VERT900 antenna~\cite{VERT900}. 
By default, the spreading factor (SF) and bandwidth (BW) are set to 10 and 125 kHz, respectively.
The sampling rate is 1 MHz.

\begin{figure}
    \centering
    \includegraphics[width=0.37\textwidth]{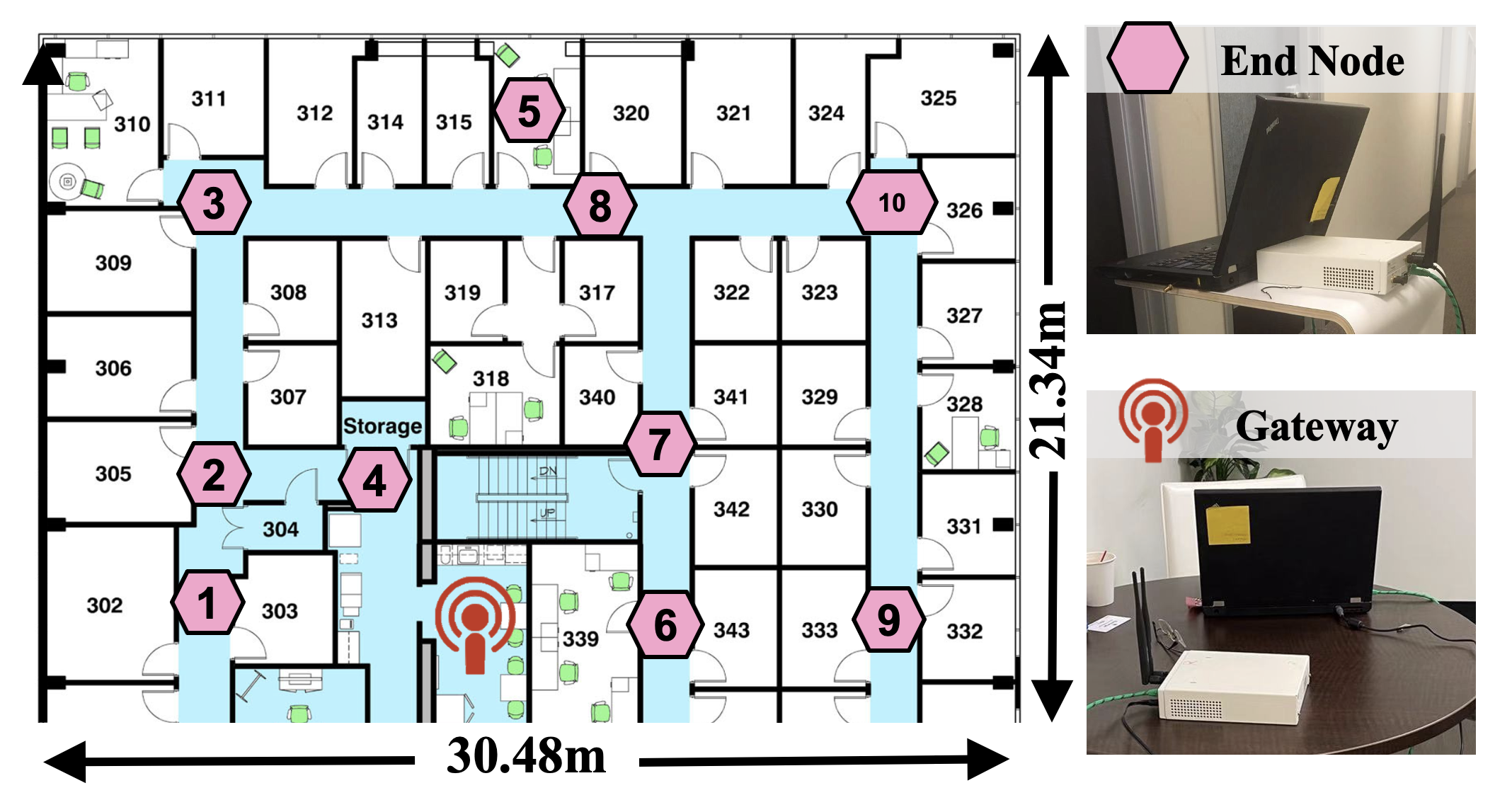}
    \caption{The indoor experimental plan and SDR devices spread out across tens of rooms.}
    \label{fig-indoor-environment}
    \vspace{-4mm}
\end{figure}

\begin{figure}
    \centering
    \includegraphics[width=0.4\textwidth]{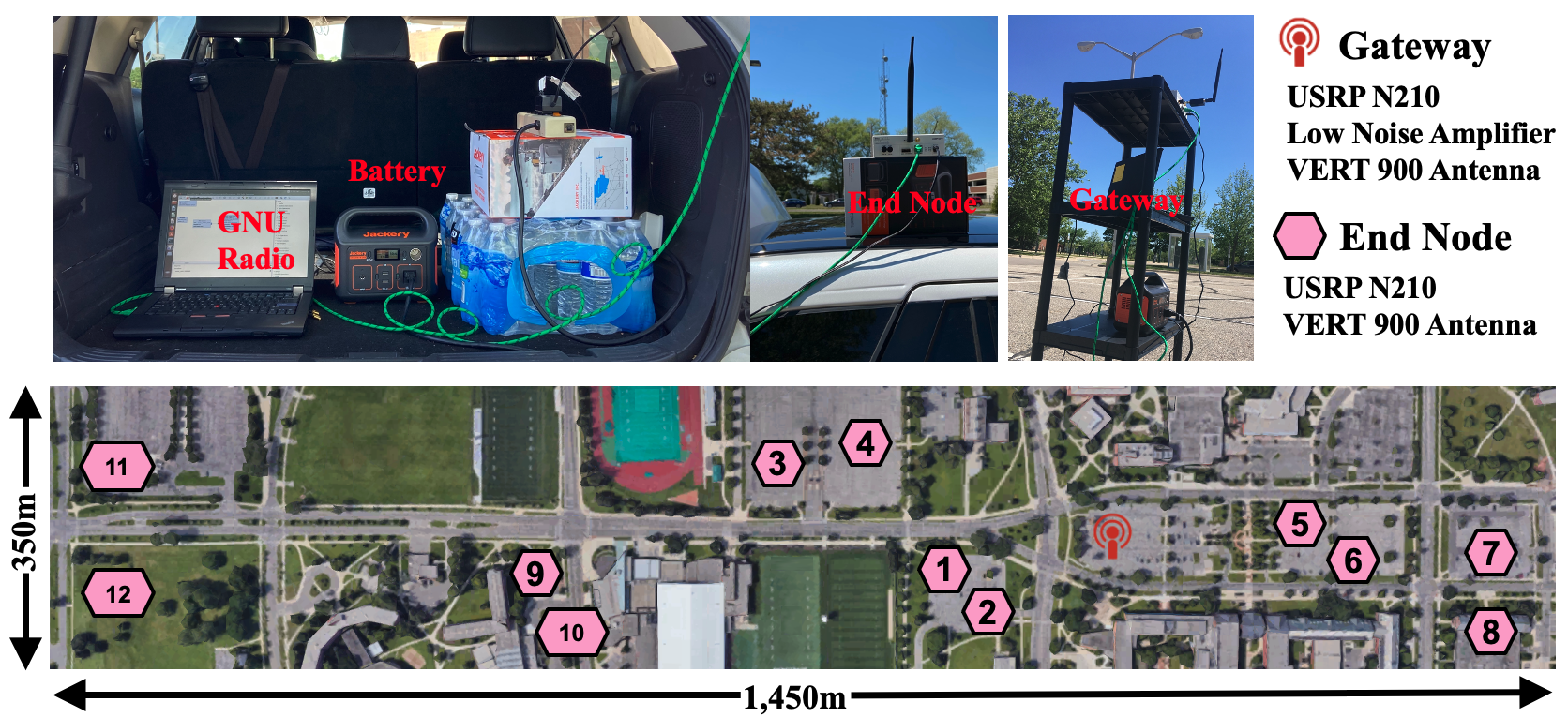}
    \vspace{-4mm}
    \caption{Bird view of the outdoor experiment field with the mobile gateway and LoRa nodes.}
    \label{fig-outdoor-environment}
    \vspace{-4mm}
\end{figure}

\begin{figure}[t]
\centering
\subfigure[Overall SER performance]{
    \includegraphics[width=0.495\columnwidth]{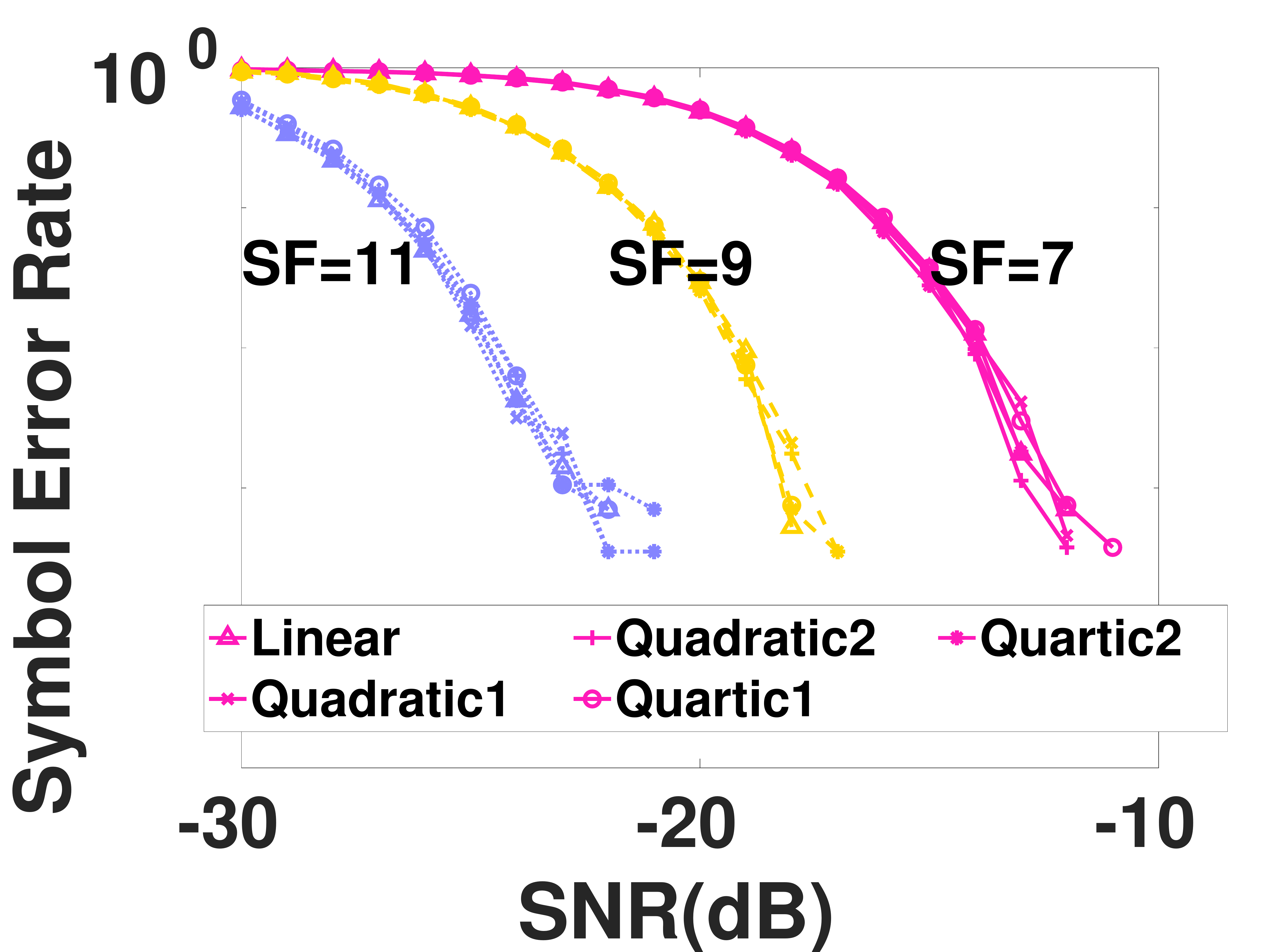}
    \label{subfig-noise-resilience-no-collisions-1}
}\hspace{-0.06\columnwidth}
\subfigure[SER distribution in code space]{
	\includegraphics[width=0.495\columnwidth]{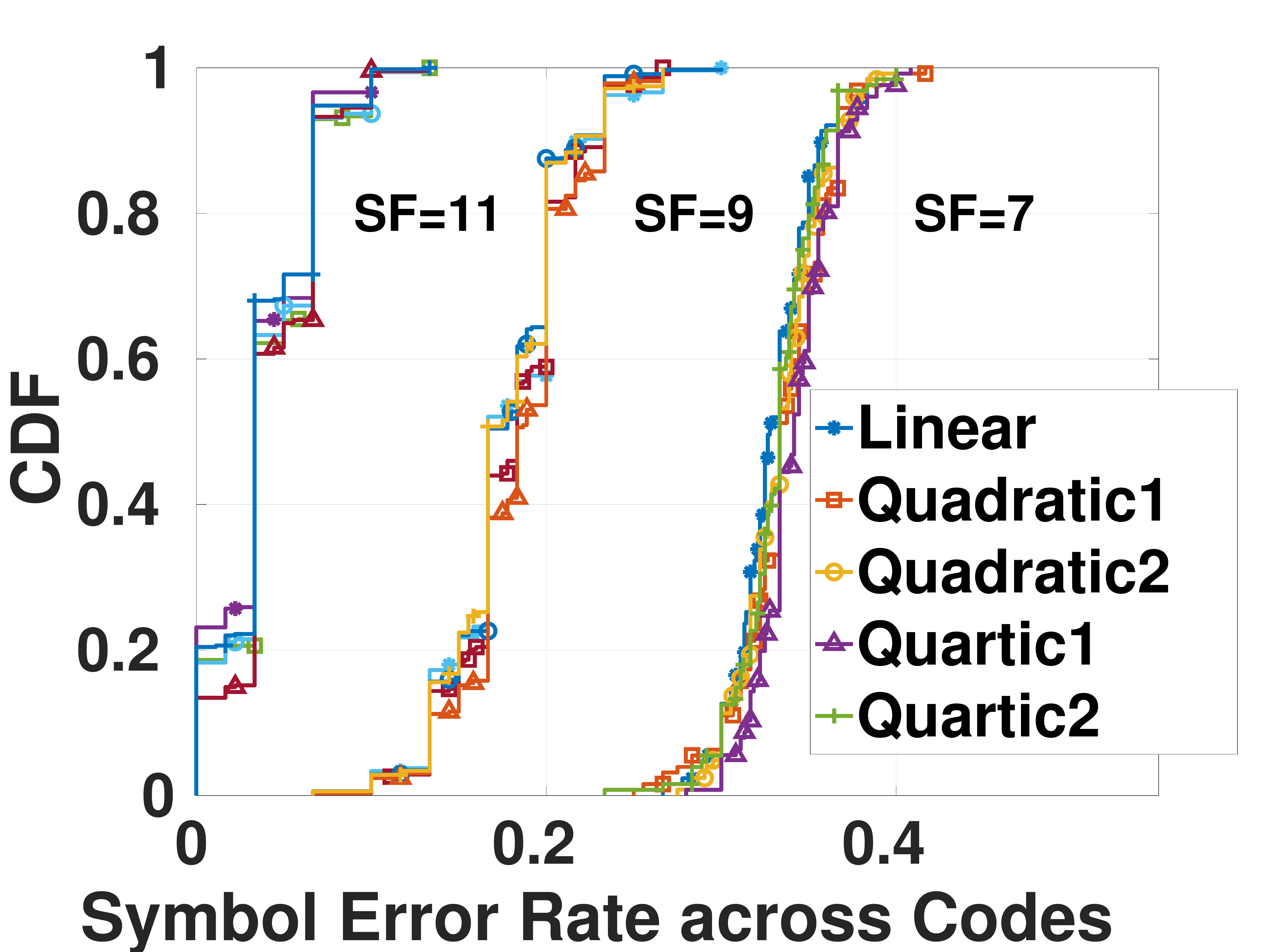}
    \label{subfig-noise-resilience-no-collisions-2}
}
\caption{Noise resilience in the absence of collisions.}
\label{fig-noise-resilience-no-collisions}
\vspace{-4mm}
\end{figure}

\begin{figure*}[t]
    \centering
    \subfigure[SER vs. SNR when SF=8]{
    \includegraphics[width=0.545\columnwidth]{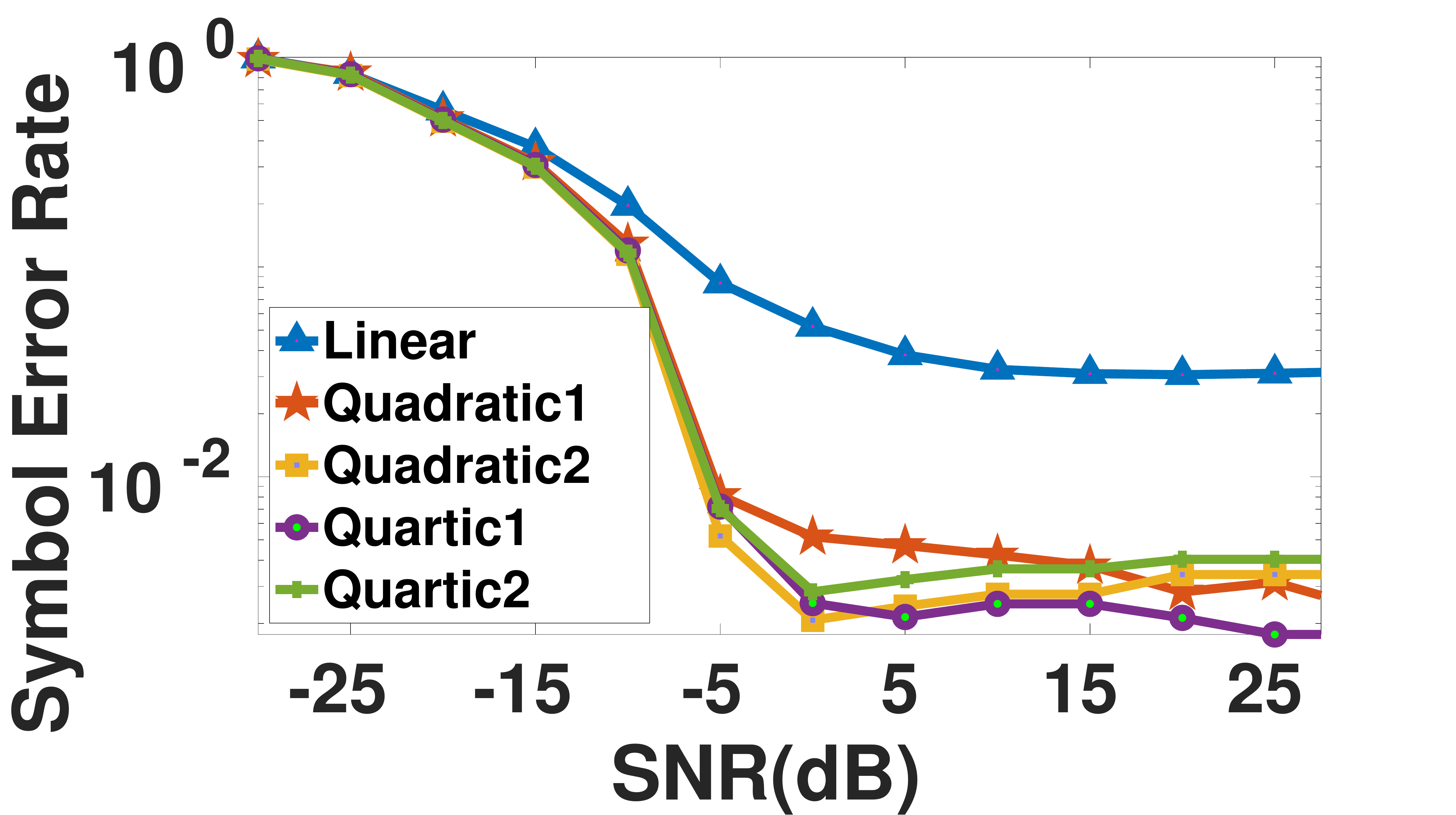}
    \label{subfig-snr-in-collisions-symbol-1}
    }
    \hspace{-0.08\columnwidth}
      \subfigure[SER vs. SNR when SF=10]{
    \includegraphics[width=0.545\columnwidth]{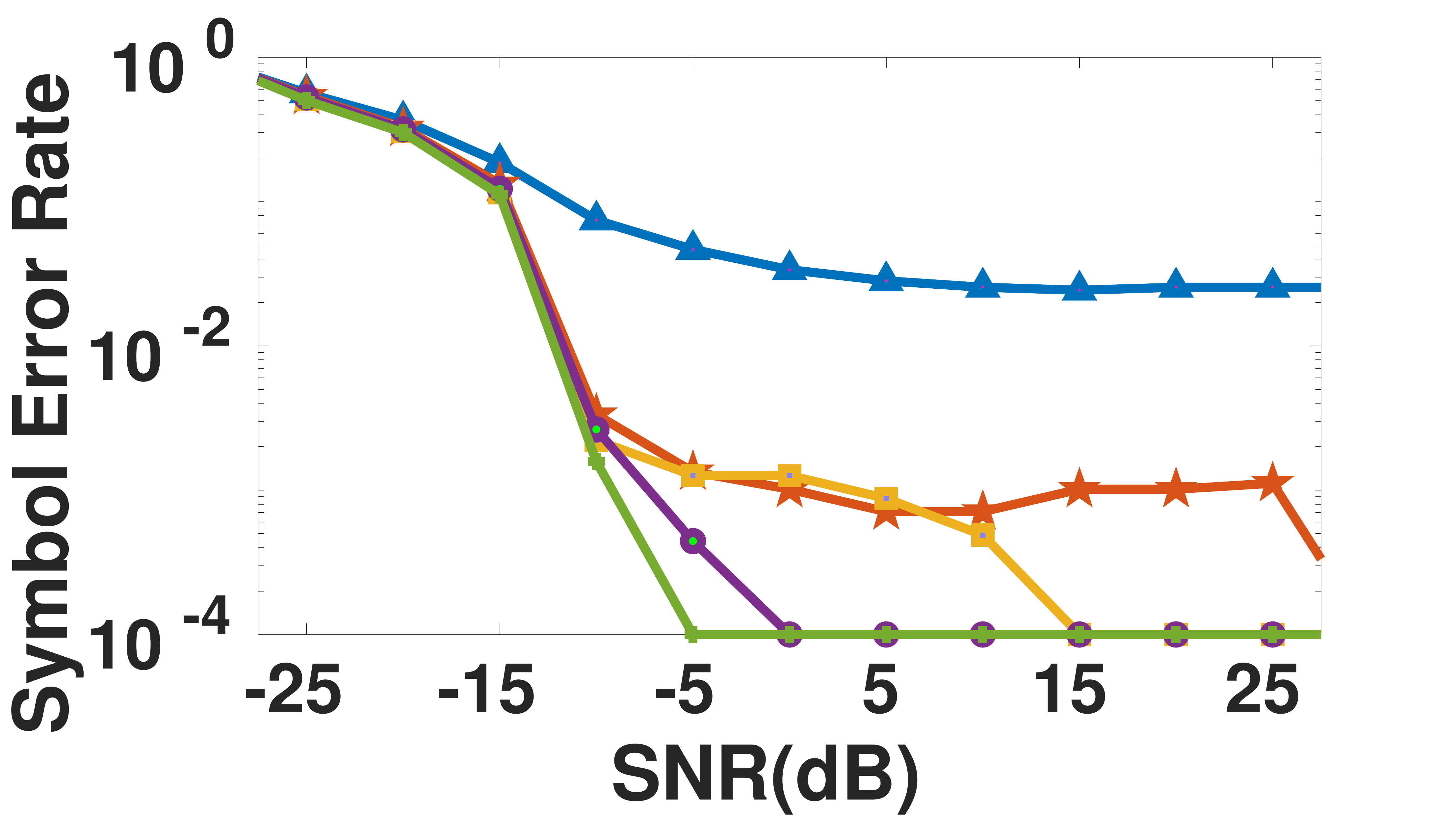}
    \label{subfig-snr-in-collisions-symbol-2}
    }
    \hspace{-0.08\columnwidth}
    \subfigure[SER vs. SNR when SF=12]{
    \includegraphics[width=0.545\columnwidth]{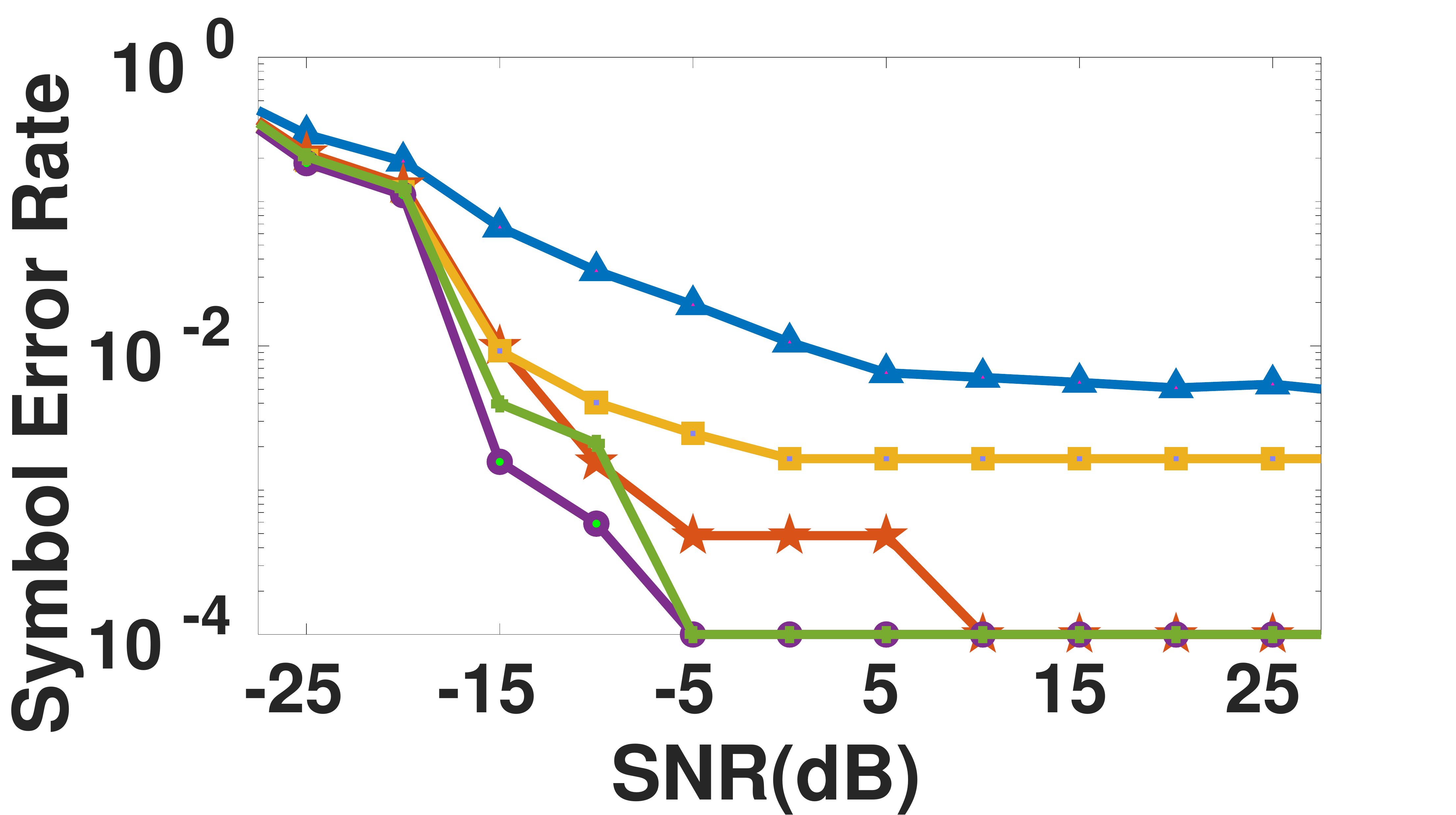}
    \label{subfig-snr-in-collisions-symbol-3}
    }
    \hspace{-0.08\columnwidth}
    \subfigure[SER vs. symbol offset]{
    \includegraphics[width=0.545\columnwidth]{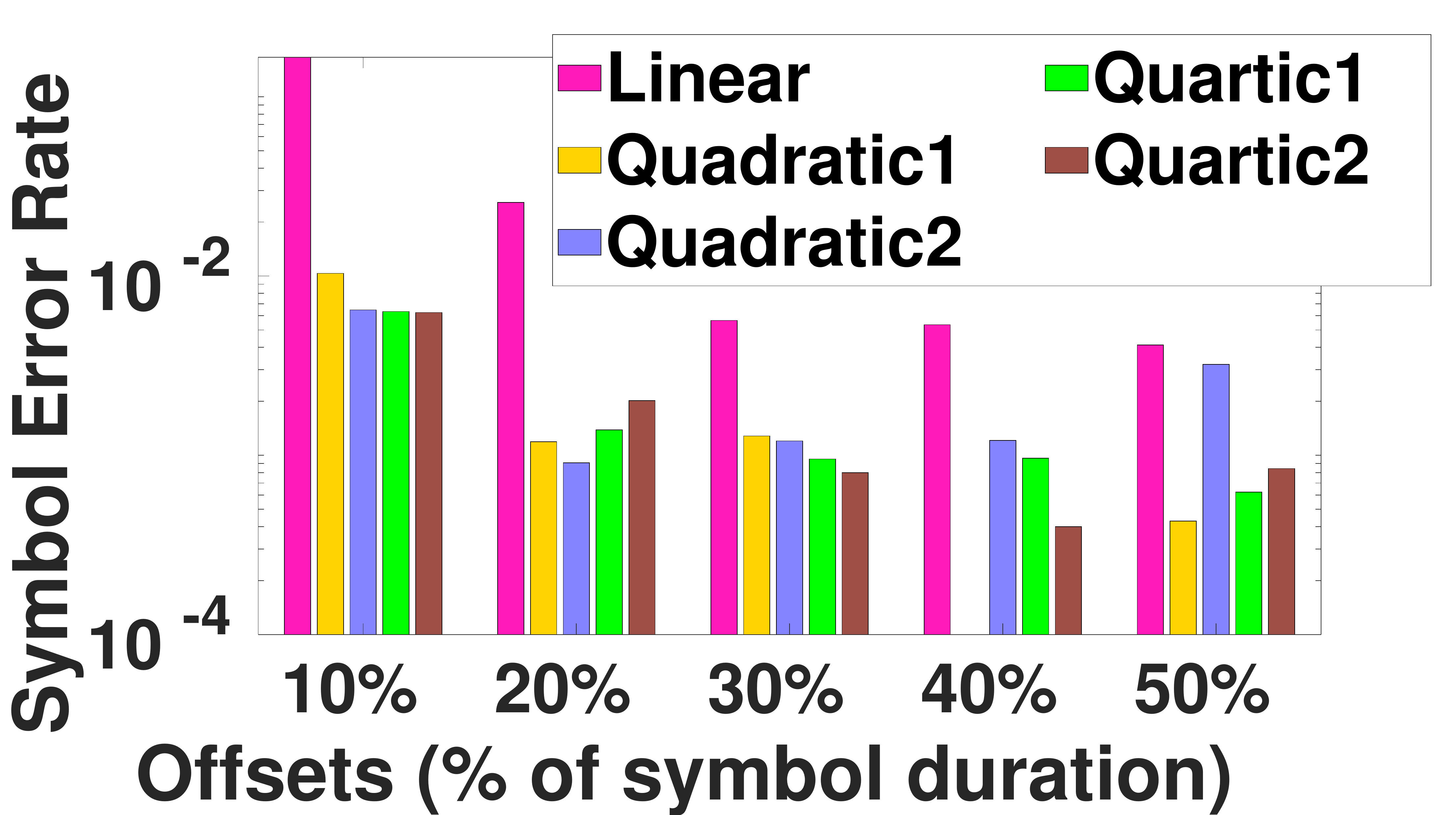}
    \label{subfig-snr-in-collisions-symbol-4}	
    }
\caption{Linear vs non-linear: symbol error rate ($SIR \approx 0dB$) in different spreading factor, SNRs and symbol offsets settings.}
\label{fig-snr-in-collisions-symbol}
\end{figure*}

\vspace{1mm}
\noindent
\textbf{Experiment setups}.
Due to LoRaWAN's constraint on channel access time (<1\% duty cycle ratio for each end node)~\cite{gamage_lmac_2020}, we conduct trace-driven emulations to evaluate our system. Specifically, we fix the gateway's location and move a transmitter to different sites. In each site, the transmitter sends packets in different spreading factors and chirp symbol settings.
We then align LoRa traces collected from different sites with varying symbol offset offline to emulate packet collisions.
The LoRa traces are collected in two different environments:

\noindent \textbf{$\bullet$} \textbf{Indoor scenario}. We place transmitters and gateway on a 30.48m$\times$21.34m office building. The offices are separated by concrete walls. Figure~\ref{fig-indoor-environment} shows the floor-plan of this office building. We place the gateway in the kitchen and move the transmitter to 10 different locations.
Due to the blockage of walls, most LoRa transmissions are under the non-line-of-sight (NLoS) condition.
    
\noindent \textbf{$\bullet$} \textbf{Outdoor scenario}. We deploy a campus-scale testbed outdoors.
The gateway powered by a UPS is placed on the parking lot. We move a transmitter to 12 locations and collect LoRa transmissions in both LoS and NLoS conditions with various link distances. The bird view of the outdoor testbed is shown in Figure~\ref{fig-outdoor-environment}. 




\vspace{1mm}
\noindent
\textbf{Large-scale packet collision emulation}. Due to the temporal diversity (e.g., the cars passing by may block the LoS path or generate a new reflection path), the LoRa traces collected from each site experience significantly different channel variations. This allows us to emulate large-scale LoRa networks by reusing each individual LoRa trace as a transmission from a new LoRa transmitter.
We further enhance the link diversity by varying the SIR of each trace at the gateway. The symbol offset is randomly chosen from [0.2, 0.8]$\times$$T_{symbol\_time}$.




\vspace{1mm}
\noindent
\textbf{Evaluation Metrics}. We adopt three metrics to evaluate \ours. $i)$: \textit{Symbol Error Rate (SER)} measures the demodulation of \ours at the symbol level, under various SNRs and SIRs~\cite{tong_combating_2020,colora}; $ii)$: \textit{Packet Delivery Rate (PDR)} computes the packet reception rate. in which 80\% of symbols can be decoded successfully.\footnote {Most error correction codes can recover 1/5 symbol errors~\cite{tse2005fundamentals}.} $iii)$:  \textit{Throughput} can be derived with the received packets and decoded symbols, denoted by Symbol/Second. 
Note that LoRa gateways are usually deployed with tethered power supplies, and thus we do not consider energy consumption at the gateway~\cite{colora,tong_combating_2020}. 

\vspace{1mm}
\noindent
\textbf{Baselines}. We compare our design with two SOTA LoRa collision decoding systems, namely, successive interference cancellation-based mLoRa~\cite{mLoRa} and spectral energy-based NScale~\cite{tong_combating_2020}.
The standard LoRaWAN is also adopted as the baseline.
As a proof of concept, we design four types of non-linear chirps to evaluate: 
\vspace{-0.15in}
\begin{table}[h]
\small
\centering
\begin{adjustbox}{width=1\columnwidth}
\begin{tabular}{ll}
\hline 
 (1): $quadratic1$---$f(t) = t^2$ & (2): $quadratic2$---$f(t) = -t^2+2t$ \\
 (3): $quartic1$---$f(t) = t^4$ & (4): $quartic2$---$f(t) = -t^4+4t^3-6t^2+4t$ \\
\hline 
\end{tabular}
\end{adjustbox}
\vspace{-0.15in}
\end{table}

\begin{figure*}[t]
    \centering
    \subfigure[SER vs. SIR when SF=8]{
    \includegraphics[width=0.535\columnwidth]{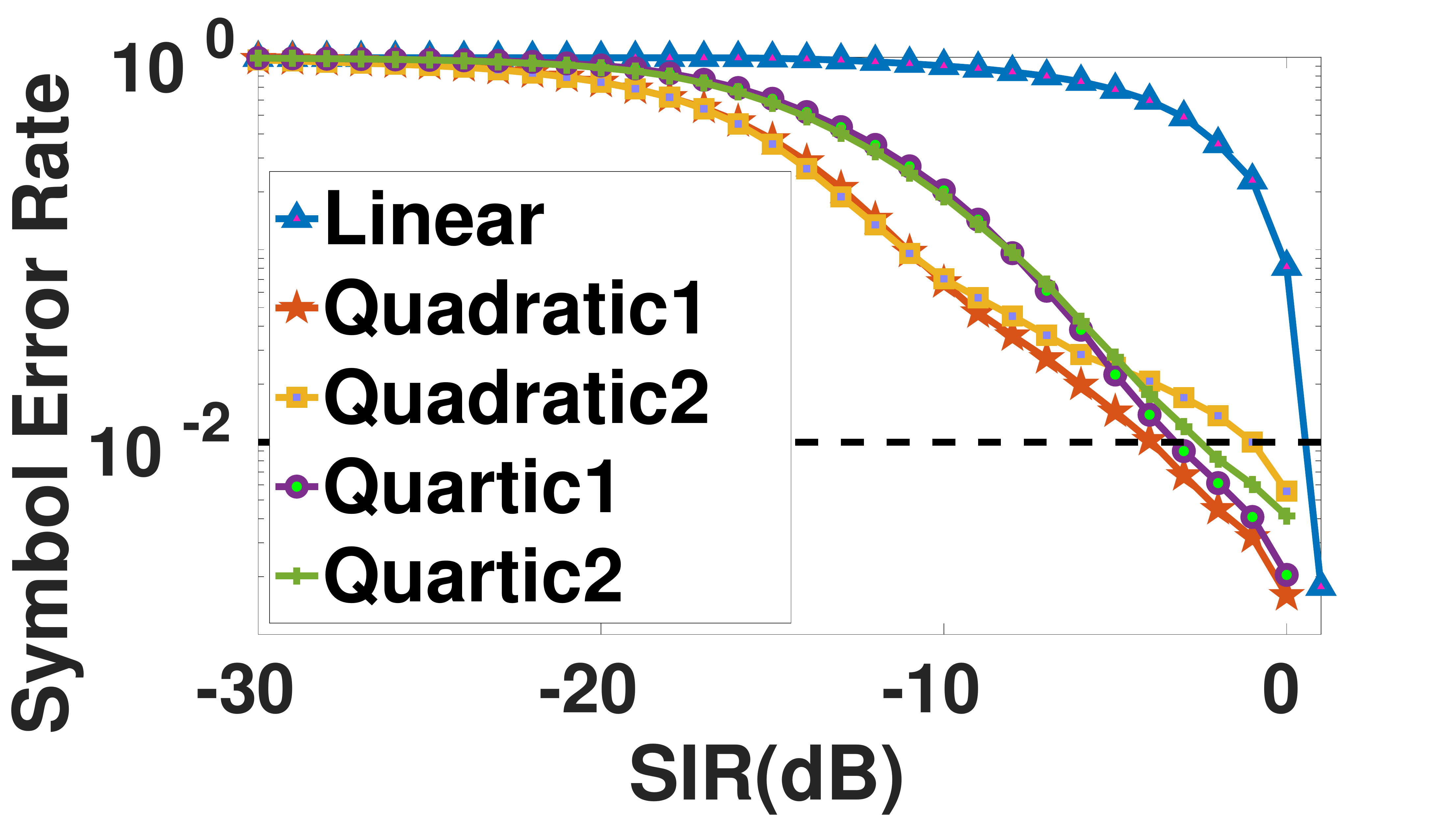}
    \label{subfig-sir-in-collisions-symbol-1}
    }
    \hspace{-0.06\columnwidth}
    \subfigure[SER vs. SIR when SF=10]{
    \includegraphics[width=0.535\columnwidth]{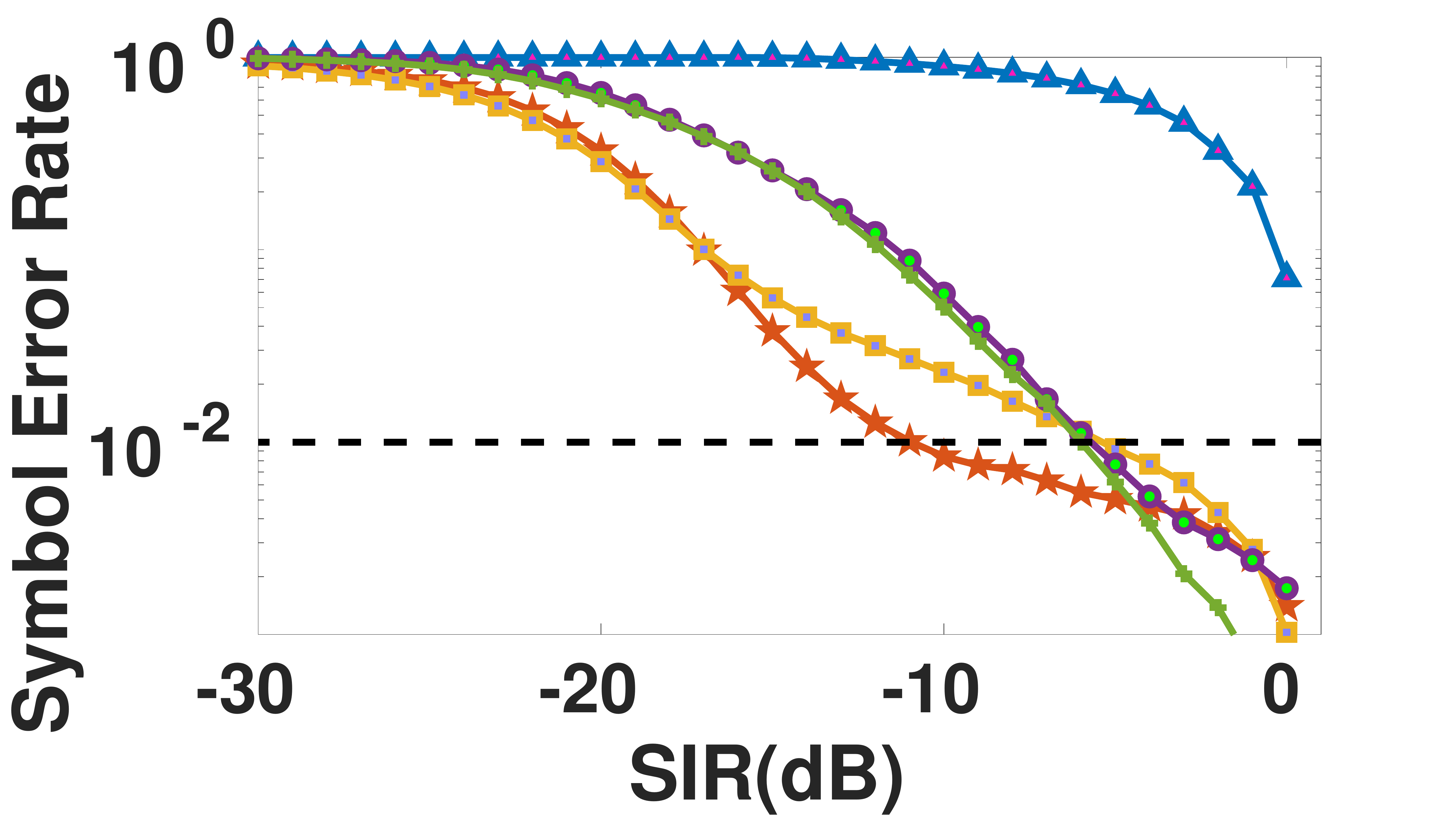}
    \label{subfig-sir-in-collisions-symbol-2}	
    }
    \hspace{-0.06\columnwidth}
    \subfigure[SER vs. SIR when SF=12]{
    \includegraphics[width=0.535\columnwidth]{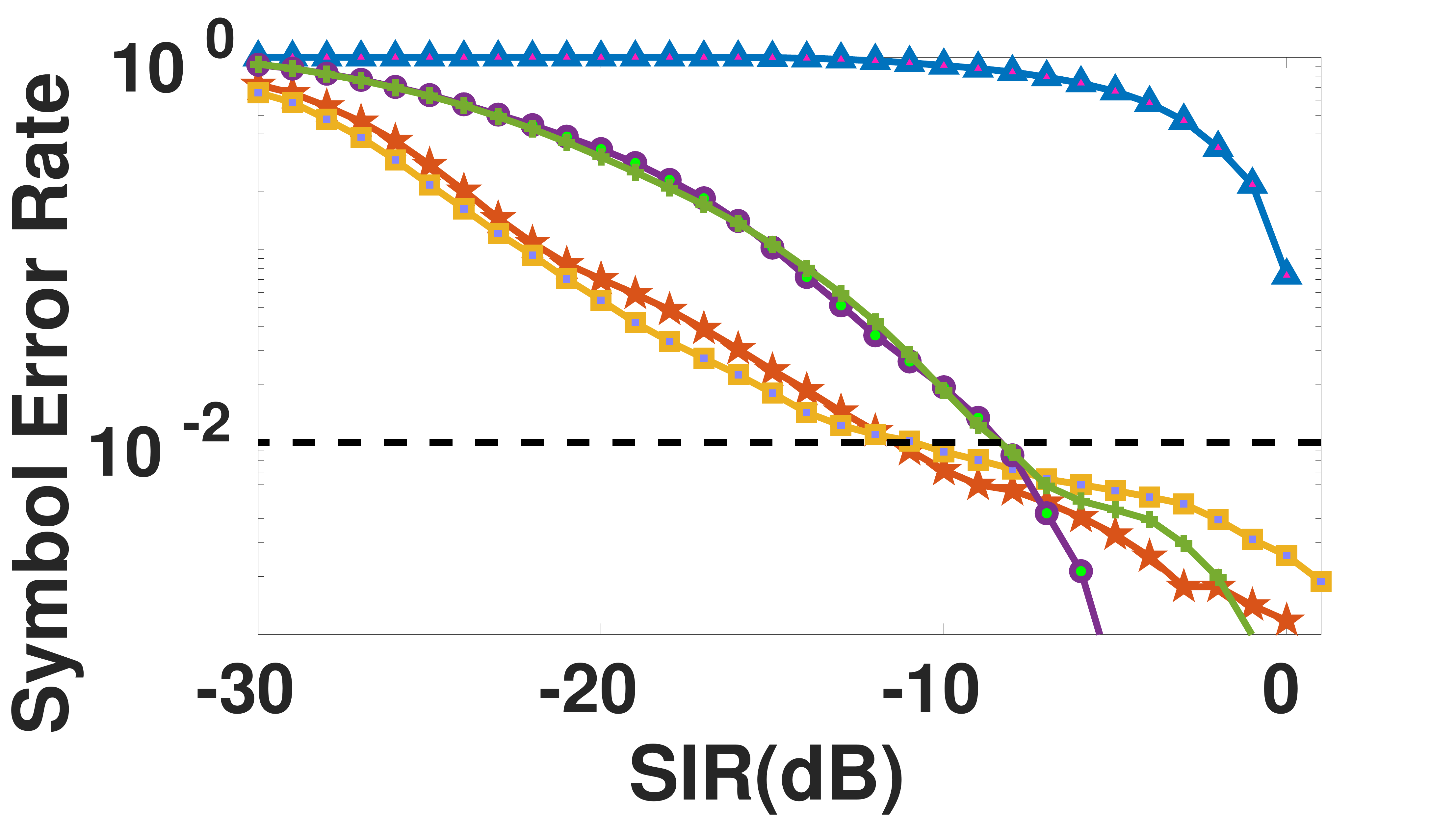}
    \label{subfig-sir-in-collisions-symbol-3}	
    }
    \hspace{-0.08\columnwidth}
    \subfigure[SER vs. symbol offset]{
    \includegraphics[width=0.535\columnwidth]{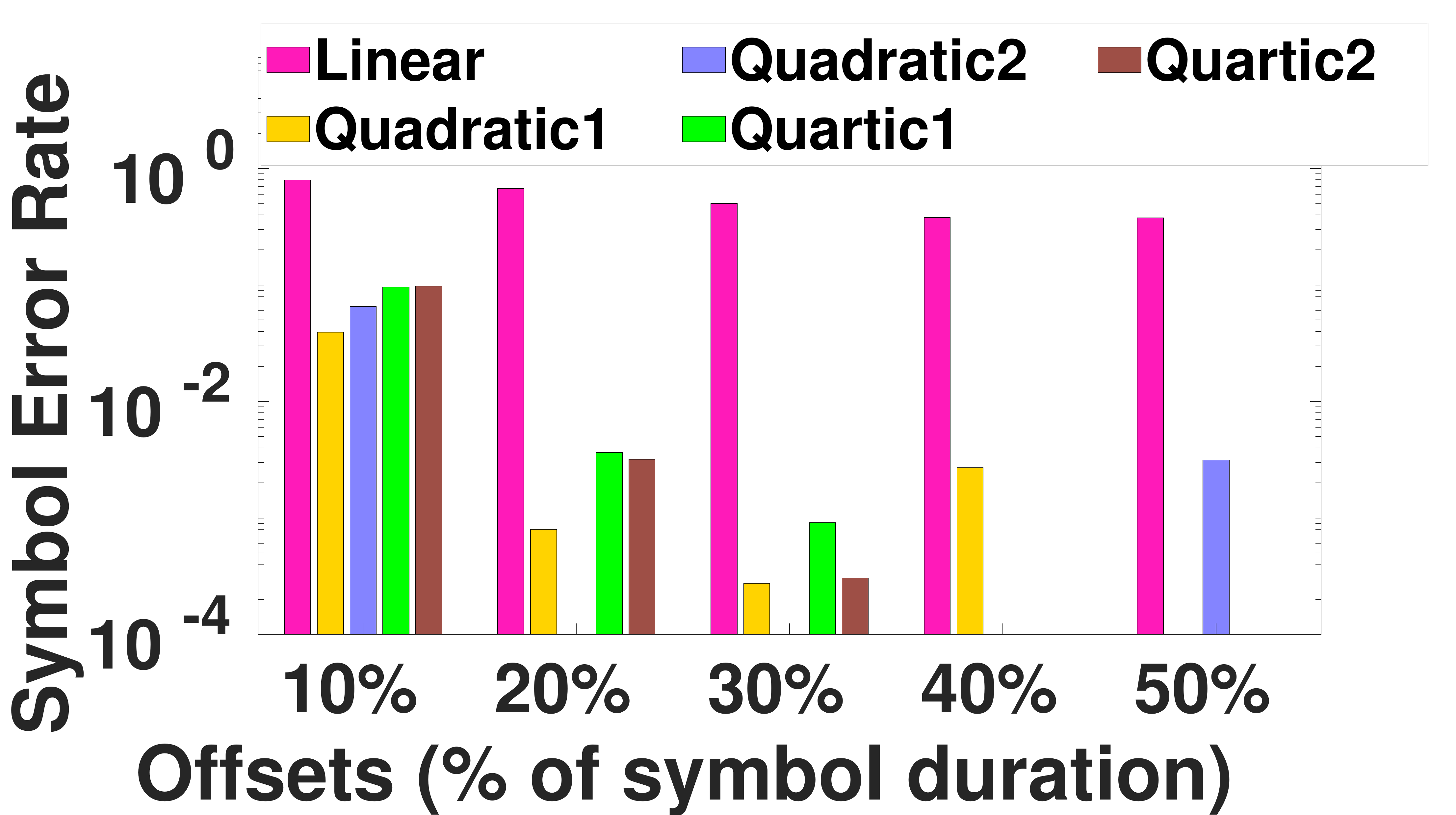}
    \label{subfig-sir-in-collisions-symbol-4}	
    }
    \caption{Linear vs. non-linear: symbol error rate ($SNR>30dB$) in different spreading factor, SIR, and symbol offset settings.}
    \label{fig-sir-in-collisions-symbol}
\end{figure*}

\section{Evaluation}
\label{sec-evaluation}

We present the experiment results in this section. \S\ref{subsec-evaluation-1} first gives an overall comparison of \ours with linear chirps at the symbol and packet level, followed by the outdoor experiments at the campus scale in \S\ref{subsec-evaluation-2}. Finally, we provide the large-scale emulation to explore the impact of concurrency on \ours in \S\ref{subsec-evaluation-3}. And indoor evaluations can be found in $\S$\ref{appendix-3}.


\subsection{Overall Comparisons with Linear Chirps}\label{subsec-evaluation-1}

\vspace{1mm}
\noindent
\textbf{Noise resilience}. We first compare the noise resilience of \ours with linear chirps in the absence of collisions.
The results are shown in Figure~\ref{subfig-noise-resilience-no-collisions-1}.
In accordance with our analysis (\S\ref{subsec-design-3}), we observe that all four types of non-linear chirps in \ours demonstrate comparable noise resilience with linear chirps across all three $SF$ settings.
Figure~\ref{subfig-noise-resilience-no-collisions-2} shows that the symbol error rate is distributed evenly over the entire code space, confirming that the non-linear chirp achieves consistent SER for different symbols.

We also compare the noise resilience of \ours with LoRaWAN in the presence of collisions.
Figure~\ref{fig-snr-in-collisions-symbol}(a)-(c) shows the SER in various SNR and spreading factor settings.
When $SF$=8, we observe that both LoRaWAN and four types of non-linear chirps fail to demodulate packets in extremely low SNR conditions (i.e., SNR$\leq$-25$dB$).
As the SNR grows to $-15dB$, the SER achieved by non-linear chirps drops dramatically to around 1\%, whereas the SER of linear chirps is still above 20\%. As the SNR grows further, we observe the SER of non-linear chirps is always 10$\times$ lower than that of the linear chirps, e.g., 0.3\% versus 3\% at SNR=30$dB$.
Similar trends hold for $SF$=10 and 12.

\vspace{1mm}
\noindent
\textbf{Symbol offset}. Next, we compare the SER of \ours and linear chirps in various symbol offset settings.
Specifically, from our collected dataset, we randomly pick up LoRa symbols with different spreading factors (SF=8,10,12) and SNRs ([$-15dB$,$15dB$]). We then vary the symbol offset between two collision symbols from 10\% to 50\% and plot their SER in Figure~\ref{fig-snr-in-collisions-symbol}(d).
In consistency with our simulation in Figure~\ref{fig-nonlinear-vs-linear-tgap}, we observe the SER of linear chirps drops with increasing symbol offsets.
In contrast, the SER achieved by non-linear chirps maintains a consistent low level, with the maximal value of 1\% when the offset of two collision symbols is merely 10\%. In contrast, the SER achieved by linear chirps spans varies from 1\% to 80\% as the symbol offset decreases.
These results demonstrate that the non-linear chirps are robust to collisions with different symbol offsets.

\vspace{1mm}
\noindent
\textbf{Resolving near-far issue}.
We also compare \ours with linear chirp-based LoRa (i.e., LoRaWAN) in the presence of the near-far issue. 
In particular, we vary the SIR of the targeting symbol and measure its SER in each SIR setting.
Figure~\ref{fig-sir-in-collisions-symbol}(a)-(c) show the results in three different spreading factor settings.
We observe the standard LoRaWAN fails to decode weak targeting signal (i.e., SIR<$0dB$) across all three $SF$ settings.
In contrast, by leveraging the power scattering effect, all four types of non-linear chirps in \ours can successively demodulate weak symbols in the presence of strong collisions.
For instance, the averaging SIR threshold for achieving less than 1\% SER is -$3.7dB$, $-7.7dB$, and $-10.2dB$ for SF=8, 10, and 12, respectively.
We further vary the symbol offset of two collision symbols and measure the SER achieved by both \ours and LoRaWAN. The SIR and SF of symbols for evaluation varies from -$10dB$ to $1dB$, and from 8 to 12, respectively.
The results are shown in Figure~\ref{subfig-sir-in-collisions-symbol-4}.
We observe \ours achieves a robust low SER (i.e., less than 1\%) in the presence of a large symbol offset. It then degrades slightly as the symbol offset decreases. In contrast, the linear chirp achieves consistently high SER (i.e., $\geq$38\%) in all five different symbol offset settings.
These results clearly demonstrate that \ours can successfully decode weak signals in the presence of strong collisions in various conditions.




\begin{figure}[t]
\centering
\subfigure{
    \includegraphics[width=0.48\columnwidth]{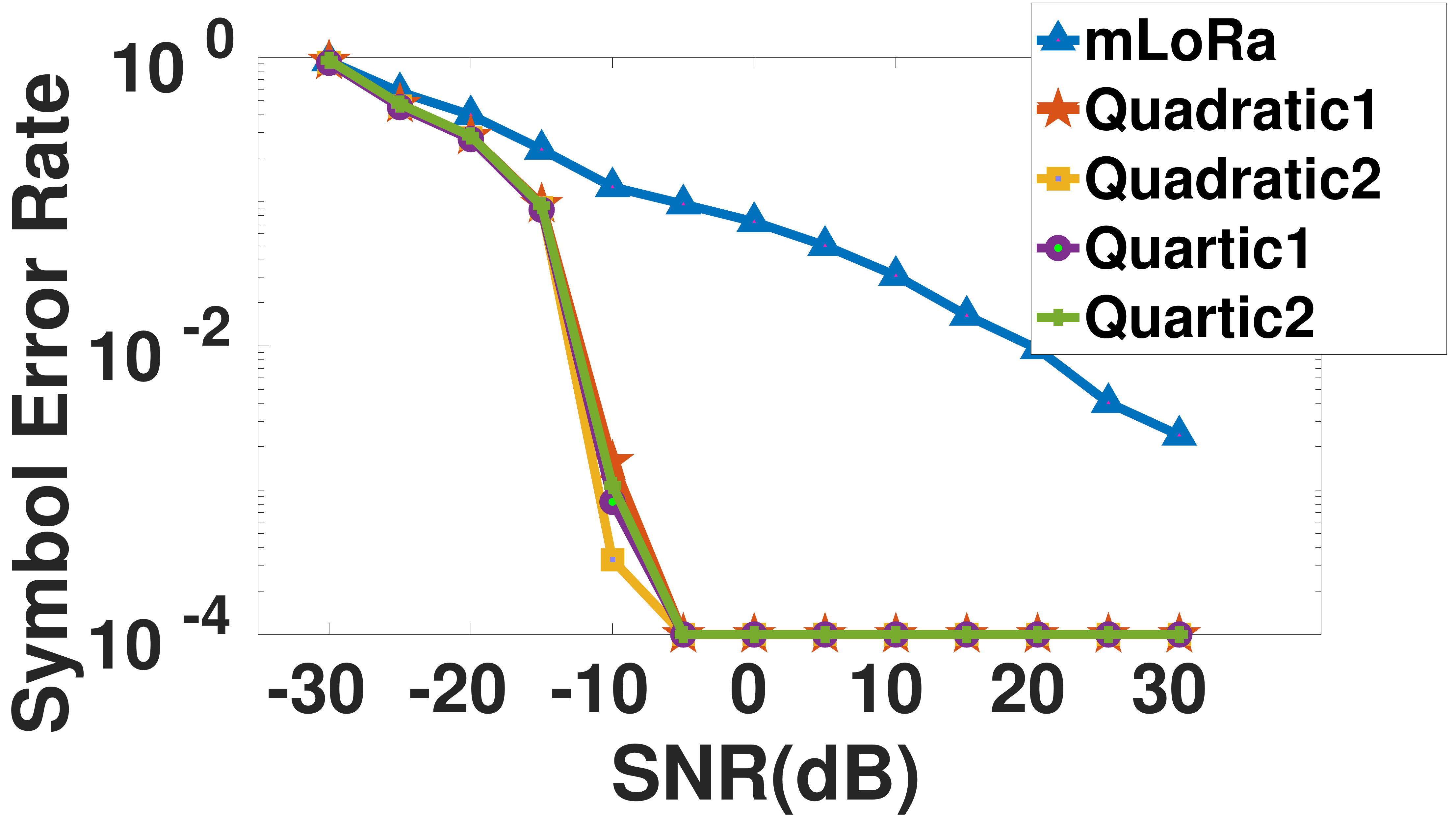}
    \label{fig-in-collisions-packet-1}
}\hspace{-0.02\columnwidth}
\subfigure{
	\includegraphics[width=0.48\columnwidth]{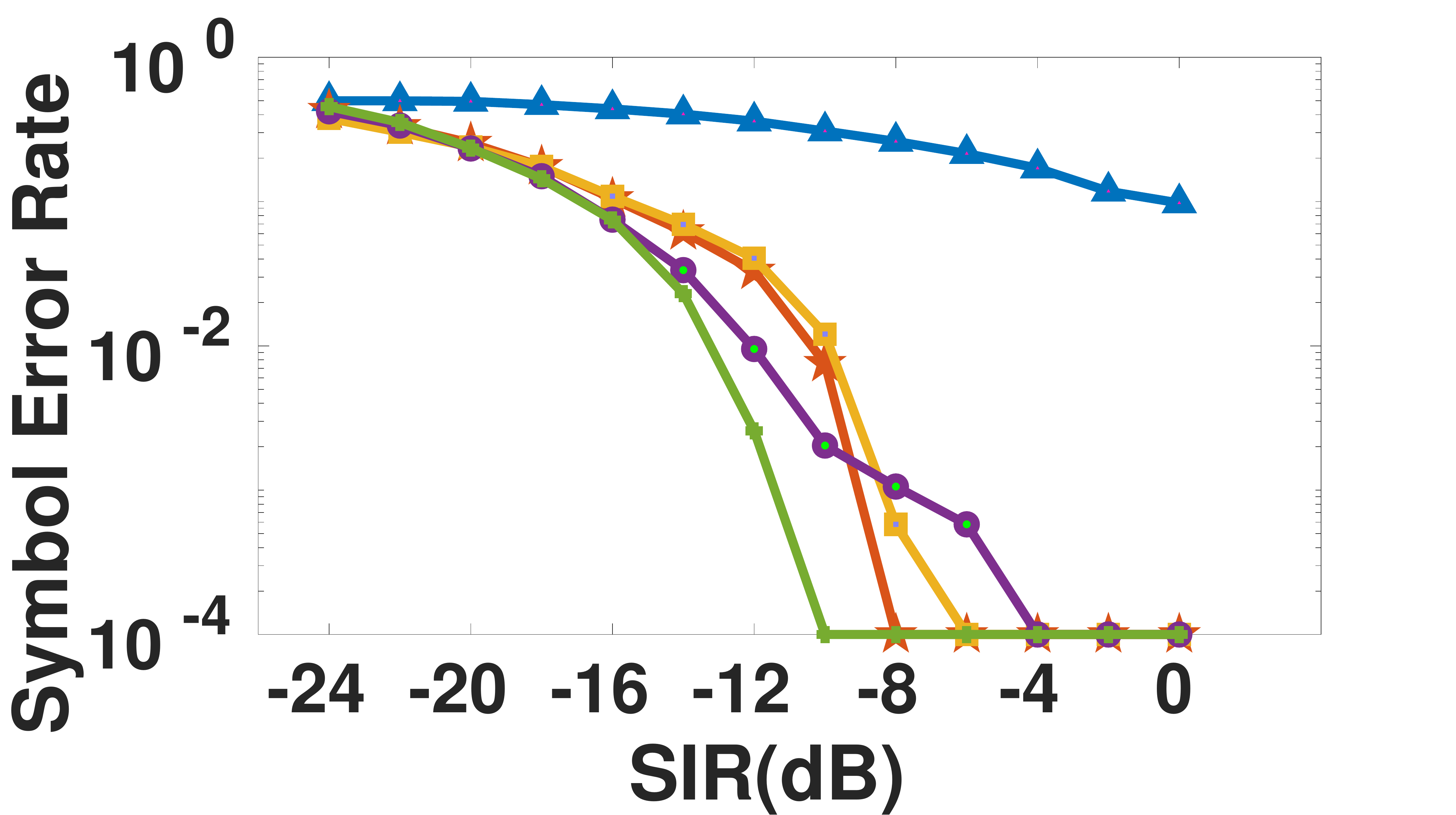}
    \label{fig-in-collisions-packet-2}
}
\caption{Head-to-head comparison with mLoRa\cite{mLoRa}.}
\label{fig-in-collisions-packet}
\vspace{-4mm}
\end{figure}

\begin{figure*}[t]
    \centering
    \subfigure{
    \includegraphics[width=0.68\columnwidth]{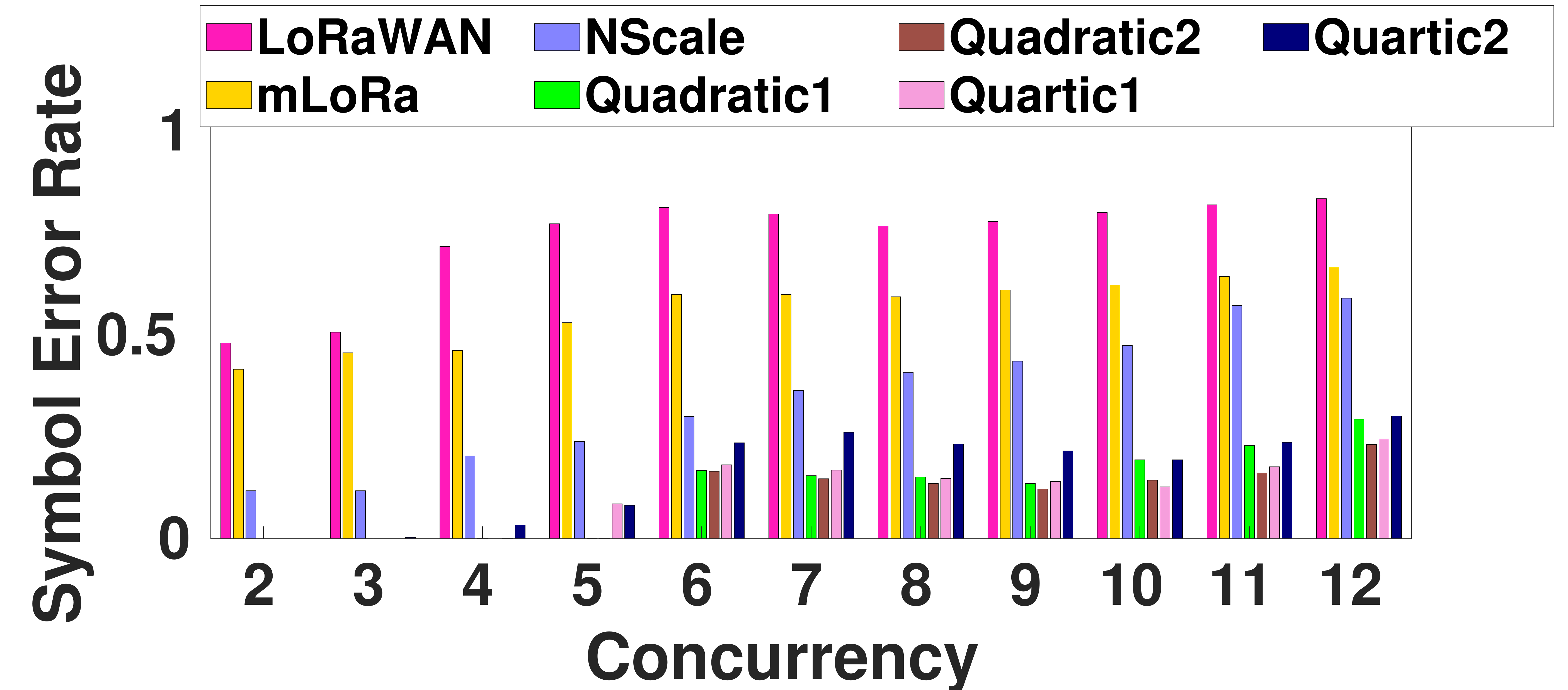}
    \label{subfig-outdoor-1}
    }
    \hspace{-0.06\columnwidth}
    \subfigure{
    \includegraphics[width=0.68\columnwidth]{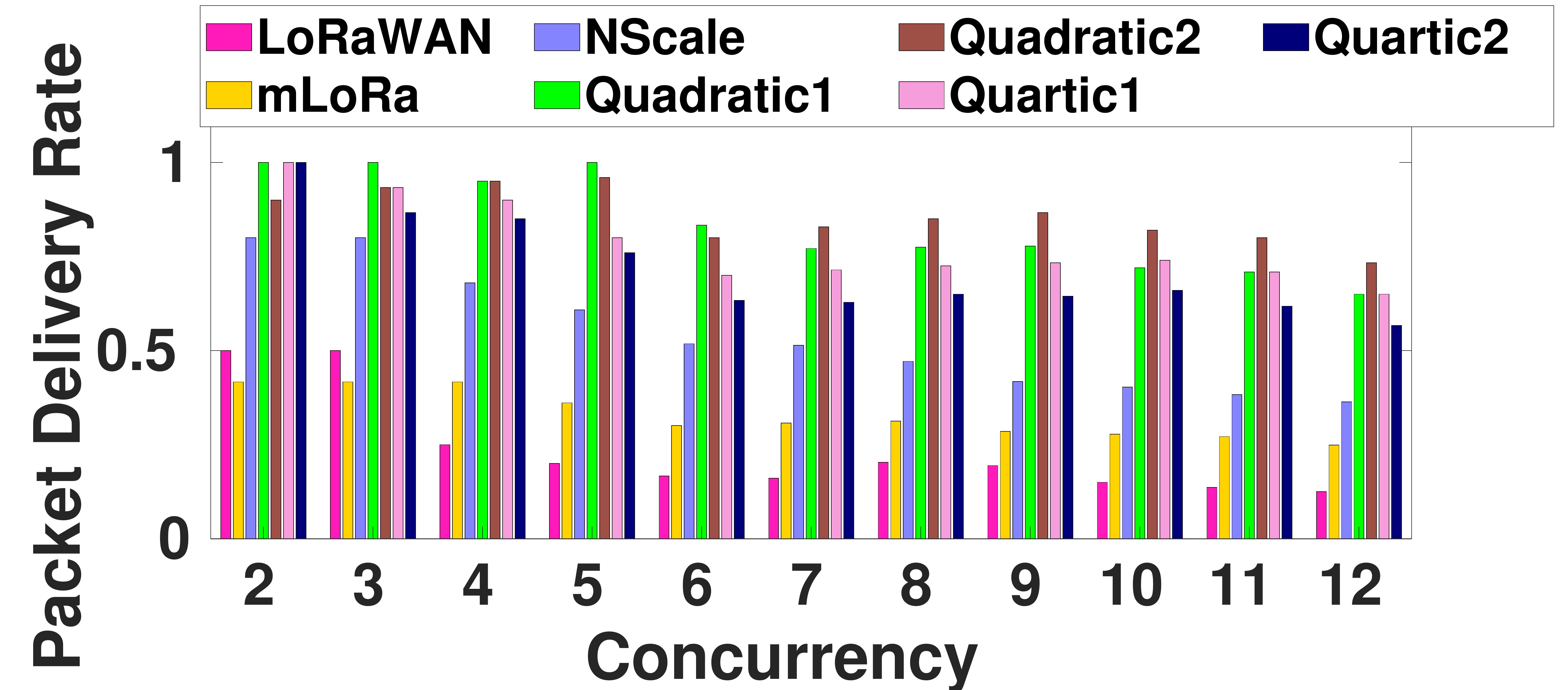}
    \label{subfig-outdoor-2}	
    }
    \hspace{-0.06\columnwidth}
    \subfigure{
    \includegraphics[width=0.68\columnwidth]{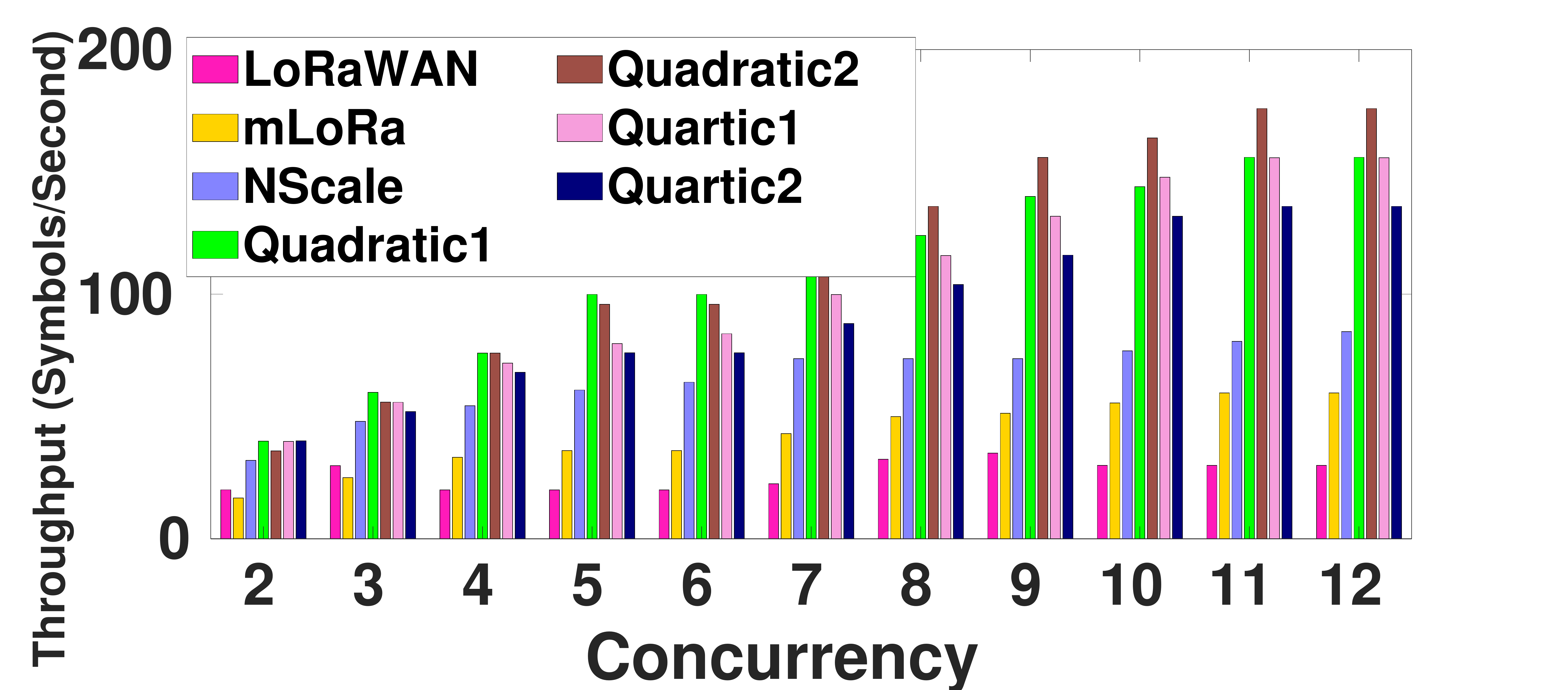}
    \label{subfig-outdoor-3}	
    }
    \caption{Outdoor experiment: examine the impact of concurrent transmissions on SER, PDR, and network throughput.}
    \label{fig-outdoor}
     \vspace{-4mm}
\end{figure*}

\vspace{1mm}
\noindent
\textbf{Head-to-head comparison with mLoRa\cite{mLoRa}}. We compare \ours with mLoRa on using our indoor dataset.
Figure~\ref{fig-in-collisions-packet-1} shows the SER of each system in the presence of two collision packets.
We observe that the SER of mLoRa drops gradually from 15\% to 1\% as the SNR grows from -$15dB$ to 15$dB$.
It finally drops to 0.3\% when SNR grows to $30dB$.
In contrast, all four types of non-linear chirps adopted by \ours achieve a consistently low SER ($\leq$0.01\%) when SNR is larger than -$15dB$.
Similarly, as shown in  Figure~\ref{fig-in-collisions-packet-2}, the SER of mLoRa is over 10\% in the presence of a strong collision (i.e., $SIR<0dB$) whereas \ours can demodulate weak signals at an $SER<1\%$ as long as the SIR is larger than -$12dB$.


\vspace{1mm}
\noindent
\textbf{Remarks}. These experiments show the advantage of \ours in dealing with near-far issues. It also reveals that the performance of \ours varies with the non-linear function being adopted.
Overall the quadratic function $f(t) = t^2$ achieves consistently better SER than the other types of non-linear functions. 
We leave the exploration of non-linear space as our future work.

\subsection{Concurrency at the Campus Scale}\label{subsec-evaluation-2}

We evaluate \ours on decoding collisions in different numbers of concurrent transmissions (termed as $N$) settings.
In particular, we measure the SER, PDR, and network throughput and compare with mLoRa~\cite{mLoRa}, NScale~\cite{tong_combating_2020}, and LoRaWAN three baselines. We repeat the experiments in indoor environments and put the results in Appendix \ref{appendix-3}.



As $N$ grows, the SER of LoRaWAN, mLoRa, and NScale all increase gradually, as shown in Figure~\ref{subfig-outdoor-1}.
Specifically, LoRaWAN can only demodulate the strongest transmission for most settings.
And mLoRa achieves a slightly better performance than LoRaWAN. However, its SER aggravates significantly ($\geq$50\%) when demodulating more than four concurrent transmissions. Besides, NScale performs better than the above two schemes, and the SER grows gradually from less than 15\% to 55\% when $N$ grows to 12.
In contrast, \ours achieves an average SER of less than 25\% in all settings.

Figure~\ref{subfig-outdoor-2} shows the packet delivery ratio achieved by these systems. We observe that as $N$ grows, the PDR achieved by \ours drops slightly from 100\% to 65\%.
In contrast, the PDR drops significantly to less than 36.5\%, 25.0\%, and 12.5\% for NScale, mLoRa, and LoRaWAN, respectively.
We further compute the network throughput and plot the results in Figure~\ref{subfig-outdoor-3}.
The overall network throughput of \ours, NScale, and mLoRa grow with the increasing $N$.
However, the network throughput of LoRaWAN manifests a converse trend due to the magnified interference as $N$ grows.
Taking a further scrutiny on this result, we find that the network throughput of \ours grows almost linearly with $N$.
In contrast, the growing trend of network throughput in both NScale and mLoRa drops gradually as $N$ grows. This is because the near-far issue grows extensively with an increasing number of concurrent transmissions. However, both NScale and mLoRa are not scaling to such circumstances. Statistically, when $N$=12, the average network throughput of \ours is 5.21$\times$, 2.61$\times$, and 1.84$\times$ higher than that of LoRaWAN, mLoRa, and NScale, respectively.

\subsection{Large-scale emulation}\label{subsec-evaluation-3} 
\begin{figure}
    \centering
    \includegraphics[width=1\columnwidth]{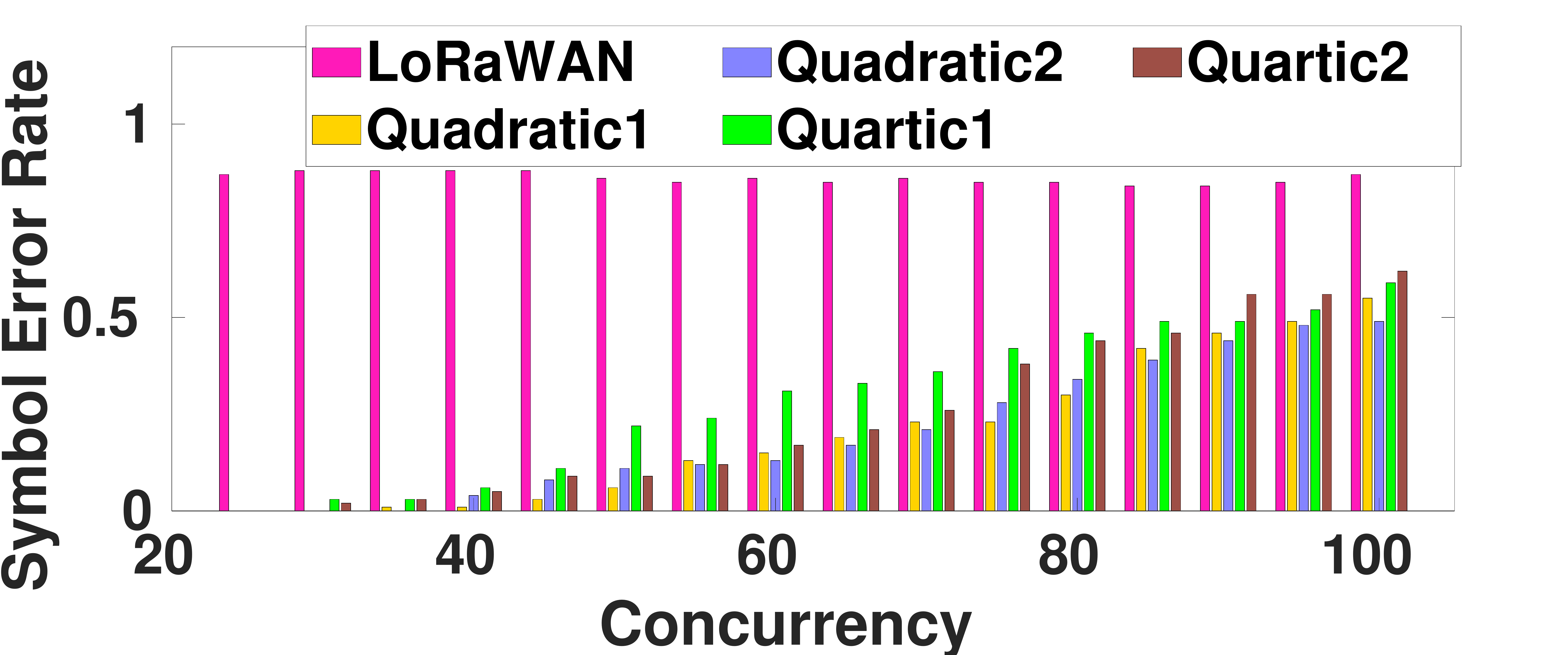}
     \vspace{-4mm}
    \caption{Emulation of large-scale collisions.}
     \vspace{-4mm}
    \label{subfig-maximum-1}
\end{figure}

We also emulate large-scale collisions using the data collected both indoors and outdoors.
Specifically, in each number of concurrent transmission settings, we only measure the SER of the weakest transmissions as those stronger transmissions are likely to be correctly demodulated.
Figure~\ref{subfig-maximum-1} shows the SER of \ours and LoRaWAN when the SIR varies randomly between [$-5dB$,$0dB$], We observe that all four types of non-linear chirps adopted by \ours can successively demodulate the weakest transmission (i.e., SER=0) when the number of concurrent transmissions is less than 30. The SER then grows up gradually as the network scales.
It peaks at 50\% when 100 transmitters work concurrently.
In contrast, the standard LoRaWAN fails to demodulate the weakest transmission when there are more than two concurrent transmissions.


\section{Conclusion}
\label{sec-conclusion}
We have presented the design, implementation, and evaluation of \ours, a PHY-layer amendment to LoRaWAN. By replacing the linear-chirp modulation on standard LoRaWAN with its non-linear chirp counterpart, the receiver can effectively demodulate large numbers of collided LoRa transmissions in extreme SNR, SIR, and symbol offset conditions.
We practice this idea by design a holistic PHY-layer and implement it on software defined radios.
The results demonstrate \ours improves the network throughput by 7.6$\times$ against the standard LoRaWAN, outperforming two state-of-the-art approaches by 1.6$\times$ and 2.8$\times$, respectively.


\let\oldbibliography\thebibliography
\renewcommand{\thebibliography}[1]{%
  \oldbibliography{#1}%
  \setlength{\itemsep}{0pt}%
}

\clearpage

\bibliographystyle{concise2}
\begin{raggedright}
\bibliography{main.bib}

\begin{thebibliography}{10}
\expandafter\ifx\csname urlstyle\endcsname\relax
  \providecommand{\doi}[1]{doi:\discretionary{}{}{}#1}\else
  \providecommand{\doi}{doi:\discretionary{}{}{}\begingroup
  \urlstyle{rm}\Url}\fi

\bibitem{abramson1970aloha}
N.~Abramson.
\newblock The aloha system: Another alternative for computer communications.
\newblock \textit{Proceedings of the November 17-19, 1970, fall joint computer
  conference}, 281--285, 1970.

\bibitem{alliance2020lorawan}
L.~Alliance.
\newblock Lorawan specification.
\newblock \url{https://lora-alliance.org/about-lorawan}.
\newblock Accessed 08-Apr-2020.

\bibitem{VERT900}
V.~Antenna.
\newblock Vert900 vertical antenna (824-960 mhz, 1710-1990 mhz) dualband.
\newblock \textit{https://www.ettus.com/all-products/vert900/}, Retrieved by
  May 10th 2021.

\bibitem{benson2015modern}
S.~R. Benson.
\newblock Modern digital chirp receiver: Theory, design and system integration,
  2015.

\bibitem{berni1973utility}
A.~Berni, W.~Gregg.
\newblock On the utility of chirp modulation for digital signaling.
\newblock \textit{IEEE Transactions on Communications}, 1973.

\bibitem{cali2000dynamic}
F.~Cal{\`\i}, M.~Conti, E.~Gregori.
\newblock Dynamic ieee 802.11: design, modeling and performance evaluation.
\newblock \textit{International Conference on Research in Networking},
  786--798. Springer, 2000.

\bibitem{LoRa_Transceiver}
C.~Chiu, Z.~Zhang, L.~T. Hsien.
\newblock The near/far effect in local aloha radio communications.
\newblock \textit{IEEE Journal of Solid-State Circuit}, 2020.

\bibitem{doerry2006generating}
A.~W. Doerry.
\newblock Generating nonlinear fm chirp waveforms for radar.
\newblock Tech. rep., Sandia National Laboratories, 2006.

\bibitem{eletreby2017empowering}
R.~Eletreby, D.~Zhang, S.~Kumar, O.~Ya{\u{g}}an.
\newblock Empowering low-power wide area networks in urban settings.
\newblock \textit{Proceedings of ACM SigComm}, 2017.

\bibitem{gamage_lmac_2020}
A.~Gamage, J.~C. Liando, C.~Gu, R.~Tan, M.~Li.
\newblock {LMAC}: efficient carrier-sense multiple access for {LoRa}.
\newblock \textit{Proceedings of ACM MobiCom}, 2020.

\bibitem{gollakota2009interference}
S.~Gollakota, S.~D. Perli, D.~Katabi.
\newblock Interference alignment and cancellation.
\newblock \textit{Proceedings of the ACM SIGCOMM}, 2009.

\bibitem{hessar2018netscatter}
M.~Hessar, A.~Najafi, S.~Gollakota.
\newblock Netscatter: Enabling large-scale backscatter networks.
\newblock \textit{Proceedings of USENIX NSDI}, 2019.

\bibitem{hosseini2019nonlinear}
N.~Hosseini, D.~W. Matolak.
\newblock Nonlinear quasi-synchronous multi user chirp spread spectrum
  signaling.
\newblock \textit{arXiv preprint arXiv:1909.09887}, 2019.

\bibitem{hosseini2021nonlinear}
---{}---.
\newblock Nonlinear quasi-synchronous multi user chirp spread spectrum
  signaling.
\newblock \textit{IEEE Transactions on Communications}, 2021.

\bibitem{hu2020sclora}
B.~Hu, Z.~Yin, S.~Wang, Z.~Xu, T.~He.
\newblock Sclora: Leveraging multi-dimensionality in decoding collided lora
  transmissions.
\newblock \textit{Proceedings of {IEEE} {ICNP}}, 2020.

\bibitem{upconverter}
T.~Instruments.
\newblock {Optimizing power consumption and power-up overshoot using
  TPS54160-Q1 family in automotive applications}.
\newblock
  \url{https://www.ti.com/lit/an/slva436a/slva436a.pdf?ts=1622474296492&ref_url=https\%253A\%252F\%252Fwww.google.com.hk\%252F}.
\newblock {Accessed 30-May-2021}.

\bibitem{khan2013performance}
M.~A. Khan, R.~K. Rao, X.~Wang.
\newblock Performance of quadratic and exponential multiuser chirp spread
  spectrum communication systems.
\newblock \textit{Proceedings of IEEE SPECTS}, 2013.

\bibitem{dds_SRA1}
K.-R. Kim, S.~Kim, C.-H. Ki, T.-H. Kim, H.~Yang, J.-H. Kim.
\newblock {Development and comparison of DDS and multi-DDS chirp waveform
  generator}.
\newblock \textit{Proceeding of IEEE International Geoscience and Remote
  Sensing Symposium (IGARSS)}, 2019.

\bibitem{lesnik2008modification}
C.~Lesnik, A.~Kawalec.
\newblock Modification of a weighting function for nlfm radar signal designing.
\newblock \textit{Acta Physica Polonica A}, \textbf{114}(6-A), 2008.

\bibitem{liando2019known}
J.~C. Liando, A.~Gamage, A.~W. Tengourtius, M.~Li.
\newblock Known and unknown facts of lora: experiences from a large-scale
  measurement study.
\newblock \textit{ACM Transactions on Sensor Networks}, \textbf{15}(2), 16,
  2019.

\bibitem{dds_fmcw2}
L.~Lou, Z.~Fang, B.~Chen, T.~Guo, Z.~Liu, Y.~Zheng.
\newblock {A DDS-driven ADPLL chirp synthesizer with ramp-interpolating
  linearization for FMCW radar application in 65nm CMOS}.
\newblock \textit{Proceeding of IEEE International Symposium on Circuits and
  Systems (ISCAS)}, 2018.

\bibitem{dds_fmcw1}
B.~Mohring, C.~Moroder, U.~Siart, T.~Eibert.
\newblock {Broadband, fast, and linear chirp generation based on DDS for FMCW
  radar applications}.
\newblock \textit{Proceeding of IEEE Radar Conference (RadarConf)}, 2019.

\bibitem{mollanoori2013uplink}
M.~Mollanoori, M.~Ghaderi.
\newblock Uplink scheduling in wireless networks with successive interference
  cancellation.
\newblock \textit{IEEE Transactions on Mobile Computing}, 2013.

\bibitem{url:gnu-radio}
G.~R. project.
\newblock Gnu radio website.
\newblock \url{http://www.gnuradio.org}.
\newblock {Accessed 07-Apr-2020}.

\bibitem{FPGA}
F.~provider.
\newblock Zynq-7000 fpga.
\newblock
  \url{https://www.renesas.com/sg/en/application/technologies/fpga-power/zynq-7000}.
\newblock {Accessed 30-May-2021}.

\bibitem{analog_frequency}
A.~Rokita.
\newblock {Direct analog synthesis modules for an X-band frequency source}.
\newblock \textit{Proceeding of 12th International Conference on Microwaves and
  Radar}, 1998.

\bibitem{shahid2021concurrent}
M.~O. Shahid, M.~Philipose, K.~Chintalapudi, S.~Banerjee, B.~Krishnaswamy.
\newblock Concurrent interference cancellation: decoding multi-packet
  collisions in lora.
\newblock \textit{Proceedings of ACM SIGCOMM}, 503--515, 2021.

\bibitem{colora}
T.~Shuai, X.~Zhenqiang, W.~Jiliang.
\newblock Colora: Enable muti-packet reception in lora.
\newblock \textit{Proceedings of IEEE INFOCOM}, 2020.

\bibitem{talla2017lora}
V.~Talla, M.~Hessar, B.~Kellogg, A.~Najafi, J.~R. Smith, S.~Gollakota.
\newblock Lora backscatter: Enabling the vision of ubiquitous connectivity.
\newblock \textit{Proceedings of the ACM on Interactive, Mobile, Wearable and
  Ubiquitous Technologies}, \textbf{1}(3), 1--24, 2017.

\bibitem{tong_combating_2020}
S.~Tong, J.~Wang, Y.~Liu.
\newblock Combating packet collisions using non-stationary signal scaling in
  {LPWANs}.
\newblock \textit{Proceedings of ACM MobiSys}, 2020.

\bibitem{tse2005fundamentals}
D.~Tse, P.~Viswanath.
\newblock \textit{Fundamentals of wireless communication}.
\newblock Cambridge university press, 2005.

\bibitem{dds_tutorials}
D.~tutorials.
\newblock {Fundamentals of direct digital synthesis (DDS)}.
\newblock
  \url{https://www.analog.com/media/en/training-seminars/tutorials/MT-085.pdf}.
\newblock {Accessed 30-May-2021}.

\bibitem{wang2015non}
Q.~Wang.
\newblock \textit{Non-Linear Chirp Spread Spectrum Communication Systems of
  Binary Orthogonal Keying Mode}.
\newblock Ph.D. thesis, The University of Western Ontario, 2015.

\bibitem{mLoRa}
X.~Wang, L.~Kong, L.~He, G.~Chen.
\newblock mlora: A multi-packet reception protocol in lora networks.
\newblock \textit{Proceedings of IEEE ICNP}, 2019.

\bibitem{xia2019ftrack}
X.~Xia, Y.~Zheng, T.~Gu.
\newblock Ftrack: parallel decoding for lora transmissions.
\newblock \textit{Proceedings of ACM Sensys}, 2019.

\bibitem{xu_fliplora_2020}
Z.~Xu, S.~Tong, P.~Xie, J.~Wang.
\newblock {FlipLoRa}: {Resolving} {Collisions} with {Up}-{Down}
  {Quasi}-{Orthogonality}.
\newblock \textit{Proceedings of {IEEE} {International} {Conference} on
  {Sensing}, {Communication}, and {Networking} ({SECON})}, 2020.

\bibitem{dds_SRA2}
H.~Yang, S.-B. Ryu, H.-C. Lee, S.-G. Lee, S.-S. Yong, J.-H. Kim.
\newblock {Implementation of DDS chirp signal generator on FPGA}.
\newblock \textit{Proceeding of International Conference on Information and
  Communication Technology Convergence (ICTC)}, 2014.

\bibitem{LoSee}
Y.~Yao, Z.~Ma, Z.~Cao.
\newblock Losee: Long-range shared bike communication system based on lorawan
  protocol.
\newblock \textit{Proceedings of ACM EWSN}, 2019.

\bibitem{oct}
W.~Zhe, K.~Linghe, X.~Kangjie, H.~Liang, W.~Kaishun, C.~Guihai.
\newblock Online concurrent transmissions at lora gateway.
\newblock \textit{Proceedings of IEEE INFOCOM}, 2020.

\end{thebibliography}
\end{raggedright}
\appendix
\begin{figure}[!t]
    \centering
    \subfigure{
    \includegraphics[width=0.68\columnwidth]{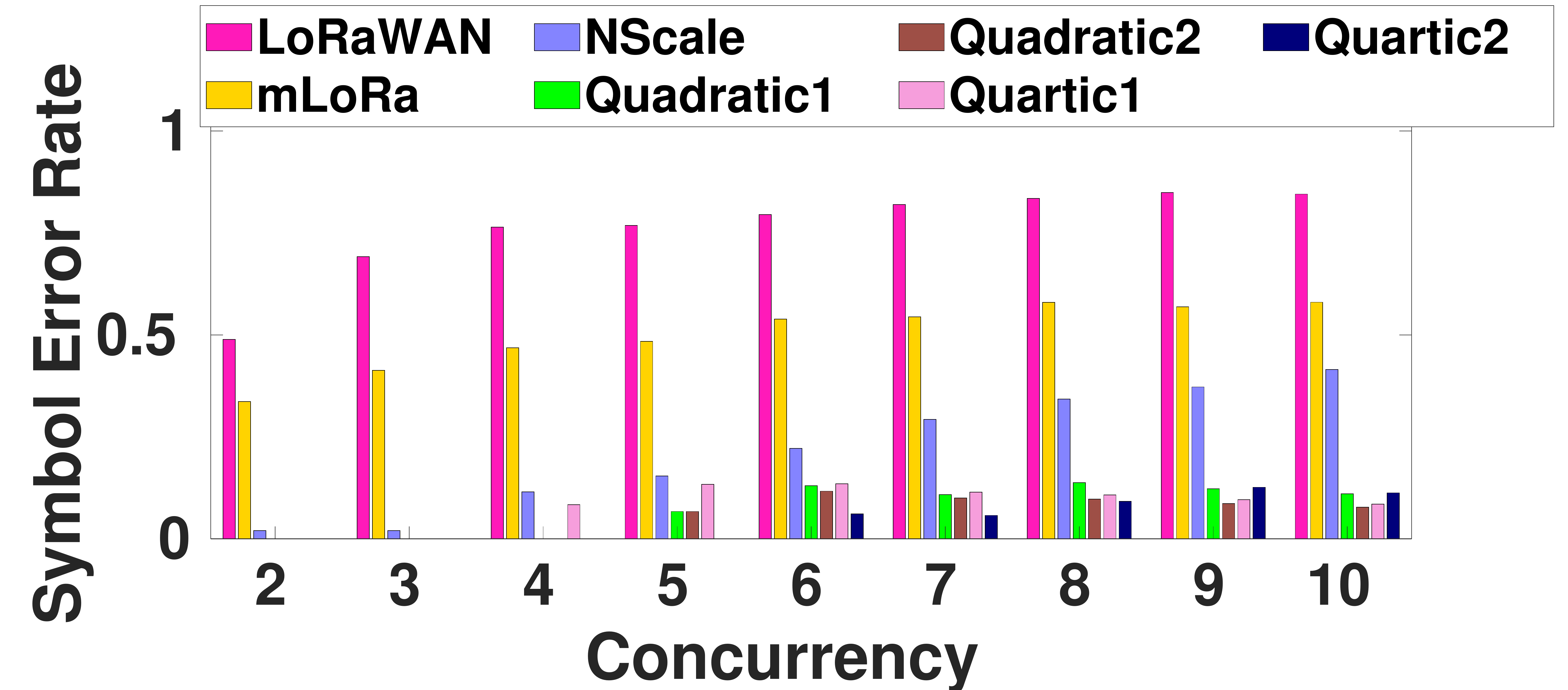}
    \label{subfig-indoor-1}
    }\\
    \hspace{-0.06\columnwidth}
    \subfigure{
    \includegraphics[width=0.68\columnwidth]{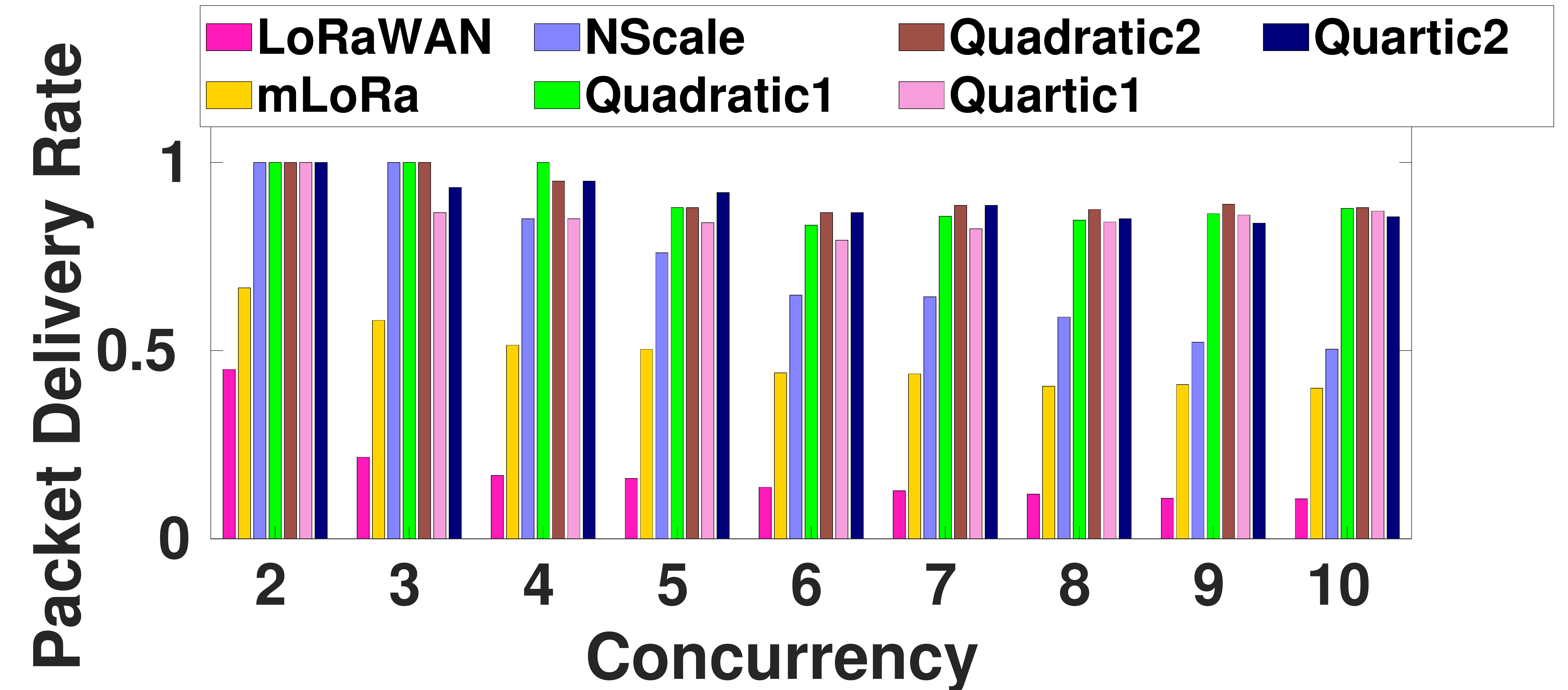}
    \label{subfig-indoor-2}	
    }\\
    \hspace{-0.06\columnwidth}
    \subfigure{
    \includegraphics[width=0.68\columnwidth]{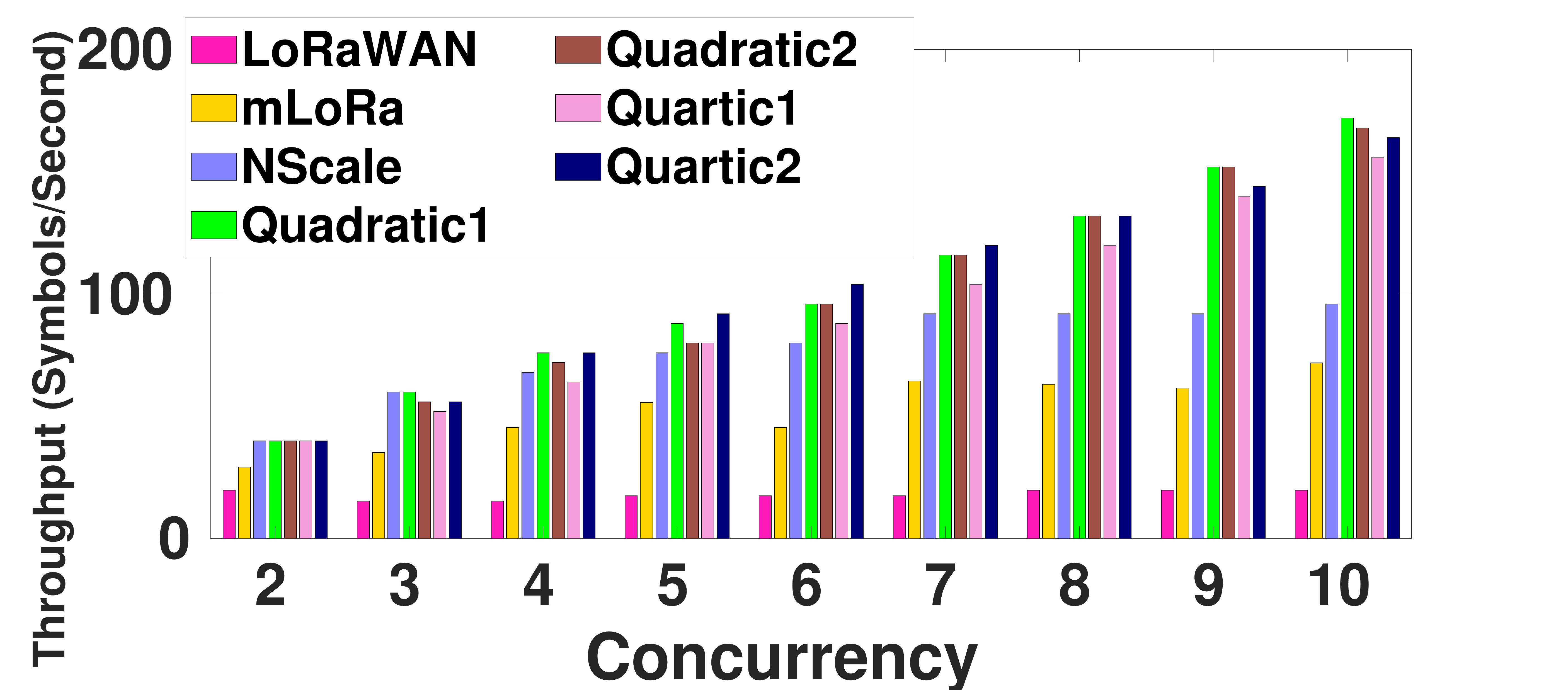}
    \label{subfig-indoor-3}	
    }
    \vspace{-4mm}
    \caption{Indoor experiment: examine the impact of concurrent transmissions on SER, PDR, and network throughput.}
    \label{fig-indoor}
    \vspace{-4mm}
\end{figure}

\section{Energy scattering effect}
\label{appendix-2}
Upon the reception of a non-linear chirp symbol, the receiver multiplies it with the base down-chirp:
\begin{equation*}
    e^{j 2 \pi (f_0+f_c(t+t_{gap})) t}*e^{-j 2 \pi f_c(t) t}=e^{j 2 \pi F(t)t}\label{equ-dechirp}\\
\end{equation*}
where $t_{gap}$ denotes the symbol offset between the incident chirp symbol and the FFT window (i.e., the base down-chirp); $f_0$ represents the initial frequency offset of this non-linear chirp.
The spectral energy is determined by the term $F(t)$.
For a non-linear chirp, e.g., $f_{c}(t) = t^2$, we have:
\begin{align}
    F(t)&=f_0+k_2(t+t_{gap})^2+k_0-(k_2t^2+k_0)\nonumber\\
& =f_0+k_2t^2_{gap}+2k_2t_{gap}\times t
\end{align}
where $k_0$ and $k_2$ represent the coefficients to fit the non-linear curve into the range of symbol time and bandwidth (Equation~\eqref{equ-nonlinear-baseline-up-chirp}).
When the incident chirp is not well aligned with the down-chirp (i.e., $t_{gap}!=0$), from the above equation we can find that the spectrum energy of this incident chirp will spread to multiple FFT bins.
In contrast, when $t_gap=0$, we have $F(t)=f_0$, indicating the spectrum energy will converge a single frequency point $f_0$.

\section{Concurrency in the Indoor Environment}\label{appendix-3}
Similar to the SER trend of outdoor experiments, we observe a huge SER gap between \ours and its competitors as $N$ grows in indoor experiments (Figure~\ref{subfig-indoor-1}).
On the other hand, compared with outdoor experiments, we find that all systems in indoor spaces achieve slightly lower SER, with up to 7.75\%-11.26\% when $N$=10.
This is because the transmitters are facing less severe near-far issue indoors.
The packet delivery rate in indoor experiments also shows the similar trend with their outdoor counterparts, as shown in Figure~\ref{subfig-indoor-2}. Specifically, the PDR achieved by \ours drops slightly from 100\% to 87.10\% on average as $N$ grows from 2 to 10.
While both mLoRa and LoRaWAN drops significantly from around 50.5\% to less than 40.0\% and 10.6\%, respectively.
Figure~\ref{subfig-indoor-3} shows the network throughput achieved by these three systems in indoor environment.
The overall network throughput of \ours grows up consistently as the number of transmitters scales up. When 10 packets collide simultaneously with significant power difference, the network throughput of \ours in average is about 1.6$\sim$7.6$\times$ higher than the network throughput achieved by NScale, mLoRa and the standard LoRaWAN.

\end{document}